\documentclass[12pt,a4paper,final]{iopart}

\usepackage{iopams}  
\usepackage{graphicx}
\usepackage{hyperref}
\usepackage{cleveref}
\begin{document}

\title[Einstein-$\Lambda$ flow]{Attractors of the `n+1' dimensional Einstein-$\Lambda$ flow}

\author{Puskar Mondal$^{*}$}

\begin{abstract}
Here we prove a global existence theorem for sufficiently small however fully nonlinear perturbations of a family of background solutions of the $`n+1$' dimensional vacuum Einstein equations in the presence of a positive cosmological constant $\Lambda$. The future stability of vacuum solutions in the small data and zero cosmological constant limit has been studied previously for both $`3+1$' and higher dimensional spacetimes. However, with the advent of dark energy driven accelerated expansion of the universe, it is of fundamental importance in mathematical cosmology to include a positive cosmological constant, the simplest form of the dark energy for the vacuum Einstein equations. Such Einsteinian evolution is here designated as the `Einstein-$\Lambda$' flow. We study the background solutions of this `Einstein-$\Lambda$' flow in $`n+1$' dimensional spacetimes in constant mean curvature spatial harmonic gauge, $n\geq3$ and establish both linear and non-linear stability of such solutions. In the cases of number of spatial dimensions being strictly greater than $3$, the finite dimensional Einstein moduli spaces form the center manifolds of the dynamics. A suitable shadow gauge condition \cite{andersson2011einstein} is implemented in order to treat these cases. In addition, the autonomous character of the suitably re-scaled Einstein flow breaks down as a consequence of including $\Lambda(>0)$. We construct a Lyapunov function (controlling a suitable norm of the small data) similar to a wave equation type energy for the non-linear non-autonomous evolution of the small data and prove its decay in the direction of cosmological expansion utilizing the structure of the non-autonomous terms and smallness assumption on the data. Our results demonstrate the future stability and geodesic completeness of the perturbed spacetimes, and show that the scale-free geometry converges to an element of the space of constant negative scalar curvature metrics sufficiently close to and containing the Einstein moduli space (a point for $n=3$ and a finite dimensional space for $n>3$), which has significant consequences for the cosmic topology while restricting to the case of $n=3$.        
\end{abstract}
Since the Einstein equations formulated as a Cauchy problem leave the spatial topology of the universe unrestricted, a natural question arises whether one may constrain the topology through studying the dynamics of the Einsteinian evolution, while also satisfying the cosmological principle. Recent articles \cite{ashtekar2015general, moncrief2019could} unfolded such a possibility i.e., a dynamical mechanism at work within the Einstein flow (both with and without $\Lambda$) which strongly suggests that many closed 3-manifolds that do not admit a locally homogeneous and isotropic metric at all (and thus are incompatible with the cosmological principle) will nevertheless evolve under Einsteinian evolution to be asymptotically compatible with the observed, approximate, spatial homogeneity and isotropy of the universe. Results of these studies, based on the monotonic decay of a weak Lyapunov function namely the reduced Hamiltonian, suggested that a 3-manifold of negative Yamabe type, if it contains parts supporting hyperbolic metrics in its connected sum decomposition, will be volume dominated by these hyperbolic components asymptotically by the Einstein flow (Einstein-$\Lambda$ flow for $\Lambda\neq0$). On the other hand, if one takes the cosmological principle literally, the so-called FLRW model restricts the choice of global spatial topology of the universe to a small set consisting of negatively curved hyperbolic space $\mathbf{H}^{3}$, flat Euclidean space $\mathbf{E}^{3}$, positively curved 3-sphere $\mathbf{S}^{3}$ with its canonical round metric and its two fold quotient $\mathbf{PR}^{3}=\mathbf{S}^{3}/Z_{2}$. However, the astronomical observations that motivate the cosmological principle are necessarily limited to a fraction (possibly small) of the entire universe and such observations are compatible with spatial metrics being locally but not globally homogeneous and isotropic. Once the restriction on the topology (a global property) is removed, numerous closed manifolds may be constructed as the quotients of $\mathbf{H}^{3}, \mathbf{E}^{3}$, and $\mathbf{S}^{3}$ by discrete, proper, and torsion free subgroups of their respective isometry groups, with each satisfying the local homogeneity and isotropy criteria but no longer being globally homogeneous or isotropic. In order for these topologically rich spatially compact spacetimes to be possible candidates for the cosmological models, it is crucial to study the asymptotic behaviour of the fully nonlinear perturbations of these models. Non-linear stability of these spacetimes clearly opens up the possibility for the universe to have an exotic spatial topology. In this paper, we study the linear (which acts as a motivation towards studying non-linear stability) and non-linear stability of the spacetimes with spatial part being negative Einstein for the cases of $n\geq 3$ (hyperbolic for $n=3$). However, we will assume certain smallness condition on the fully nonlinear perturbations.\\ 
The family of background solutions (a class of fixed points of the Einstein-$\Lambda$ flow) designated as \textit{`conformal spacetimes'} (due to the fact that each of these spacetimes admits a timelike conformal Killing field) with the spacetime topology $\mathbf{R}\times M$ ($M$ being the spatial manifold) in constant mean curvature spatial harmonic gauge (CMCSH), may be written in the following warped product form 
\begin{eqnarray}
\label{eq:background}
\hat{g}=-\frac{n^{2}}{(\tau^{2}-\frac{2n\Lambda}{n-1})^{2}}d\tau\otimes d\tau+\frac{1}{(\tau^{2}-\frac{2n\Lambda}{n-1})}\gamma_{ij}dx^{i}\otimes dx^{j},
\end{eqnarray}
where $\gamma$ is an Einstein metric satisfying $R_{ij}(\gamma)=-\frac{n-1}{n^{2}}\gamma_{ij}$ and $\tau\in (-\infty,-\sqrt{\frac{2n\Lambda}{n-1}})$ is the mean extrinsic curvature of $M$ in the globally hyperbolic spacetime on $\mathbf{R}\times M$. For the special case of $n=3$, the negative Einstein spaces are hyperbolic i.e., the Einstein moduli space reduces to a point. In the limit of $\Lambda=0$, \cite{andersson2011einstein} calls these spacetimes the Lorentz cone space times (these are constructed by taking quotient of the interior of the future light cone in Minkowski space by $SO^{+}(3,1)$ in case of $n=3$). Stability of these `3+1' Lorentz cone spacetimes was proven by \cite{andersson2004future} utilizing the Bel-Robinson energy. In the more general setting of $n>3$, a finite dimensional space of Einstein metrics provides the `center manifold' towards which the re-scaled spatial metric is flowing in the limit of infinite cosmological expansion. \cite{andersson2011einstein} proved the stability of these background solutions by invoking a so called shadow gauge condition and later utilized a wave equation type of energy for the sufficiently small however fully nonlinear perturbations. This energy acts like a Lyapunov function for these perturbations (and, in particular vanishes at fixed points), which is defined to control the desired norm of perturbations. In the case of the `2+1' Einstein flow on $\mathbf{R}\times \Sigma_{genus}$ with $genus>1$, Teichm\"uller space plays the role of the Einstein moduli space and special techniques \cite{anderson1997global, moncrief2007relativistic} were used to study the global existence (by utilizing the properness of the Dirichlet energy functional defined on the Teichm\"uller space). Recently \cite{fajman2015stable} studied the Lyapunov stability of these background solutions including a positive cosmological constant. However, in order to establish the `attractor' property of certain solutions, it is necessary to prove their asymptotic stability. \cite{fischer2002hamiltonian, moncrief2019could} constructed a reduced phase space as the cotangent bundle of the higher dimensional analogue of the Teichm\"uller space and obtained the following true Hamiltonian of the dynamics while expressed as a functional of the reduced phase space variables through a conformal technique
\begin{eqnarray}
\label{eq:reduced}
H_{reduced}&:=&\frac{2(n-1)}{n}\int_{M}\frac{\partial \tau}{\partial t}\mu_{g}.
\end{eqnarray}
They have shown that the reduced Hamiltonian acting as a Lyapunov function decays along certain solutions of the Einstein equations and achieves its infimum precisely for the background Lorentz cone spacetimes (conformal spacetimes for $\Lambda\neq0$). Such a property provides a notion of the stability of these background solutions for arbitrarily large perturbations. However, the Lyapunov function only controls the $H^{1}\times L^{2}$ norm of the reduced data and therefore, such a notion of stability is weak. Motivated by these results, we intend to study the stability of these background solutions for sufficiently small fully nonlinear perturbations in the case when a positive consmological constant is included in the Einstein equations. A subtlety is that the properties of the wave equation type energy exploit the information about the lowest eigenvalue of the Lichnerowicz type Laplacian (acting on the space of symmetric (0,2) tensors on $M$, that is, $S^{0}_{2}(M)$) which enters into the evolution equation.\\   
In this paper, we consider the complete generality of the problem in a framework of sufficiently small however fully nonlinear perturbations of the background solutions. However, the inclusion of a positive cosmological constant introduces several seemingly restrictive features of the field equations. A few examples may be seen as follows. In the CMCSH gauge, the vacuum Einstein equations with $\Lambda=0$ are non-autonomous due to the fact that the mean extrinsic curvature acting as time explicitly appears in the equations. However, after a suitable re-scaling, the equations can be made to be autonomous. In the presence of a nonzero $\Lambda$, such a property is lost. This is one of the major differences from the case of $\Lambda=0$. Nevertheless, one may still obtain estimates necessary to prove the decay property of a suitably defined Lyapunov function by introducing a Newtonian like time co-ordinate. In particular, we take the advantage of the structure of the Einstein's equations. The explicitly time dependent terms appear in the field equation in such a way as to help drive the flow towards a class of fixed points. Roughly speaking, the potentially problematic terms in the expression for the time derivative of the energy either gets cancelled point-wise or are multiplied by asymptotically decaying factor $e^{-T}$ ($T\in(-\infty,\infty)$). Therefore, one needs to keep track of the explicitly time dependent terms accurately to reach the conclusion of energy decay. This subtlety does not arise in the case of vanishing cosmological constant and one may easily obtain the decay of the suitably defined Lyapunov function by introducing a correction factor (see \cite{andersson2011einstein} for detail). In the process of controlling the energy decay, invoking the shadow gauge in order to handle the nontrivial moduli space (such moduli space is assumed to have a smooth structure and be stable) becomes necessary. In the $\Lambda=0$ case, the scale-free geometry converges to the Einstein structure and therefore the attractors of the Einstein flow is identified to be the Einstein spaces. However, in the presence of $\Lambda>0$, the scale-free geometry exponentially converges to space of metrics with constant negative scalar curvature sufficiently close to and containing the Einstein structure. In other words, if we start the evolution with an arbitrary metric $g$ sufficiently close to the Einstein moduli space such that the difference is $L^{2}$ orthogonal to the moduli space then the conformal geometry converges in infinite time to an element of the space of metrics with constant negative scalar curvature sufficiently close to and containing the Einstein structure. On the other hand perturbations tangential to the Einstein moduli space exponentially decays such that the metric converges to another element of the moduli space, that is, tangential perturbations exhibit trivial asymptotic stability. Therefore, the attractor of the Einstein-$\Lambda$ flow is identified to be the space of constant negative scalar curvature metrics lying sufficiently close to and containing the Einstein moduli space. In addition, the only requirement for the asymptotic stability of these spacetimes is the stability of the negative Einstein structure i.e., the non-negativity of the eigen spectrum of the associated Lichnerowicz type Laplacian operator. These are some major differences between the $\Lambda=0$ and $\Lambda>0$ cases. The asymptoically stable physical conformal spacetimes are given by the following warp product metric 
\begin{eqnarray}
\label{eq:background2}
\hat{g}=-\frac{n^{2}}{(\tau^{2}-\frac{2n\Lambda}{n-1})^{2}}d\tau\otimes d\tau+\frac{1}{(\tau^{2}-\frac{2n\Lambda}{n-1})}\gamma^{\dag}_{ij}dx^{i}\otimes dx^{j},
\end{eqnarray}
where $R(\gamma^{\dag})=-\frac{n-1}{n}$. Our strategy would be to perturb the fixed point solutions described by the negative Einstein spaces and employ the shadow gauge to treat the evolution of the perturbations orthogonal and tangential to the Einstein moduli space. Using necessary estimates, we then prove that the spatial metric exponentially converges (in appropriate norm topology) to a point of the space of constant negative scalar curvature metrics sufficiently close to and containing the moduli space. 
In summary, the structure of the paper is the following. We start with the gauge fixed Einstein-$\Lambda$ equations and state the necessary theorems ensuring local well-posedness of the Cauchy problem. Next, we move on to computing background solutions and study their linear stability. After obtaining a series of estimates utilizing the elliptic equations arising as a result of gauge fixing and imposing the shadow gauge condition, we construct a Lyapunov function for the small data. In the last part, utilizing the obtained estimates, we prove the decay of the constructed energy functional (which vanishes only for the spacetimes of type (\ref{eq:background}) and remains constant for the spacetimes of type (\ref{eq:background2})) thereby establishing the stability of the background solutions.

\section{Notations and facts}
We denote the `$n+1$' dimensional ($n\geq3$) spacetime manifold by $\tilde{M}$ with its topology being $\mathbf{R}\times M$, $M$ being the $n$-dimensional spatial slice diffeomorphic to a Cauchy hypersurface. The space of Riemannian metrics on $M$ is denoted by $\mathcal{M}$. $\mathcal{M}_{-\frac{n-1}{n}}$ is defined as follows 
\begin{eqnarray}
\mathcal{M}_{-\frac{n-1}{n}}:=\{g\in \mathcal{M}| R(g)=-\frac{n-1}{n}\},\nonumber
\end{eqnarray}
with $R(g)$ being the scalar curvature associated with $g$. $R_{ijkl}[g]$ and $\Gamma[g]^{i}_{jk}$ denote the Riemann curvature and connection coefficients with respect to the metric $g$, respectively. In terms of the function space of fields (metric, second fundamental form etc.), we work in the $L^{2}$ (with respect to a given metric) Sobolev space $W^{s,2}$ for $s>\frac{n}{2}+2$, also denoted by $H^{s}$. We denote the $L^{2}$ inner product between two 2-tensors on $M$ with respect to a background metric $\gamma$ as 
\begin{eqnarray}
<u|v>_{L^{2}}=\int_{M}u_{ij}v_{kl}\gamma^{ik}\gamma^{jl}\mu_{g}\nonumber
\end{eqnarray}
and the inner product on derivatives as
\begin{eqnarray}
<\nabla[\gamma]u|\nabla[\gamma]v>_{L^{2}}&=&\int_{M}\nabla[\gamma]_{m}u_{ij}\nabla[\gamma]_{n}v_{kl}g^{mn}\gamma^{ik}\gamma^{jl}\mu_{g}\nonumber,
\end{eqnarray} 
where $\mu_{g}$ is the volume form associated with $g\in\mathcal{M}$
\begin{eqnarray}
\mu_{g}=\sqrt{\det(g_{ij})}dx^{1}\wedge dx^{2}\wedge dx^{3}\wedge.........\wedge dx^{n}.\nonumber
\end{eqnarray}
Abusing notation, we use $\mu_{g}$ to denote both the volume form as well as $\sqrt{\det(g_{ij})}$. The rough Laplacian $\Delta_{g,\gamma}$ acting on a vector bundle (symmetric covariant 2-tensors are sections of this bundle) over $(M,g)$ is defined as 
\begin{eqnarray}
\Delta_{g,\gamma}h_{ij}:=-\frac{1}{\mu_{g}}\nabla[\gamma]_{m}(g^{mn}\mu_{g}\nabla[\gamma]_{n}h_{ij}).
\end{eqnarray}
This rough Laplacian is self-adjoint with respect to the $L^{2}$ inner product on covariant 2-tensors. Using the rough Laplacian, a self-adjoint Lichnerowicz type Laplacian which will be crucial later is defined as follows
\begin{eqnarray}
\mathcal{L}_{g,\gamma}h_{ij}:=\Delta_{g,\gamma}h_{ij}-2R[\gamma]_{i}~^{k}~_{j}~^{l}h_{kl}.
\end{eqnarray}
We may sometimes drop the Sobolev index to simplify the notation. The reader is expected to assume the function space for ($g,k,N,X$) to be $H^{s}\times H^{s-1}\times H^{s+1}\times H^{s+1}$ with $s>\frac{n}{2}+2$.
The Laplacian $\Delta_{g}$ is defined so as to have a non-negative spectrum i.e., \begin{eqnarray}
\Delta_{g}\equiv-g^{ij}\nabla_{i}\nabla_{j}.
\end{eqnarray}
For $a,b \in R_{>0}$, $a\lesssim b$ is defined to be $a\leq Cb$ for some constant $0<C<\infty$. The spaces of symmetric covariant 2-tensors and vector fields on $M$ are denoted by $S^{0}_{2}(M)$ and $\mathfrak{X}(M)$, respectively.
\section{Field equations and gauge fixing} 
The ADM formalism splits the spacetime described by an `n+1' dimensional Lorentzian manifold $\tilde{M}$ into $\mathbf{R}\times M$ with each level set $\{t\}\times M$ of the time function $t$ being an orientable n-manifold diffeomorphic to a Cauchy hypersurface (assuming the spacetime to be globally hyperbolic) and equipped with a Riemannian metric. Such a split may be implemented by introducing a lapse function $N$ and shift vector field $X$ belonging to suitable function spaces and defined such that
\begin{eqnarray}
\partial_{t}&=&N\hat{n}+X
\end{eqnarray}
with $t$ and $\hat{n}$ being time and a hypersurface orthogonal future directed timelike unit vector i.e., $\hat{g}(\hat{n},\hat{n})=-1$, respectively. The above splitting puts the spacetime metric $\hat{g}$ in local coordinates $\{x^{\alpha}\}_{\alpha=0}^{n}=\{t,x^{1},x^{2},....,x^{n}\}$ into the form 
\begin{eqnarray}
\label{eq:spacetime}
\hat{g}&=&-N^{2}dt\otimes dt+g_{ij}(dx^{i}+X^{i}dt)\otimes(dx^{j}+X^{j}dt)
\end{eqnarray} 
where $g_{ij}dx^{i}\otimes dx^{j}$ is the induced Riemannian metric on $M$. In order to describe the embedding of the Cauchy hypersurface $M$ into the spacetime $\tilde{M}$, one needs the information about how the hypersurface is curved in the ambient spacetime. Thus, one needs the second fundamental form $k$ defined as 
\begin{eqnarray}
k_{ij}&=-\frac{1}{2N}(\partial_{t}g_{ij}-(L_{X}g)_{ij}),
\end{eqnarray} 
the trace of which ($\tau=g^{ij}k_{ij}$, $g^{ij}\frac{\partial}{\partial x^{i}}\otimes \frac{\partial}{\partial x^{j}}:=g^{-1}$) is the mean extrinsic curvature of $M$ in $\tilde{M}$.  Here $L$ denotes the Lie derivative operator.
 The vacuum Einstein equations with a cosmological constant $\Lambda$ 
 \begin{equation}
 \label{eq:ein}
 R_{\mu\nu}(\hat{g})-\frac{1}{2}R(\hat{g})\hat{g}_{\mu\nu}+\Lambda \hat{g}_{\mu\nu}=0 
 \end{equation}
may now be expressed as the evolution and Gauss and Codazzi constraint equations of $g$ and $k$
\begin{eqnarray}
\label{eq:evol1}
\partial_{t}g_{ij}=-2Nk_{ij}+(L_{X}g)_{ij},\\
\partial_{t}k_{ij}=-\nabla_{i}\nabla_{j}N+N(R_{ij}+\tau k_{ij}-2k^{k}_{i}k_{jk}-\frac{2\Lambda}{n-1}g_{ij})+(L_{X}k)_{ij},\\
\label{eq:HC}
2\Lambda=R(g)-|k|^{2}_{g}+(tr_{g}k)^{2},\\
0=\nabla^{i}k_{ij}-\nabla_{j}\tau,
\label{eq:cons}
\end{eqnarray}
where $\tau=tr_{g}k$.
A solution to the Einstein evolution and constraint equations is a curve $t\mapsto (g(t),k(t),N(t),X(t))$ in $H^{s}\times H^{s-1}\times H^{s+1}\times H^{s+1}$ (at least in our case where the local existence theorem holds in this function space)   satisfying equations (\ref{eq:evol1})-(\ref{eq:cons}). The spacetime metric $\tilde{g}$ given in terms of $(g,N,X)$ by (\ref{eq:spacetime}) solves the Einstein equation (\ref{eq:ein}) if and only if $(g,k,N,X)$ solves the evolution and constraint equations (\ref{eq:evol1})-(\ref{eq:cons}). However, the system (\ref{eq:evol1})-(\ref{eq:cons}) is not hyperbolic. We may reduce the system to a canonical hyperbolic evolution equation for $g$ by fixing gauge. The physical concept of gauge fixing (spatial and temporal) may be described as follows. Let us first consider the spatial gauge. With the spacetime topology of $\mathbf{R}\times M$, one has the freedom to choose the spatial slice as long as it is diffeomorphic to a Cauchy hypersurface. Let $M$ be a Cauchy hypersurface with an induced metric $g$ which together with $(k,N,X)$ satisfies the Einstein evolution and constraint equations (\ref{eq:evol1}-\ref{eq:cons}). Now let $\phi$ ($t-$independent) be an element of the identity component of the diffeomorphism group ($\mathcal{D}_{0}$) of $M$. Then $(\phi^{-1})^{*}g, (\phi^{-1})^{*}k, (\phi^{-1})^{*}N=N\circ \phi^{-1}$, and $\phi_{*}X$ solves the Einstein equations as well, where $^{*}$, and $_{*}$ denote the pullback and push-forward operations on the cotangent and tangent bundles of $M$, respectively. This is obvious due to the spatial covariant nature of the evolution and constraint equations. More generally, let the identity component of the diffeomorphism group act on $M$ by a time dependent element $\phi_{t}$. The evolution equation (\ref{eq:evol1}) under the action of $\phi_{t}$ reads 
\begin{eqnarray}
\partial_{t}((\phi^{-1}_{t})^{*}g)_{ij}=-2(\phi^{-1}_{t})^{*}(Nk)_{ij}+(L_{\phi_{t*}X}(\phi^{-1}_{t})^{*}g)_{ij},\\\nonumber
((\phi^{-1}_{t})^{*}\partial_{t}g)_{ij}+(\partial_{t}(\phi^{-1}_{t})^{*}g)_{ij}=-2(\phi^{-1}_{t})^{*}(Nk)_{ij}\\\nonumber
~~~~~~~~~~~~~~~~~~~~~~~~~~~~~~~~~~~~~~~~+\frac{\partial}{\partial s}((\phi^{-1}_{t}\Psi^{X}_{s}\phi_{t})^{*}(\phi^{-1}_{t})^{*}g)_{ij}|_{s=0},\\\nonumber
((\phi^{-1}_{t})^{*}\partial_{t}g)_{ij}+(\partial_{s}(\phi^{-1}_{t+s})^{*}g)_{ij}|_{s=0}=-2(\phi^{-1}_{t})^{*}(Nk)_{ij}+(\phi^{-1}_{t})^{*}(L_{X}g)_{ij},\\\nonumber
(\phi^{-1}_{t})^{*}\partial_{t}g_{ij}+(\phi^{-1}_{t})^{*}(L_{Y}g)_{ij}=-2(\phi^{-1}_{t})^{*}(Nk)_{ij}+(\phi^{-1}_{t})^{*}(L_{X}g)_{ij},\\\nonumber
(\phi^{-1}_{t})^{*}\left\{\partial_{t}g_{ij}=-2Nk_{ij}+(L_{X-Y}g)_{ij}\right\}.
\end{eqnarray}
Here $Y$ is the vector field associated with the flow $\phi_{t}$ and $\Psi^{X}_{s}$ is the flow of the shift vector field $X$.
A similar calculation for the evolution equation for the second fundamental form shows that if we make a trasformation $X\mapsto X+Y$, the Einstein evolution and constraint (due to their natural spatial covariance nature) equations are satisfied by the transformed fields. 
The choice of spatial hypersurface is fixed by choosing constant mean extrinsic curvature spatial harmonic gauge. Constant mean extrinsic curvature gauge defines a time function and therefore it is the temporal gauge choice. We briefly describe the gauge fixing below starting with spatial harmonic gauge. Let $\phi: (M,g)\to (M,\gamma)$ be a harmonic map. Clearly it satisfies the Euler-Lagrange equations arising from criticality of the associated Dirichlet energy $\frac{1}{2}\int_{M}g^{ij}\frac{\partial \phi^{k}}{\partial x^{i}}\frac{\partial \phi^{l}}{\partial x^{j}}\gamma_{kl}\mu_{g}$ i.e., 
\begin{eqnarray}
g^{ij}\left(\frac{\partial^{2}\phi^{k}}{\partial x^{i}\partial x^{j}}-\Gamma[g]_{ij}^{l}\frac{\partial \phi^{k}}{\partial x^{l}}+\Gamma[\gamma]_{\alpha\beta}^{k}\frac{\partial \phi^{\alpha}}{\partial x^{i}}\frac{\partial \phi^{\beta}}{\partial x^{j}}\right)=0.
\end{eqnarray}
Now, we fix the gauge by imposing the condition that $\phi=id$, which leads to the following equation 
\begin{eqnarray}
\label{eq:sh}
-g^{ij}(\Gamma[g]_{ij}^{k}-\hat{\Gamma}[\gamma]_{ij}^{k})&=&0.
\end{eqnarray}
where $\hat{\Gamma}[\gamma]_{ij}^{k}$ is the connection with respect to some arbitrary background Riemannian metric $\gamma$. Choice of this spatial harmonic gauge yields an elliptic equation for the shift vector field $X$ after time differentiating equation (\ref{eq:sh}). The spatial harmonic slice is chosen to have uniform mean extrinsic curvature i.e.,  
\begin{eqnarray}
\partial_{i}\tau&=&0,
\end{eqnarray}
and thus $\tau$ may play the role of time i.e.,
\begin{eqnarray}
t=monotonic~function~of~\tau,
\end{eqnarray}
and in this case, we choose $t=\tau$.
Choice of the Constant mean extrinsic curvature gauge (CMC) yields an elliptic equation for the lapse $N$. Note that we do not have evolution equations for the lapse and shift. However, they are constrained by the elliptic equations obtained through gauge fixing which together with the evolution equations for $g$ and $k$ comprises the full `Einstein-$\Lambda$' system 
\begin{eqnarray}
\label{eq:full1}
\partial_{t}g_{ij}=-2Nk_{ij}+(L_{X}g)_{ij},\\
\label{eq:full2}
\partial_{t}k_{ij}=-\nabla_{i}\nabla_{j}N+N(R_{ij}+\tau k_{ij}-2k^{k}_{i}k_{jk}-\frac{2\Lambda}{n-1}g_{ij}-\alpha_{ij})\\\nonumber{}+(L_{X}k)_{ij},\\
\label{eq:full3}
\frac{\partial \tau}{\partial t}=\Delta_{g}N+(k_{ij}k^{ij}-\frac{2\Lambda}{n-1})N,\\
\label{eq:full4}
\Delta_{g}X^{i}-R^{i}_{j}X^{j}+L_{X}V^{i}=(\nabla^{i}N)\tau-2\nabla^{j}Nk^{i}_{j}+(2Nk^{jk}\\\nonumber
-2\nabla^{j}X^{k})(\Gamma[g]^{i}_{jk}-\hat{\Gamma}[\gamma]^{i}_{jk})-g^{jk}\partial_{t}\hat{\Gamma}[\gamma]^{i}_{jk},
\end{eqnarray}
where $\alpha_{ij}=\frac{1}{2}(\nabla_{i}V_{j}+\nabla_{j}V_{i})$ and $-V^{i}$ is the tension field defined as 
\begin{eqnarray}
V^{i}&=&g^{jk}(\Gamma[g]_{jk}^{i}-\hat{\Gamma}[\gamma]_{jk}^{i}).
\end{eqnarray}
 In addition, we also have the constraints 
\begin{eqnarray}
2\Lambda&=&R(g)-|k|^{2}_{g}+(tr_{g}k)^{2},\\
0&=&\nabla^{i}k_{ij},
\end{eqnarray}  
which are conserved throughout the term of evolution as a consequence of the Bianchi identity. $V^{i}=0$ essentially corresponds to the spatial harmonic gauge.
This Cauchy problem with constant mean extrinsic curvature and spatially harmonic gauge is referred to as `\textbf{CMCSH Cauchy}' problem. 

\subsection{\textbf{local well-posedness and gauge conservation}}
\cite{andersson2003elliptic} proved a well-posedness theorem for the Cauchy problem for a family of elliptic-hyperbolic systems that included the `$n+1$' dimensional vacuum Einstein equations in CMCSH gauge. \cite{moncrief2019could} sketched how to apply the theorem of \cite{andersson2003elliptic} to a gauge fixed system of `Einstein-$\Lambda$' field equations. Since the `Einstein-$\Lambda$' field  equations only differ from the vacuum equations by the addition of some rather innocuous linear terms, most of the technicalities of this extended application of their theorem are straightforward to verify. There are however a couple of subtle points involving the elliptic equations for the lapse function and the shift vector field. Firstly, inclusion of $\Lambda>0$~seemingly creates an obstruction to achieving a trivial kernel for the lapse equation (\ref{eq:full3}). However, note that we are primarily interested in negative Yamabe manifolds (see \cite{ashtekar2015general} and \cite{moncrief2019could} for the relevant definitions) and therefore the scalar curvature $R(g)$ can never be positive everywhere on $M$ yielding the following range of allowed mean extrinsic curvature (for cosmologically expanding solutions) by virtue of the Hamiltonian constraint
\begin{eqnarray}
\label{eq:conditionlapse}
-\infty<\tau<-\sqrt{\frac{2n\Lambda}{n-1}}.
\end{eqnarray}     
This condition indeed guarantees a unique positive solution of the lapse equation (\ref{eq:full3}). Secondly, allowing for the time dependent behavior of the background metric (a negative Einstein metric in our case) introduces the extra term `$-g^{jk}\partial_{t}\hat{\Gamma}[\gamma]^{i}_{jk}$' in the elliptic equation for the shift vector field. However, our primary concern is the small perturbations about the background and the term `$-g^{jk}\partial_{t}\hat{\Gamma}[\gamma]^{i}_{jk}$' acts as small perturbation (see lemma (3) and (4) for the relevant estimates). Therefore, the extra term in the shift equation due to time dependence of the background spatial metric does not affect the existence and uniqueness results. In a sense, these previous studies together complete the desired local well-posedness for the `Einstein-$\Lambda$' system. For this reason we shall mostly refer the reader to the relevant sections of \cite{andersson2003elliptic}, \cite{andersson2011einstein}, and \cite{moncrief2019could} rather than reiterate the detailed arguments herein. The most important point to note is that the local existence theorem provides the time of existence in terms of the size of the initial data. Therefore, in the global existence argument, if the appropriate norm of the perturbation is bounded, one may immediately use to local existence to obtain the desired result.\\  
In addition to proving the local well-posedness of the `Einstein-$\Lambda$' quasi hyperbolic evolution equations, we also need to ensure the conservation of gauges and constraints i.e., whenever $(g,k,N,X)$ solve the `Einstein-$\Lambda$' equations (\ref{eq:full1}-\ref{eq:full4}), the following entities, if vanishing initially, are zero along the solution curve
\begin{eqnarray}
A&=&\tau-t,\\
V^{i}&=&g^{jk}(\Gamma[g]^{i}_{jk}-\hat{\Gamma}[\gamma]^{i}_{jk}),\\
F&=&R(g)+\tau^{2}-|k|^{2}-\nabla_{i}V^{i}-2\Lambda,\\
D_{i}&=&\nabla_{i}\tau-2\nabla^{k}k_{ki}.
\end{eqnarray}
One may show by direct calculation using the modified evolution equations (\ref{eq:full1}-\ref{eq:full2}) that the set of constraint and gauge entities $(A, V^{i}, F, D_{i})$ satisfy exactly the same induced evolution equations as those given in equations (4.4a-d) in \cite{andersson2003elliptic}. Thus the energy argument in section 4 of this reference goes through unchanged and shows that if $(A, F, V^{i}, D_{i})=0$ for the initial data ($g(t_{0}), k(t_{0})$), then $(A, F, V^{i}, D_{i})\equiv0$ along the solution curve $(g(t),k(t),N(t),X(t))$. This completes the analysis of the desired local well-posedness and gauge conservation criteria.      

\section{Re-scaled equations}
In this section, we convert the evolution and constraint equations to scale free equations after rescaling the dimensionful entities by suitable powers of the conformal factor $\phi^{2}=\tau^{2}-\frac{2n\Lambda}{n-1}$ (which is strictly positive according to the condition (\ref{eq:conditionlapse})). Before rescaling, we observe that the solution of the momentum constraint 
\begin{eqnarray}
\nabla^{j}k_{ij}=0,
\end{eqnarray}
may be written as 
\begin{eqnarray}
k=K^{TT}+\frac{\tau}{n}g,
\end{eqnarray}
where $K^{TT}$ is traceless with respect to g. We will obtain equations in terms of $K^{TT}$. We denote the dimensional entities by a $\tilde{ }$ sign, while dimensionless entities are written simply without $\tilde{ }$ sign for convenience. The re-scaled entities are given as follows
\begin{eqnarray}
\label{eq:scaling}
\tilde{g}_{ij}=\frac{1}{\phi^{2}}g_{ij},~\tilde{N}=\frac{1}{\phi^{2}}N,~\tilde{X}^{i}=\frac{1}{\phi}X^{i},~\tilde{K}^{TT}_{ij}=\frac{1}{\phi}K^{TT}_{ij},
\end{eqnarray}
where $\phi=-\sqrt{\tau^{2}-\frac{2n\Lambda}{n-1}}$ such that $\frac{\phi}{\tau}>0$.
In CMCSH gauge, the re-scaled evolution and constraint equations may be written as
\begin{eqnarray}
\label{eq:timedef}
\partial_{T} \tau=-\frac{\phi^{2}}{\tau},\\
\label{eq:gd1}
\partial_{T}g_{ij}=\frac{2\phi(\tau)}{\tau}NK^{TT}_{ij}-2(1-\frac{N}{n})g_{ij}-\frac{\phi(\tau)}{\tau}(L_{X}g)_{ij},\\
\label{eq:fd1}
\partial_{T}K^{TT}_{ij}=-(n-1)K^{TT}_{ij}-\frac{\phi(\tau)}{\tau}N(R_{ij}+\frac{n-1}{n^{2}}g_{ij}-\alpha_{ij})\\\nonumber
+\frac{\phi(\tau)}{\tau}\nabla_{i}\nabla_{j}N+\frac{2\phi(\tau)}{\tau}\nonumber NK^{TT}_{im}K^{Tm}_{j}\\
-\frac{\phi(\tau)}{n\tau}(\frac{N}{n}-1)g_{ij}-(n-2)(\frac{N}{n}-1)K^{TT}_{ij}-\frac{\phi(\tau)}{\tau}(L_{X}K^{TT})_{ij},\nonumber
\end{eqnarray}
\begin{eqnarray}
\label{eq:con1}
R+\frac{n-1}{n}-|K^{TT}|^{2}&=&0,\\
\label{eq:con2}
\nabla_{j}K^{TTij}&=&0.
\end{eqnarray}
Here the new time coordinate is defined as
\begin{eqnarray}
\label{eq:time}
\partial_{T}&=&-\frac{\phi^{2}}{\tau}\partial_{\tau}
\end{eqnarray}
which may be integrated explicitly to yield 
\begin{eqnarray}
\label{eq:ex1}
\phi(\tau)&=&-e^{-T},\\
\label{eq:ex2}
\tau&=&-\sqrt{e^{-2T}+\frac{2n\Lambda}{n-1}}
\end{eqnarray}
with $T$ being Newtonian like i.e., $-\infty<T<\infty$. Note an important fact that $0<\frac{\phi(\tau)}{\tau}<1$ and $\frac{\phi(\tau)}{\tau}\approx \sqrt{\frac{n-1}{2n\Lambda}}e^{-T}$ as $T\to\infty$. 
The re-scaled elliptic equations for the lapse function and the shift vector field may be expressed as follows
\begin{eqnarray}
\label{eq:lapseE}
\Delta_{g}N+(|K^{TT}|^{2}+\frac{1}{n})N=1,\\\nonumber
\label{eq:shiftE}
\frac{\phi(\tau)}{\tau}(\Delta_{g}X^{i}-R^{i}_{j}X^{j}+L_{X}V^{i})=\frac{\phi(\tau)}{\tau}(2NK^{Tjk}-2\nabla^{j}X^{k})(\Gamma[g]^{i}_{jk}\\\nonumber{}-\Gamma[\gamma]^{i}_{jk})-(2-n)\nabla^{i}(\frac{N}{n}-1)
-\frac{2\phi(\tau)}{\tau}\nabla^{j}NK^{Ti}_{j}+g^{jk}\partial_{T}\Gamma[\gamma]^{i}_{jk}.
\end{eqnarray}  

\subsection{\textbf{Background solutions: conformal spacetimes}}
The fixed point solutions are computed as the solutions of the following set of equations in CMCSH gauge i.e., by setting $V^{i}=0$ and taking $\tau=$a monotonic function of t ($t=\tau$ in this case)
\begin{eqnarray}
\label{eq:gd}
0=\frac{2\phi(\tau)}{\tau}NK^{TT}_{ij}-2(1-\frac{N}{n})g_{ij}-\frac{\phi(\tau)}{\tau}(L_{X}g)_{ij},\\
\label{eq:fd}
0=-(n-1)K^{TT}_{ij}-\frac{\phi(\tau)}{\tau}N(R_{ij}+\frac{n-1}{n^{2}}g_{ij})+\frac{\phi(\tau)}{\tau}\nabla_{i}\nabla_{j}N\\\nonumber 
+\frac{2\phi(\tau)}{\tau}\nonumber NK^{TT}_{im}K^{Tm}_{j}
-\frac{\phi(\tau)}{n\tau}(\frac{N}{n}-1)g_{ij}-(n-2)(\frac{N}{n}-1)K^{TT}_{ij}\\\nonumber 
-\frac{\phi(\tau)}{\tau}(L_{X}K^{TT})_{ij},\\
\label{eq:lapse}
1=\Delta_{g}N+(|K^{TT}|^{2}+\frac{1}{n})N,\\
\label{eq:shift}
\frac{\phi(\tau)}{\tau}(\Delta_{g}X^{i}-R^{i}_{j}X^{j})=\frac{\phi(\tau)}{\tau}(2NK^{Tjk}-2\nabla^{j}X^{k})(\Gamma[g]^{i}_{jk}-\Gamma[\gamma]^{i}_{jk})\\\nonumber
-(2-n)\nabla^{i}(\frac{N}{n}-1)-\frac{2\phi(\tau)}{\tau}\nabla^{j}NK^{Ti}_{j}+g^{jk}\partial_{T}\Gamma[\gamma]^{i}_{jk}.
\end{eqnarray}  
Contracting equation (\ref{eq:gd}) with $K^{TT}$, using the momentum constraint $\nabla_{j}K^{Tij}=0$, and integrating over $M$, we obtain
\begin{eqnarray}
\label{eq:intg}
\int_{M}\frac{2\phi(\tau)}{\tau}N|K^{TT}|^{2}_{g}\mu_{g}&=&0.
\end{eqnarray} 
Standard maximum principle arguments for the elliptic equation (\ref{eq:lapseE}), yield estimates for the re-scaled lapse 
\begin{eqnarray}
\label{eq:lapseestimate}
0<\frac{1}{\sup(|K^{TT}|^{2})+\frac{1}{n}}\leq N\leq n
\end{eqnarray}
which, together with equation (\ref{eq:intg}) implies 
\begin{equation}
\label{eq:pitzero}
K^{TT}\equiv0
\end{equation}
on $M$. From the lapse equation (\ref{eq:lapseE}), we immediately obtain 
\begin{equation}
\label{eq:lapsevalue}
N=n
\end{equation}
on $M$. 
Substituting equations (\ref{eq:pitzero}) and (\ref{eq:lapsevalue}) into equation (\ref{eq:gd}) leads to 
\begin{eqnarray}
L_{X}g|_{K^{TT}=0}&=&0,
\end{eqnarray}
which implies that the shift vector field is a generator of the isometry group of $\mathbf{M}$. After substituting the available variables into the fixed point equation of the transverse traceless second fundamental form (\ref{eq:fd}), we obtain the re-scaled metric to be a negative Einstein metric 
\begin{eqnarray}
R_{ij}(g)=-\frac{n-1}{n^{2}}g_{ij}.
\end{eqnarray}  
Now, the isometry group of compact manifold $\mathbf{M}$ with negative Ricci curvature is discrete. Therefore, the existing Killing fields are only trivial i.e., $X=0$. A sketch of the proof is as follows. The divergence of the Killing equation along with commutation of covariant derivative yields 
\begin{eqnarray}
\label{eq:kill}
-\Delta_{g}X^{i}+R^{i}_{j}X^{j}+\nabla^{i}\nabla_{j}X^{j}=0.
\end{eqnarray}
The trace of the Killing equation provides $\nabla_{i}X^{i}=0$ and therefore, after multiplying both sides of equation (\ref{eq:kill}) with $X_{i}$ and integrating over $M$, the following expression is obtained 
\begin{eqnarray}
\int_{M}[\nabla_{j}X_{i}\nabla^{j}X^{i}+\frac{n-1}{n^{2}}g_{ij}X^{i}X^{j}]\mu_{g}=0
\end{eqnarray}
which implies 
\begin{eqnarray}
X\equiv0
\end{eqnarray}
everywhere on $M$.
One observes that the unknowns obtained from the momentum constraint and the dynamical equations satisfy the Hamiltonian constraint. Therefore, we have proved the following theorem.\\
\textbf{Theorem 1.} \textit{Let $\mathbf{M}$ be a closed (compact without boundary) connected orientable n-manifold, $n\geq3$, of negative Yamabe type. Then the fixed point solutions of the re-scaled `Einstein-$\Lambda$' flow (\ref{eq:gd}-\ref{eq:shift})  on $(T_{-},T_{+})\times M,$ $-\infty\leq T_{-}<T_{+}\leq \infty$, have the Cauchy data ($g, K^{TT}, N, X$)=($g_{0},K^{TT}_{0}, N_{0}, X_{0}$) which satisfy the following equations:\\
$R_{ij}(g_{0})=-\frac{n-1}{n^{2}}g_{0},~~K^{TT}=0,~~N_{0}=n,~~X_{0}=0.
$
}\\
For convenience, we may replace $g_{0}$ by $\gamma$ with 
\begin{eqnarray}
R_{ij}(\gamma)&=&-\frac{n-1}{n^{2}}\gamma_{ij},
\end{eqnarray} 
The physical variables are then given by
\begin{eqnarray}
\tilde{g}_{ij}&=&\frac{1}{\phi^{2}}g_{ij}=\frac{1}{\tau^{2}-\frac{2n\Lambda}{n-1}}\gamma_{ij},\\
\tilde{N}&=&\frac{N}{\tau^{2}-\frac{2n\Lambda}{n-1}}=\frac{n}{\tau^{2}-\frac{2n\Lambda}{n-1}},\\
\tilde{X}^{i}&=&\frac{X^{i}}{\sqrt{\tau^{2}-\frac{2n\Lambda}{n-1}}}=0.
\end{eqnarray}
If $M$ admits an Einstein metric $\gamma$, then the corresponding re-scaled variables ($\gamma, 0, n, 0$) provide constant mean extrinsic curvature Cauchy data ($g,K, N, X$) through equations (50-53) for a vacuum spacetime with a positive cosmological constant on $(-\infty, -\sqrt{\frac{2n\Lambda}{n-1}})\times \mathbf{M}$, locally expressible as 
\begin{eqnarray}
\label{eq:model}
ds^{2}=-\frac{n^{2}}{(\tau^{2}-\frac{2n\Lambda}{n-1})^{2}}d\tau^{2}+\frac{1}{(\tau^{2}-\frac{2n\Lambda}{n-1})}\gamma_{ij}dx^{i}dx^{j}.
\end{eqnarray}
This so called `trivial' evolution exists for $n=3$ if and only if the spatial manifold $\mathbf{M}$ is hyperbolizable (by the Mostow rigidity theorem). For $n>3$, the existence of a negative Einstein space is sufficient to guarantee the existence of this Cauchy data. This is the isolated fixed point for $n=3$ and is in general non-isolated (by virtue of non-trivial Einstein moduli spaces) for $n>3$. The spacetime (\ref{eq:model}) admits a globally defined time-like conformal Killing field $Y=\sqrt{\tau^{2}-\frac{2n\Lambda}{n-1}}\partial_{\tau}$ i.e.,
 \begin{eqnarray}
 L_{Y}g^{n+1}&=&-\frac{2\tau}{\sqrt{\tau^{2}-\frac{2n\Lambda}{n-1}}}g^{n+1}.
 \end{eqnarray}
 We therefore designate these spacetimes as `\textit{conformal spacetimes'}. A summary of the results obtained so far yields the following theorem \\
  \textbf{Theorem 2.} \textit{Let $M$ be a closed connected oriented n-manifold of negative Yamabe type. The fixed points of the non-autonomous re-scaled `Einstein-$\Lambda$' evolution and constraint equations on $(-\infty, -\sqrt{\frac{2n\Lambda}{n-1}})\times M$ with the gauge condition $t=\tau$ and spatial harmonic slice gauge condition are the `trivial' spacetimes given by (\ref{eq:model}). Such spacetimes admit $Y=\sqrt{\tau^{2}-\frac{2n\Lambda}{n-1}}\partial_{\tau}$ as a globally defined time-like conformal Killing vector field.}\\  
A second class of solutions of the re-scaled equations lying sufficiently close to and containing the one described by the negative Einstein metrics will be of particular importance to us. These solutions are described by 
\begin{eqnarray}
R(g)=-\frac{n-1}{n}, X=0, N=0, K^{TT}=0
\end{eqnarray}
provided that $||g-\gamma||_{H^{s}}<\epsilon, \gamma\in Ein_{-\frac{n-1}{n^{2}}}, s>\frac{n}{2}+2$ and $\epsilon<<||\gamma||_{H^{s}}$. Small calculation similar to the previous case together with lemma 5 shows that the space described by the set $\{R(g)=-\frac{n-1}{n}, X=0, N=0, K^{TT}=0\}$ satisfies the scale-free Einstein's equations (\ref{eq:timedef}-\ref{eq:con2}) in CMCSH gauge in the limit $\frac{\phi}{\tau}\to 0$. Clearly the centre manifold described by the negative Einstein spaces in theorem 1 is a subset of the solutions $\{R(g)=-\frac{n-1}{n}, X=0, N=0, K^{TT}=0\}$ since $Ein_{-\frac{n-1}{n^{2}}}\subset \mathcal{M}_{-\frac{n-1}{n}}$. We will designate the space of solutions $\{R(g)=-\frac{n-1}{n}\}$ lying sufficiently close to the Einstein structure with metric $g$ satisfying the spatial harmonic gauge condition the `\textbf{extended centre manifold}' and denote it by $\mathcal{M}^{\epsilon}_{-\frac{n-1}{n}}\cap \mathcal{S}_{\gamma}$ (the reason for such a notation will be clear in section 4.1). Our intention is to prove that this extended centre manifold is indeed the attractor of the Einstein-$\Lambda$ flow.      
\subsection{\textbf{Linear stability of the conformal spacetimes}}
\cite{moncrief2019could} constructed the following reduced Hamiltonian (while expressed as a functional of the reduced phase space variables)
\begin{eqnarray}
\label{eq:RH}
H_{reduced}&=&\frac{2(n-1)}{n}\int_{M}\frac{\partial \tau}{\partial t}\mu_{g}>0
\end{eqnarray}
 of the dynamics which was shown to decay monotonically along the solution curves and achieve its infimum precisely for the conformal spacetimes described by equation (\ref{eq:model}). Such a reduced Hamiltonian plays the role of a weak Lyapunov function of the reduced dynamics, which indicates that these background solutions may be stable against perturbations. Motivated by this notion we conduct a linear stability analysis of the re-scaled equations about the background solutions. 
 Let the perturbation be $(h_{ij}=\delta g_{ij},\delta K^{TT}_{ij}=K^{TT}_{ij}, \delta N, \delta X^{i})$. However, $\delta N$ and $\delta X^{i}$ satisfy elliptic equations from which we prove that they vanish if the background metric is negative Einstein (which is the case here). We state and prove the following lemma regarding the vanishing of the perturbations to the lapse function and the shift vector field.\\
\textbf{Lemma 1:}\textit{ Let $M$ be a closed connected oriented n-manifold of negative Yamabe type. The fixed points of the non-autonomous re-scaled `Einstein-$\Lambda$' evolution and constraint equations on $(-\infty, -\sqrt{\frac{2n\Lambda}{n-1}})\times M$ with the gauge condition $t=\tau$ and spatial harmonic slice gauge condition are the `trivial' spacetimes given by (\ref{eq:model}). Let the perturbations about these background solutions be $(h_{ij}, \delta \pi^{Tij}, \delta N, \delta X^{i})$. Then $\delta N\equiv0$ and $\delta X^{i}\equiv 0$ everywhere on M.}\\
Proof: Perturbation of the lapse equation leads to 
\begin{eqnarray}
\Delta_{\gamma}\delta N+\frac{1}{n}\delta N&=&0.
\end{eqnarray}
Application of the standard maximum principle immediately yields $\delta N\equiv0$ everywhere on M, which completes the proof of the first part.  
Perturbation of the shift equation yields 
\begin{eqnarray}
\Delta_{\gamma}\delta X_{i}-R[\gamma]_{ij}\delta X^{j}&=&0.
\end{eqnarray} 
Now, multiplying both sides with $\delta X^{i}$ and integrating over closed $M$, we obtain 
\begin{eqnarray}
\int_{M}\left(\delta X^{i}\Delta_{\gamma}\delta X_{i}-\delta X^{i}R[\gamma]_{ij}\delta X^{j}\right)\mu_{\gamma}&=&0,\\
\int_{M}\left(\nabla[\gamma]^{i}\delta X^{j}\nabla[\gamma]_{i}\delta X_{j}-R[\gamma]_{ij}\delta X^{i}\delta X^{j}\right)\mu_{\gamma}&=&0.
\end{eqnarray}
Substituting $R_{ij}[\gamma]=-\frac{n-1}{n^{2}}\gamma_{ij}$ yields 
\begin{eqnarray}
\int_{M}\left(\nabla[\gamma]^{i}\delta X^{j}\nabla[\gamma]_{i}\delta X_{j}+\frac{n-1}{n^{2}}g_{ij}\delta X^{i}\delta X^{j}\right)\mu_{\gamma}&=&0,
\end{eqnarray}
which implies $\delta X^{i}\equiv0$ everywhere on $M$, completing the second part of the proof. 
With the results $\delta N=0=\delta X^{i}$, the spacetime perturbations (at the linear level) reduce to $(h_{ij}, K^{TT}_{ij}, 0, 0)$. In the next lemma, we prove that utilizing the constraint and gauges, the perturbations $h_{ij}$ to the spatial metric may be reduced to pure transverse-traceless form with respect to the background metric. \\
\textbf{Lemma 2:}\textit{ Let the perturbations describing the dynamics be $(h_{ij},K^{TT}_{ij})$. Full-filling the Hamiltonian constraint by the perturbed data $(g_{ij}=\gamma_{ij}+h_{ij}, K^{TT}_{ij}=0+K^{TT}_{ij})$ through the constant mean extrinsic curvature spatial harmonic gauge implies the transverse-traceless property of the metric perturbations.}\\
Proof: 
Variation of Hamiltonian constraint (\ref{eq:con1}) reads
\begin{eqnarray}
DR\cdot h=0,
\end{eqnarray}
which upon using the expression of the Frechet derivative of $R$ yields
\begin{eqnarray}
\Delta_{\gamma}tr_{\gamma}h+\nabla^{i}\nabla^{j}h_{ij}-R[\gamma]_{ij}h^{ij}=0.
\end{eqnarray}
Following the Hodge-like decomposition, we may write the symmetric 2-tensor $h_{ij}$ as
\begin{eqnarray}
\label{eq:decm}
h_{ij}&=&h^{TT}_{ij}+f\gamma_{ij}+(L_{W}\gamma)_{ij},
\end{eqnarray}
where $h^{TT}$ is a symmetric transverse-traceless (w.r.t the background metric $\gamma_{ij}$) 2 tensor and $f$ and $W$ are a function and vector field lying in suitable function spaces. 
Upon substituting decomposition (\ref{eq:decm}) into the variation of the Hamiltonian constraint and noticing $\Delta_{\gamma}tr_{\gamma}(L_{W}\gamma)+\nabla^{i}\nabla^{j}(L_{W}\gamma)_{ij}-R[\gamma]_{ij}(L_{W}\gamma)^{ij}\equiv0$ one arrives at 
\begin{eqnarray}
3\Delta_{\gamma}f+\gamma^{ij}\nabla_{i}\nabla_{j}f-R[\gamma]f=0,\\\nonumber
2\Delta_{\gamma}f-R[\gamma]f=0,
\end{eqnarray}
where $R[\gamma]=-\frac{n-1}{n}$. Noting that the operator $(2\Delta_{\gamma}+\frac{n-1}{n})$ is invertible on compact $M$, we immediately obtain
\begin{eqnarray}
f\equiv 0
\end{eqnarray}
throughout $M$.
The vector field $W$ may be obtained in terms of $f$ through the following elliptic equation, which follows from the variation of the spatial harmonic gauge condition (i.e., variation of the tension field). At the linear level, the gauge condition `$id: (M,\gamma+h) \to (M,\gamma)$ is harmonic' yields
\begin{eqnarray}
\gamma^{jk}\left(\Gamma[\gamma+h]^{i}_{jk}-\Gamma[\gamma]^{i}_{jk}\right)=0\\\nonumber
2\nabla_{j}h^{ij}-\nabla^{i}tr_{\gamma}h=0\\\nonumber
-\nabla^{i}f-2\Delta_{\gamma}W^{i}+2R[\gamma]^{i}_{j}W^{j}=0\\
\Delta_{\gamma}W^{i}-R[\gamma]^{i}_{j}W^{j}=-\frac{1}{2}\nabla^{i}f.
\end{eqnarray}
Now substituting $f=0$ and $R[\gamma]_{ij}=-\frac{n-1}{n^{2}}\gamma_{ij}$, yields 
\begin{eqnarray}
\Delta_{\gamma}W^{i}+\frac{n-1}{n^{2}}W^{i}&=&0.
\end{eqnarray}
Again, invertibility of the operator $(\Delta_{\gamma}+\frac{n-1}{n^{2}})$ in compact $M$ implies $W\equiv 0$ throughout M. Therefore, $h_{ij}=h^{TT}_{ij}$, which concludes the proof.\\
Following the previous lemma, we need only consider the perturbations of the transverse-traceless type and no information is lost doing so. We will denote $h^{TT}_{ij}$ simply by $h_{ij}$ if there is no confusion. Now the linearized equations of motion around the background solution take the following forms
\begin{eqnarray}
\partial_{T}h_{ij}&=&\frac{2n\phi(\tau)}{\tau}K^{TT}_{ij},\\
\partial_{T}K^{TT}_{ij}&=&-(n-1)K^{TT}_{ij}-\frac{n\phi(\tau)}{\tau}\left(\delta R_{ij}+\frac{n-1}{n^{2}}h_{ij}\right),
\end{eqnarray}
where we have dropped $\delta N$ and $\delta X^{i}$ following lemma 1. We may now obtain the wave equation for the metric perturbation as follows. Using the perturbation to the Ricci tensor 
\begin{eqnarray}
\delta R_{ij}=\frac{1}{2}[-\frac{2(n-1)}{n^{2}}h_{ij}+(R[\gamma]_{kijm}+R[\gamma]_{kjim})h^{km}\\\nonumber
-\nabla[\gamma]^{m}\nabla[\gamma]_{m}h_{ij}]
\end{eqnarray}
along with $\gamma^{ij}h_{ij}=0$, $\nabla[\gamma]_{m}h^{ml}=0$, and $h^{ij}=\gamma^{ik}\gamma^{jl}h_{kl}\neq\delta g^{ij}$, the wave equation for the metric perturbation takes the form
\begin{eqnarray}
\frac{\partial^{2}h_{ij}}{\partial T^{2}}+\left[(n-1)+\frac{2n\Lambda}{(n-1)\tau^{2}}\right]\frac{\partial h_{ij}}{\partial T}\\\nonumber
+\frac{\phi^{2}(T)}{\tau^{2}(T)}n^{2}(\Delta_{\gamma}h_{ij}+(R[\gamma]_{kijm}+R[\gamma]_{kjim})h^{km})&=&0.
\end{eqnarray}\\
Let the Laplacian be defined as $\Delta_{\gamma}=-\nabla[\gamma]^{i}\nabla[\gamma]_{i}$. Utilizing the eigenvalue equation of the linear differential operator on the right hand side of the previous equation i.e.,  
 \begin{equation}
 \Delta_{\gamma}h_{ij}+(R[\gamma]_{kijm}+R[\gamma]_{kjim})h^{km}=\lambda h_{ij},
 \end{equation}
 the wave equations for the transverse-traceless metric perturbations reduce to the following set of ordinary differential equations
 \begin{eqnarray}
 \label{eq:2ndorder}
 \frac{\partial^{2}h_{ij}}{\partial T^{2}}+\left[(n-1)+\frac{2n\Lambda}{(n-1)\tau^{2}}\right]\frac{\partial h_{ij}}{\partial T}+\frac{\phi^{2}(T)}{\tau^{2}(T)}\lambda n^{2}h_{ij}&=&0.
 \end{eqnarray}
Here, we assume that the negative Einstein spaces are stable, that is, $\lambda\geq 0$. No example of an unstable, compact, negative Einstein space is known \cite{andersson2011einstein}. Therefore we will assume $\lambda\geq 0$. This leads to the result of \cite{fischer2002hamiltonian} in the limit of $\Lambda=0$.
The time coordinate $T$ satisfies the following from equation (\ref{eq:time})
 \begin{eqnarray}
 -\frac{\phi^{2}}{\tau}\frac{\partial}{\partial \tau}&=&\frac{\partial}{\partial T},
 \end{eqnarray}
which yields the range of $T$ as $(-\infty,\infty)$ for $\tau\in(-\infty,-\sqrt{\frac{2n\Lambda}{n-1}})$.
Note that for $h$ in the kernel of the differential operator i.e., for $-\Delta_{\gamma}h_{ij}-(R[\gamma]_{kijm}+R[\gamma]_{kjim})h^{km}=0,$ (i.e., $\lambda=0$) the equation reduces trivially to 
\begin{eqnarray}
\frac{\partial^{2}h_{ij}}{\partial T^{2}}+\left[(n-1)+\frac{2n\Lambda}{(n-1)\tau^{2}}\right]\frac{\partial h_{ij}}{\partial T}&=&0,
\end{eqnarray}
which yields asymptotic stability following the fact that the damping coefficient is strictly positive. Asymptotic stability of perturbations lying in the space orthogonal to the kernel of $\mathcal{L}$ may be shown by explicitly constructing a Lyapunov function and deriving its monotonic decay property in the time future direction. Let us convert the second order equation (\ref{eq:2ndorder}) to a system of first order equations by substituting $h_{ij}=u$ and $\frac{\partial h_{ij}}{\partial T}=v$
\begin{eqnarray}
\label{eq:odenew1}
\partial_{T}u&=&\frac{\phi}{\tau}v,\\
\label{eq:odenew2}
\partial_{T}v&=&-(n-1)v-\frac{\phi(\tau)}{\tau}n^{2}\lambda u,
\end{eqnarray}
Let the Lyapunov function be defined as 
\begin{eqnarray}
E&=&\frac{1}{2}v^{2}+\frac{1}{2}n^{2}\lambda u^{2},
\end{eqnarray}
the time derivative of which reads 
\begin{eqnarray}
\frac{dE}{dT}&=&-(n-1)v^{2}\\\nonumber 
&<&0
\end{eqnarray} 
i.e., energy decays monotonically and $\frac{dE}{dT}=0$ only when $v\equiv0$. The equations (\ref{eq:odenew1}) and (\ref{eq:odenew2}) together with the decay of the Lyapunov function yield $\partial_{T}u\lesssim e^{-2T}, v\lesssim e^{-T}$ as $T\to\infty$ and therefore $u$ remains bounded and limits to $u^{*}<\infty$. The asymptotic solution is as follows 
\begin{eqnarray}
|u-u^{*}|\lesssim e^{-2T},\\
|v|\lesssim e^{-T}.
\end{eqnarray}
 We may prove the future completeness of these perturbed spacetimes at the linear level as follows.\\
In order to establish the future completeness of the spacetime, we need to show that the length of a timelike geodesic goes to infinity. For a family (homotopy class) of rectifiable timelike curves $c:[a,b]\to \mathbf{R}\times M$, the length of its geodesic representative is computed as follows
\begin{eqnarray}
\sup_{c}\int_{a}^{b}\sqrt{(-\hat{g}(\dot{c},\dot{c}))}dt=d_{\hat{g}}(a,b)=l_{\hat{g}}(\mathcal{C}),
\end{eqnarray}
where $\mathcal{C}$ is the geodesic representative of the family $c$ and $\hat{g}$ is the spacetime metric. 
 Note that the problem arises due to the fact that these family of curves are not necessarily uniformly timelike. Here we will use a theorem proved in \cite{choquet2009general}, which serves as sufficient condition for the geodesic completeness. Such condition is satisfied in this particular case and therefore future completeness holds. \\
\textbf{Theorem \cite{choquet2009general}} Sufficient conditions for future timelike and null geodesic completeness of a regularly sliced spacetime are that\\ 
1. $|\nabla N|_{g}$ is bounded by a function of $t$ which is integrable on $[a,\infty)$\\
2. $|K|_{g}$ is bounded by a function of $t$ which is integrable on $[a,\infty)$ for some $a<\infty$.
Note that both of these conditions are trivially satisfied in our case of linearized perturbations. We will obtain a separate proof of future completeness in the fully nonlinear case.
\textbf{Theorem 3:}\textit{ Let $M$ be a closed connected oriented n-manifold of negative Yamabe type. The background solutions (\ref{eq:model}) of the Einstein-$\Lambda$ evolution and constraint equations on $(-\infty,-\sqrt{\frac{2n\Lambda}{n-1}})\times M$ with the gauge condition $t=\tau$ and spatial harmonic slice gauge condition are stable and future complete against linear perturbations.}  

\section{Fully nonlinear perturbations}
Theorem 3 provides us with a notion of stability of the background solutions. However, it leaves out the fully non-linear evolution of the small perturbations. The reduced Hamiltonian described in equation (\ref{eq:RH}) is always at hand and may be used to study fully non-linear and arbitrarily large perturbations. But, it seems to control only the $H^{1}\times L^{2}$ norm of the reduced phase space data (see \cite{fischer2002hamiltonian} and \cite{moncrief2019could} for a definition of reduced phase space). However, following the local existence theorem developed here, we need to control the $H^{s}\times H^{s-1}$ norm with $s>\frac{n}{2}+2$, $n\geq3$. Therefore we construct a Lyapunov function of the dynamics which indeed controls the required Sobolev norm ($L^{2}$ norm of the $s>\frac{n}{2}+2$ spatial derivatives) of the fully non-linear small perturbations. We show that it decays along the solution curve if we start sufficiently close to the background spacetime. In addition, as mentioned previously, finite dimensionality of the Einstein moduli space in the case of higher dimensions ($n>3$) has to be addressed carefully in order to show the attractor property of the centre manifold. However, before doing so, a few important geometric notions are to be discussed which have substantial impact on the stability analysis in the case of the spatial dimension $n>3$.   
\subsection{\textbf{Deformation space}}
Here we provide a brief description of the deformation space of the Einstein structure necessary for the nonlinear analysis. Details can be found in several studies \cite{anderson1992thel, andersson2011einstein}.
The background solutions of the Einstein-$\Lambda$ flow in CMCSH gauge are the conformal spacetimes as described in the previous section. The spatial metric component of these conformal spacetimes is a negative Einstein metric i.e., the spatial metric satisfies 
\begin{eqnarray}
\label{eq:ein}
R_{ij}(\gamma)=-\frac{n-1}{n^{2}}\gamma_{ij}.
\end{eqnarray}
Let us denote the space of metrics satisfying equation (\ref{eq:ein}) by $\mathcal{E}in_{-\frac{n-1}{n^{2}}}$ and consider the following map
\begin{eqnarray}
F: \mathcal{M}&\to& \mathcal{M}^{'},\\\nonumber
\gamma&\mapsto& R_{ij}(\gamma)+\frac{n-1}{n^{2}}\gamma_{ij},
\end{eqnarray} 
where $\mathcal{M}^{'}$ is a subspace of the space of symmetric $(0,2)$ tensors.  
$\mathcal{E}in_{-\frac{n-1}{n^{2}}}=F^{-1}(0)$ will be a submanifold of $\mathcal{M}$ provided that the $T_{F^{-1}(0)}F$ is surjective (i.e., $F$ is a submersion). The differential $D(Ric+\frac{n-1}{n^{2}}\gamma)\cdot h$ may be reduced to a second order elliptic differential operator acting on the variation $h$, whose adjoint is injective and thus $T_{F^{-1}(0)}F$ is surjective. The tangent space of $\mathcal{E}in_{-\frac{n-1}{n^{2}}}$ may be calculated as the kernel of $T_{\gamma}F$. The operator $T_{\gamma}F=D(Ric+\frac{n-1}{n^{2}}\gamma)$ may be decomposed in terms of the Lichnerowicz type Laplacian $\mathcal{L}$, via
\begin{eqnarray}
T_{\gamma}F\cdot h=\mathcal{L}h-2\delta^{*}\delta h-\nabla[\gamma] d(tr h),
\end{eqnarray}
where $\mathcal{L}h_{ij}=\Delta_{\gamma} h_{ij}-2R[\gamma]_{ikjl}h^{kl}$, $(\delta h)_{i}=\nabla[\gamma]^{j}h_{ij}$, and $(\delta^{*}Y)_{ij}=-\frac{1}{2}(L_{Y}\gamma)_{ij}$. The space of symmetric 2 tensors $S^{2}M$ may be decomposed as 
\begin{eqnarray}
S^{2}M=C^{TT}(S^{2}M)\oplus (\mathcal{F}(M)\otimes\gamma)\oplus\mathcal{IM}(L),
\end{eqnarray}
where $C^{TT}(S^{2}M)$, $\mathcal{F}(M)$, and $\mathcal{IM}(L)$ are the space of symmetric transverse-traceless 2-tensors, the space of functions on $M$, and the image of the Lie derivative $L$ acting on $\gamma$ with respect to a vector field $Y\in TM$ (a section of $TM$ to be precise), respectively. In local co-ordinates, this decomposition may be represented as 
\begin{eqnarray}
h_{ij}=h^{TT}_{ij}+f\gamma_{ij}+(L_{Y}\gamma)_{ij}.
\end{eqnarray}
The kernel $\mathcal{K}(T_{\gamma}F)$ at $\gamma\in F^{-1}(0)$ may be obtained through the solution of the following equation
\begin{eqnarray}
\mathcal{L}(h^{TT}+f\gamma+L_{Y}\gamma)-2\delta^{*}\delta (h^{TT}+f\gamma+L_{Y}\gamma)\\\nonumber
-\nabla d(tr (h^{TT}+f\gamma+L_{Y}\gamma))=0,\nonumber
\end{eqnarray}
where we notice that $\mathcal{L}(L_{Y}\gamma)-2\delta^{*}\delta (L_{Y}\gamma)-\nabla d(tr (L_{Y}\gamma))\equiv0$ and thus $L_{Y}\gamma\in\mathcal{K}(T_{\gamma\in F^{-1}(0)}F)$. The remaining terms lead to 
\begin{eqnarray}
(\mathcal{L}h^{TT})_{ij}+(\Delta_{\gamma}f)\gamma_{ij}-2fR_{ij}+\nabla_{i}\nabla_{j}f+\nabla_{j}\nabla_{i}f-n\nabla_{i}\nabla_{j}f&=&0,
\end{eqnarray}
which upon taking the trace yields 
\begin{eqnarray}
2(n-1)\Delta_{\gamma}f-2R(\gamma)f&=&0.
\end{eqnarray}
For $\gamma\in F^{-1}(0)$, the scalar curvature $R(\gamma)=-\frac{n-1}{n}$ and therefore $f=0$ everywhere on $M$. The resulting tangent space of $\mathcal{E}in_{-\frac{n-1}{n^{2}}}=F^{-1}(0)$ consists of the kernel of the operator $\mathcal{L}$ (a subspace of the space of transverse-traceless symmetric 2-tensors) and the image of the Lie derivative operator with respect to a vector field
\begin{eqnarray}
\label{eq:tangent}
T_{\gamma}\mathcal{E}in_{-\frac{n-1}{n^{2}}}=\ker(\mathcal{L})\oplus \mathcal{IM}(L).
\end{eqnarray}

Let $\gamma_{0}\in \mathcal{E}in_{-\frac{n-1}{n^{2}}}$ and $\mathcal{V}$ be its connected component. Also consider $\mathcal{S}_{\gamma}$ to be the harmonic slice of the identity diffeomorphism i.e., the set of $\gamma\in\mathcal{E}in_{-\frac{n-1}{n^{2}}}$ for which the identity map $Id: (M,\gamma)\to (M,\gamma_{0})$ is harmonic. This condition is equivalent to the vanishing of the tension field $-V^{k}$ that is
\begin{eqnarray}
\label{eq:harmonic}
-V^{k}=-\gamma^{ij}(\Gamma[\gamma]^{k}_{ij}-\Gamma[\gamma_{0}]^{k}_{ij})=0.
\end{eqnarray} 
For $\gamma\in \mathcal{E}in_{-\frac{n-1}{n^{2}}}$, $\mathcal{S}_{\gamma}$ is a submanifold of $\mathcal{M}$ for $\gamma$ sufficiently close to $\gamma_{0}$ \cite{andersson2011einstein, moncrief2019could}. The deformation space $\mathcal{N}$ of $\gamma_{0}\in \mathcal{E}in_{-\frac{n-1}{n^{2}}}$ is defined as the intersection of the $\gamma_{0}-$connected component $ \mathcal{V}\subset \mathcal{E}in_{-\frac{n-1}{n^{2}}}$ and the harmonic slice $\mathcal{S}_{\gamma}$ i.e.,
\begin{eqnarray}
\label{eq:ds}
\mathcal{N}= \mathcal{V}\cap \mathcal{S}_{\gamma}.
\end{eqnarray}
$\mathcal{N}$ is assumed to be smooth. 
In the case of $n=3$, the negative Einstein structure is rigid which follows from the Mostow rigidity theorem \cite{agard1983geometric,lebrun1994einstein} and this structure corresponds to the hyperbolic structure up to isometry. For higher genus surfaces $\Sigma_{genus}$ ($genus>1$) in two dimensions, the deformation space modulo isotopy diffeomorphisms is the Teichm\"uller space diffeomorphic to $\mathbf{R}^{6genus-6}$. For $n>3$, it is a finite dimensional submanifold of $\mathcal{M}$. Examples of higher dimensional ($n>3$) negative Einstein spaces with non-trivial deformation spaces include Kahler-Einstein manifolds \cite{besse2007einstein}. \cite{andersson2011einstein} provides the details of constructing numerous negative Einstein spaces through a product operation. Therefore, we do not repeat the same here. Readers are referred to the aforementioned paper.\\
Following equations (\ref{eq:tangent}), (\ref{eq:harmonic}), and (\ref{eq:ds}), the tangent space $T_{\gamma}\mathcal{N}$ in local coordinates is represented as 
\begin{eqnarray}
\label{eq:tangentspace}
\frac{\partial \gamma}{\partial q^{a}}&=&h^{TT||}+L_{Y^{||}}\gamma,
\end{eqnarray}
where $h^{TT||}\in \ker(\mathcal{L})=C^{TT||}(S^{2}M)\subset C^{TT}(S^{2}M)$, $Y\in \mathfrak{X}(M)$ satisfy
\begin{eqnarray}
-[\nabla[\gamma]^{m}\nabla[\gamma]_{m}Y^{i}+R[\gamma]^{i}_{m}Y^{m}]+(h^{TT||}+L_{Y^{||}}\gamma)^{mn}(\Gamma[\gamma]^{i}_{mn}-\Gamma[\gamma_{0}]^{i}_{mn})=0,\nonumber 
\end{eqnarray} 
and, $\{q^{a}\}_{a=1}^{dim(\mathcal{N})}$ is a local chart on $\mathcal{N}$, $\mathfrak{X}(M)$ is the space of vector fields on $M$ (in a suitable function space setting). Also note that the space of transverse-traceless tensors may be decomposed as follows
\begin{eqnarray}
C^{TT}(S^{2}M)=C^{TT||}(S^{2}M)\oplus C^{TT\perp}(S^{2}M),
\end{eqnarray}
where $C^{TT\perp}$ is the orthogonal complement of $\ker(\mathcal{L})$ in $C^{TT}(S^{2}M)$. An important thing to note is that all known examples of closed negative Einstein spaces have integrable deformation spaces. Note that the extended centre manifold is correctly defined as the intersection of the $\epsilon-$neighbourhood of deformation space $\mathcal{N}$ in $\mathcal{M}_{-\frac{n-1}{n}}\cap \mathcal{S}_{\gamma}$, that is, $\mathcal{M}^{\epsilon}_{-\frac{n-1}{n}}\cap \mathcal{S}_{\gamma}, \gamma\in Ein_{-\frac{n-1}{n^{2}}}$ (since the spatial harmonic gauge is imposed, the metric must lie in the submanifold $\mathcal{S}_{\gamma}$ as well). Obviously, $\mathcal{N}\subset \mathcal{M}^{\epsilon}_{-\frac{n-1}{n}}\cap \mathcal{S}_{\gamma}$ and $T\mathcal{N}\subset T(\mathcal{M}^{\epsilon}_{-\frac{n-1}{n}}\cap \mathcal{S}_{\gamma})$. This concludes the business of describing the centre (and extended centre) manifold of the dynamics. 

\subsection{\textbf{Perturbations and shadow gauge}}
The previous section describes the non-trivial deformation space of a negative Einstein structure. The presence of a non-trivial deformation space necessitates the consideration of perturbations $L^{2}$ orthogonal to the deformation space of the Einstein structures forming the center manifold of the re-scaled dynamics. Perturbations tangent to the deformation space are trivially stable at the linear level, which was shown earlier and will be repeated later as well. For the treatment of the orthogonal perturbations, we invoke the shadow gauge introduced by \cite{andersson2011einstein} in addition to the constant mean curvature spatial harmonic gauge (CMCSH). 
Let $\gamma\in\mathcal{N}$ and consider metric $g\in\mathcal{M}$ close to $\gamma_{0}$ such that $||g-\gamma_{0}||_{H^{s}}<\delta$, and  $||\gamma-\gamma_{0}||_{H^{s}}<\delta$ imply that  $||g-\gamma||_{H^{s}}<2\delta$ through the triangle inequality. The shadow gauge is defined by requiring that the perturbation $(g-\gamma)$ be $L^{2}$ orthogonal to the deformation space $\mathcal{N}$. Adopting the local coordinate system $\{q^{a}\}_{a=1}^{N}$ on $\mathcal{N}$ with $N=dim(\mathcal{N})<\infty$, the local basis for the tangent space of $\mathcal{N}$ may be written as  $\frac{\partial \gamma_{ij}}{\partial q^{a}}=h^{TT||}+L_{Y^{||}}\gamma$. The shadow gauge condition is equivalent to
\begin{eqnarray}
<(g-\gamma),\frac{\partial \gamma}{\partial q^{a}}>_{L^{2}}&=&0,\\\nonumber
\int_{\mathbf{M}}(g_{ij}-\gamma_{ij})\frac{\partial \gamma^{ij}}{\partial q^{a}}\mu_{\gamma}&=&0,
\end{eqnarray}
where $\frac{\partial \gamma^{ij}}{\partial q^{a}}=-\gamma^{im}\gamma^{jn}\frac{\partial \gamma_{mn}}{\partial q^{a}}$. 
Technically, this shadow condition is equivalent to $\gamma$ being a projection of g onto $\mathcal{N}$. In other words, there is a projection map $\mathcal{P}:\mathcal{M}\to \mathcal{N}$ such that 
\begin{eqnarray}
\label{eq:projection}
\gamma=\mathcal{P}[g].
\end{eqnarray}
Noting that the space $\mathcal{N}$ is assumed to have smooth structure, this projection, in a sense, is a smoothing operation. We call $\gamma$ the shadow of $g$. The time derivative of $\gamma$ in $\mathcal{N}$ may be obtained as
\begin{eqnarray}
\frac{\partial \gamma}{\partial T}&=&D\mathcal{P}[g]\cdot \frac{\partial g}{\partial T}.
\end{eqnarray}
The expression for $D\mathcal{P}[g]$ may be obtained by time differentiating the shadow metric condition i.e., 
\begin{eqnarray}
\frac{d}{dT}\int_{\mathbf{M}}(g_{ij}-\gamma_{ij})\frac{\partial \gamma^{ij}}{\partial q^{a}}\mu_{\gamma}=0,
\end{eqnarray}
\begin{eqnarray}
\int_{M}\partial_{T}g_{ij}\frac{\partial \gamma^{ij}}{\partial q^{a}}\mu_{\gamma}+\int_{M}(g_{ij}-\gamma_{ij})\frac{\partial}{\partial q^{b}}(\frac{\partial \gamma^{ij}}{\partial q^{a}})\frac{\partial q^{b}}{\partial T}\mu_{\gamma}\\\nonumber +\frac{1}{2}\int_{M}(g_{ij}-\gamma_{ij})\frac{\partial \gamma^{ij}}{\partial q^{a}}\gamma^{lm}\frac{\partial \gamma_{lm}}{\partial q^{b}}\frac{\partial q^{b}}{\partial T}\mu_{\gamma}
-\int_{M}\frac{\partial \gamma_{ij}}{\partial q^{b}}\frac{\partial \gamma^{ij}}{\partial q^{a}}\frac{\partial q^{b}}{\partial T}\mu_{\gamma}=0,
\end{eqnarray}
where the matrix $-\int_{M}\frac{\partial \gamma_{ij}}{\partial q^{b}}\frac{\partial \gamma^{ij}}{\partial q^{a}}\mu_{\gamma}=\int_{M}\frac{\partial \gamma_{ij}}{\partial q^{b}}\gamma^{im}\gamma^{jn}\frac{\partial \gamma_{mn}}{\partial q^{a}}\mu_{\gamma}$ is invertible due to $\frac{\partial \gamma_{ij}}{\partial q^{b}}$ being a basis for $T_{\gamma}\mathcal{N}$. For the small data limit i.e., $(g-\gamma)<2\delta, \delta>0$, the combined matrix
\begin{eqnarray}
\mathcal{B}=\int_{M}(g_{ij}-\gamma_{ij})\left(\frac{\partial}{\partial q^{b}}(\frac{\partial \gamma^{ij}}{\partial q^{a}})+\frac{\partial \gamma^{ij}}{\partial q^{a}}\gamma^{lm}\frac{\partial \gamma_{lm}}{\partial q^{b}}\right)\mu_{\gamma}\\\nonumber
+\int_{M} \mu_{\gamma}\frac{\partial \gamma_{ij}}{\partial q^{b}}\gamma^{im}\gamma^{jn}\frac{\partial \gamma_{mn}}{\partial q^{a}}\mu_{\gamma}
\end{eqnarray}
is invertible as well yielding the following time evolution of the shadow metric $\gamma$ in the deformation space 
\begin{eqnarray}
\frac{\partial q^{b}}{\partial T}&=&-(\mathcal{B}^{-1})^{ba}\int_{M}\partial_{T}g_{ij}\frac{\partial \gamma^{ij}}{\partial q^{a}}\mu_{\gamma},\\\nonumber
\frac{\partial \gamma_{ij}}{\partial T}&=&-\frac{\partial \gamma_{ij}}{\partial q^{b}}(\mathcal{B}^{-1})^{ba}\int_{M}\partial_{T}g_{mn}\frac{\partial \gamma^{mn}}{\partial q^{a}},
\end{eqnarray} 
where $\frac{\partial g_{mn}}{\partial T}$ may be obtained from the re-scaled field equation (\ref{eq:gd1}). This is again equivalent to the following projection operation $D\mathcal{P}: T_{g}\mathcal{M}\to T_{\gamma}\mathcal{N}$
\begin{eqnarray}
\frac{\partial \gamma}{\partial T}=D\mathcal{P}[g]\cdot \frac{\partial g}{\partial T}
\end{eqnarray}
  In a sense, the following estimate holds 
\begin{eqnarray}
||\frac{\partial \gamma}{\partial T}||_{H^{s}}\leq C ||\frac{\partial g}{\partial T}||_{H^{s-1}}, 
\end{eqnarray}
for some constant $C=C(\delta)>0$. More generally, one may have the following estimate for $z\in T_{g}\mathcal{U}$ ($g\in\mathcal{U}\subset\mathcal{M}$) while considering the projection operation $\mathcal{P}: T_{g}\mathcal{U} \to T_{\gamma}\mathcal{N}$ 
\begin{eqnarray}
\label{eq:projectionestimate1}
||D\mathcal{P}\cdot z||_{H^{s}}\leq C||z||_{H^{s-1}}.
\end{eqnarray}
In a sense, the projection is a smoothing operation.
We are primarily interested in studying the evolution of sufficiently small however fully nonlinear perturbations under Einstein-$\Lambda$ flow. In order to do so, we need to first define the small data. The background solutions (conformal space-times (\ref{eq:model})) may be expressed in terms of the re-scaled variables ($\gamma, N=n, X^{i}=0$) after dimensionalization by suitable factors of $\phi^{2}=(\tau^{2}-\frac{2n\Lambda}{n-1})>0$. In addition to these three entities, we also have the corresponding background re-scaled transverse-traceless second fundamental form $K^{TT}=0$. Therefore, the complete set of small data is defined as the difference between the background and perturbed solutions i.e., $(g-\gamma, K^{TT}, \frac{N}{n}-1, X)$ whose norm is sufficiently small in the suitable function space.\\
Now we state and prove a series of important lemmas which shall be required later to obtain several important estimates.\\
\textbf{Lemma 3:} \textit{Let $s>\frac{n}{2}+2$, $\gamma_{0}\in \mathcal{E}in_{-\frac{n-1}{n^{2}}}$ and $\mathcal{N}$ be its integrable deformation space and $\gamma\in \mathcal{N}$. Also assume $g\in \mathcal{U}\subset \mathcal{M}$ and $\mathcal{U}$ is a neighborhood of $\mathcal{N}$ in $\mathcal{M}$ and $g\in\mathcal{U}$ satisfies $||g-\gamma_{0}||_{H^{s}}<\delta>0$. Let $z\in T_{g}\mathcal{U}$. Then
\begin{eqnarray}
\gamma^{mn}D\Gamma^{i}_{mn}[\gamma]D\mathcal{P}\cdot z: H^{s-1}\to H^{s}
\end{eqnarray}
 satisfies the following estimate
 \begin{eqnarray}
 ||\gamma^{mn}D\Gamma^{i}_{mn}[\gamma]D\mathcal{P}\cdot z||_{H^{s}}\leq C(\delta)||\gamma-\gamma_{0}||_{H^{s}}||z||_{H^{s-1}}.
 \end{eqnarray}}
 with $C(\delta)>0.$\\
\textbf{Proof:} Following equation (\ref{eq:tangentspace}), any element belonging to $T_{\gamma}\mathcal{N}$ may be written as 
\begin{eqnarray}
h&=&h^{TT||}+L_{Y^{||}}\gamma,
\end{eqnarray}
with $Y^{||}$ satisfying 
\begin{eqnarray}
-[\nabla[\gamma]^{m}\nabla[\gamma]_{m}Y^{||i}+R[\gamma]^{||i}_{m}Y^{||m}]+(h^{TT||}+L_{Y^{||}}\gamma)^{mn}(\Gamma[\gamma]^{i}_{mn}\\\nonumber 
-\Gamma[\gamma_{0}]^{i}_{mn})=0.
\end{eqnarray}
$\gamma^{mn}D\Gamma^{i}_{mn}[\gamma]h$ for $h\in T_{\gamma}\mathcal{N}$ may be written as 
\begin{eqnarray}
\label{eq:dgamma}
 \gamma^{mn}D\Gamma[\gamma]^{i}_{mn} h&=&\gamma^{mn}\gamma^{ik}\left(\nabla[\gamma]_{m}h_{kn}+\nabla[\gamma]_{n}h_{mk}-\nabla[\gamma]_{k}h_{mn}\right)\\\nonumber
 &=&2\gamma^{ik}\nabla[\gamma]^{n}h_{kn}-\nabla[\gamma]^{i}(tr_{\gamma}h),
\end{eqnarray}
from which $\gamma^{mn}D\Gamma^{i}_{mn}[\gamma]h^{TT||}=0$ follows immediately and the remaining terms lead to the following equation
\begin{eqnarray}
 \gamma^{mn}D\Gamma[\gamma]^{i}_{mn} h=2\gamma^{ik}\nabla[\gamma]^{n}\left(\nabla[\gamma]_{k}Y_{n}+\nabla[\gamma\nonumber]_{n}Y_{k}\right)-2\nabla[\gamma]^{i}(\nabla[\gamma]_{m}Y^{m})\\\nonumber
 =2\gamma^{ik}(\nabla[\gamma]_{k}\nabla[\gamma]_{n}Y^{n}+R[\gamma]_{mk}Y^{m})+2\nabla[\gamma]^{n}\nabla[\gamma]_{n}Y^{i}-2\nabla[\gamma]^{i}(\nabla[\gamma]_{m}Y^{m})\\\nonumber
 =2(\nabla[\gamma]^{n}\nabla[\gamma]_{n}Y^{i}+R[\gamma]^{i}_{m}Y^{m})\\\nonumber
 =2(L_{Y^{||}}\gamma+h^{TT})^{mn}(\Gamma[\gamma]^{i}_{mn}-\Gamma[\gamma_{0}]^{i}_{mn})\\\nonumber
 =2h^{mn}(\Gamma[\gamma]^{i}_{mn}-\Gamma[\gamma_{0}]^{i}_{mn}).
\end{eqnarray} 
Now, using the previous expression, for $\gamma$ close to $\gamma_{0}$, we have the following estimate
\begin{eqnarray}
||\gamma^{mn}D\Gamma[\gamma]^{i}_{mn} h||_{H^{s}}\leq C||\gamma-\gamma_{0}||_{H^{s}}||h||_{H^{s-1}}
\end{eqnarray}
which upon substituting $T_{\gamma}\mathcal{N}\ni h=D\mathcal{P}\cdot z$ together with ($\ref{eq:projectionestimate1}$) yields the required estimate 
\begin{eqnarray}
\label{eq:projectionestimate2}
||\gamma^{mn}D\Gamma^{i}_{mn}[\gamma]D\mathcal{P}\cdot z||_{H^{s}}\leq C||\gamma-\gamma_{0}||_{H^{s}}||z||_{H^{s-1}}.
\end{eqnarray}
We have therefore proved the lemma.\\
\textbf{Lemma 4}: \textit{let $s>\frac{n}{2}+2, \gamma_{0}\in\mathcal{E}in_{-\frac{n-1}{n^{2}}}$, and $\mathcal{N}$ be the integrable deformation space of $\gamma_{0}$, and let $\gamma\in\mathcal{N}$ be the shadow of $g$} i.e., $\mathcal{P}[g]=\gamma$, $g\in \mathcal{U}\subset \mathcal{M}$ with $\mathcal{U}$ being a neighborhood of $\mathcal{N}$ in $\mathcal{M}$ and $g\in\mathcal{U}$ satisfying $||g-\gamma_{0}||_{H^{s}}<\delta$ for some $\delta>0$, then 
\begin{eqnarray}
\label{eq:CRES}
||g^{mn}D\Gamma^{i}_{mn}[\gamma]D\mathcal{P}|_{g}z ||_{H^{s}}\leq C(\delta)\left(||g-\gamma||_{H^{s}}+||\gamma-\gamma_{0}||_{H^{s}}\right)||z||_{H^{s-1}},
\end{eqnarray}
for $z\in T_{g}\mathcal{U}$. \\
Proof: 
\begin{eqnarray}
g^{mn}D\Gamma^{i}_{mn}[\gamma]D\mathcal{P}|_{g}z=(g^{mn}-\gamma^{mn})D\Gamma^{i}_{mn}|_{\gamma}\cdot\nonumber D\mathcal{P}|_{g}z+\gamma^{mn}D\Gamma^{i}_{mn}|_{\gamma}\cdot D\mathcal{P}|_{g}z,\\\nonumber
||g^{mn}D\Gamma^{i}_{mn}[\gamma]D\mathcal{P}|_{g}z||_{H^{s}}=||(g^{mn}-\gamma^{mn})D\Gamma^{i}_{mn}|_{\gamma}\cdot D\mathcal{P}|_{g}z\\\nonumber +\gamma^{mn}D\Gamma^{i}_{mn}|_{\gamma}\cdot D\mathcal{P}|_{g}z||_{H^{s}}\\\nonumber  
\leq||(g^{mn}-\gamma^{mn})D\Gamma^{i}_{mn}|_{\gamma}\cdot D\mathcal{P}|_{g}z||_{H^{s}}+||\gamma^{mn}D\Gamma^{i}_{mn}|_{\gamma}\cdot D\mathcal{P}|_{g}z||_{H^{s}}.
\end{eqnarray}
Application of lemma 1 in conjunction with (\ref{eq:projectionestimate2}) yields the desired estimate 
\begin{eqnarray}
||g^{mn}D\Gamma^{i}_{mn}[\gamma]D\mathcal{P}|_{g}\cdot z ||_{H^{s}}\leq C\left(||g-\gamma||_{H^{s}}+||\gamma-\gamma_{0}||_{H^{s}}\right)||z||_{H^{s-1}}.
\end{eqnarray}\\
\textbf{Lemma 5:} \textit{let $s>\frac{n}{2}+2, \gamma_{0}\in\mathcal{E}in_{-\frac{n-1}{n^{2}}}$, and $\mathcal{N}$ be the integrable deformation space of $\gamma_{0}$, and let $\gamma\in\mathcal{N}$ be the shadow of $g$ i.e., $\mathcal{P}[g]=\gamma$,  $g\in \mathcal{U}\subset \mathcal{M}$ with $\mathcal{U}$ being a neighborhood of $\mathcal{N}$ in $\mathcal{M}$ and $g\in\mathcal{U}$ satisfying $||g-\gamma_{0}||_{H^{s}}<\delta$ for some $\delta>0$, then the following map 
\begin{eqnarray}
P: H^{s+1}(\mathfrak{X}(M))&\to& H^{s-1}(\mathfrak{X}(M)),\\
X&\mapsto&\Delta_{g}X^{i}-R[g]^{i}_{j}X^{j}+(2\nabla^{j}X^{k})(\Gamma[g]^{i}_{jk}-\Gamma[\gamma]^{i}_{jk})
\end{eqnarray}
is an isomorphism.}\\
\textbf{Proof:} Let us say $\psi_{s}$ is the flow of the shift vector field $X$ and thus a one parameter group of diffeomorphism of $M$. Therefore, by $\psi_{s}$ at a fixed time, we may push forward and pull back the sections of the tangent and the co-tangent bundles, respectively. The negative of the tension vector field is defined as a section of the tangent bundle $TM$ and locally expressible as 
\begin{eqnarray}
V^{i}&=&g^{jk}(\Gamma[g]^{i}_{jk}-\Gamma[\gamma]^{i}_{jk}).
\end{eqnarray}    
The co-vector counterpart of $V$ may be pulled back and dualized as
\begin{eqnarray}
(\psi^{*}_{s}V)^{i}&=&(\psi^{*}_{s}g)^{jk}\psi_{s}^{*}((\Gamma[g]-\Gamma[\gamma])^{i}_{jk},\\\nonumber
&=&(\psi^{*}_{s}g)^{jk}\left(\Gamma[\psi^{*}_{s}g]^{i}_{jk}-\Gamma[\psi^{*}_{s}\gamma]^{i}_{jk}\right).
\end{eqnarray} 
The right hand side follows as a consequence of the tensor transformation property of the difference of connection coefficients ($\Gamma^{i}_{jk}$). Differentiation with respect to $s$ at $s=0$ yields
\begin{eqnarray}
(\frac{d}{ds}(\psi^{*}_{s}V)^{i})|_{s=0}&=&\frac{d}{ds}((\psi^{*}_{s}g)^{jk}\left(\Gamma[\psi^{*}_{s}g]^{i}_{jk}-\Gamma[\psi^{*}_{s}\gamma]^{i}_{jk}\right))|_{s=0},\\\nonumber
L_{X}V^{i}&=&\frac{d}{ds}((\psi^{*}_{s}g)^{jk})|_{s=0}\left(\Gamma[g]^{i}_{jk}-\Gamma[\gamma]^{i}_{jk}\right)\\\nonumber&&+g^{jk}\frac{d}{ds}\left(\Gamma[\psi^{*}g]^{i}_{jk}-\Gamma[\psi^{*}\gamma]^{i}_{jk}\right)|_{s=0},\\\nonumber
L_{X}V^{i}&=&L_{X}g^{jk}\left(\Gamma[g]^{i}_{jk}-\Gamma[\gamma]^{i}_{jk}\right)+g^{jk}\frac{d}{ds}\left(\Gamma[\psi^{*}g]^{i}_{jk}-\Gamma[\psi^{*}\gamma]^{i}_{jk}\right)|_{s=0}.
\end{eqnarray} 
 Using the formula for the Fr\'echet derivative of the connection coefficients, we may obtain 
 \begin{eqnarray}
 \frac{d}{ds}\Gamma^{i}_{jk}[\psi^{*}_{s}g]|_{s=0}&=&\frac{1}{2}g^{im}[\nabla_{j}(\frac{d}{ds}((\psi^{*}_{s}g)_{mk})|_{s=0})+\nabla_{k}\nonumber(\frac{d}{ds}((\psi^{*}_{s}g)_{jm})|_{s=0})\\\nonumber
 &&-\nabla_{m}(\frac{d}{ds}((\psi^{*}_{s}g)_{jk})|_{s=0})]\\\nonumber
&=&\frac{1}{2}g^{im}\left(\nabla_{j}(L_{X}g_{mk})+\nabla_{k}\nonumber(L_{X}g_{jm})-\nabla_{m}(L_{X}g_{jk})\right)\\\nonumber
&=&\frac{1}{2}g^{im}(\nabla_{j}(\nabla_{m}X_{k}+\nabla_{k}X_{m})+\nabla_{k}(\nabla_{m}X_{j}+\nabla_{j}X_{m})\\\nonumber&& -\nabla_{m}(\nabla_{k}X_{j}+\nabla_{j}X_{k}))\\\nonumber
&=&\frac{1}{2}\left(\nabla_{(j}\nabla_{k)}X^{i}+g^{im}(R_{knjm}+R_{jnkm})X^{n}\right)\\\nonumber
&=&\frac{1}{2}\left(\nabla[g]_{(j}\nabla[g]_{k)}X^{i}+(R[g]^{i}_{jnk}+R[g]^{i}_{knj})X^{n}\right),
 \end{eqnarray}
where $\nabla_{(i}X_{j)}$ is the symmetrization of $\nabla_{i}X_{j}$ i..e, $\nabla_{(i}X_{j)}:=\nabla_{i}X_{j}+\nabla_{j}X_{i}$. Similarly the following holds 
\begin{eqnarray}
\frac{d}{ds}\Gamma^{i}_{jk}[\psi^{*}_{s}\gamma]|_{s=0}&=&\frac{1}{2}\left(\nabla[\gamma]_{(j}\nabla[\gamma]_{k)}X^{i}+(R[\gamma]^{i}_{jnk}+R[\gamma]^{i}_{knj})X^{n}\right).
\end{eqnarray}
 The previous expressions altogether yield
\begin{eqnarray}
L_{X}V^{i}&=&(-2\nabla[g]^{j}X^{k})\left(\Gamma[g]^{i}_{jk}-\Gamma[\gamma]^{i}_{jk}\right)+g^{jk}[\nabla[g]_{j}\nabla[g]_{k}X^{i}\nonumber+R[g]^{i}_{knj}X^{n}\\\nonumber
&&-\nabla[\gamma]_{j}\nabla[\gamma]_{k}X^{i}-R[\gamma]^{i}_{knj}X^{n}]
\end{eqnarray}
which leads to
\begin{eqnarray}
\Delta_{g}X^{i}-R[g]^{i}_{j}X^{j}+(2\nabla^{j}X^{k})(\Gamma[g]^{i}_{jk}-\Gamma[\gamma]^{i}_{jk})+L_{X}V^{i}\\\nonumber =-g^{jk}(\nabla[\gamma]_{j}\nabla[\gamma]_{k}X^{i}+R[\gamma]^{i}_{knj}X^{n}),
\end{eqnarray}
 or 
\begin{eqnarray}
\label{eq:isom}
\Delta_{g}X^{i}-R[g]^{i}_{j}X^{j}+(2\nabla^{j}X^{k})(\Gamma[g]^{i}_{jk}-\Gamma[\gamma]^{i}_{jk})\\\nonumber
=-g^{jk}(\nabla[\gamma]_{j}\nabla[\gamma]_{k}X^{i}+R[\gamma]^{i}_{knj}X^{n}).
\end{eqnarray}
upon imposing the spatial harmonic gauge condition $V^{i}=0$. Given that we've established the relation (\ref{eq:isom}), it is sufficient to show the injectivity of the map $X\mapsto-g^{jk}(\nabla[\gamma]_{j}\nabla[\gamma]_{k}X^{i}+R[\gamma]^{i}_{knj}X^{n})$ (a map from $H^{s+1}(\mathfrak{X}(M))$ to $H^{s-1}(\mathfrak{X}(M))$) in order to prove the isomorphism property (then surjectivity will follow from self-adjointness) of the map $P$. Let $Z\in\ker(X\mapsto-g^{jk}(\nabla[\gamma]_{j}\nabla[\gamma]_{k}X^{i}+R[\gamma]^{i}_{knj}X^{n}))$ i.e., 
\begin{eqnarray}
-g^{jk}(\nabla[\gamma]_{j}\nabla[\gamma]_{k}Z^{i}+R[\gamma]^{i}_{knj}Z^{n})&=&0,
\end{eqnarray}
which upon multiplying both sides by $Z_{i}$ and integrating over $M$ after imposing $V^{i}=0$ yields
\begin{eqnarray}
\int_{M}(-g^{jk}\nabla[\gamma]_{j}Z_{i}\nabla_{k}Z^{i}+g^{jk}R[\gamma]_{iknj}Z^{i}Z^{n})\mu_{g}&=&0.
\end{eqnarray}  
Now $g$ is sufficiently close to $\gamma$ and $Ric_{\gamma}(Z,Z)=-\frac{n-1}{n^{2}}\gamma(Z,Z)<0$. Therefore $g^{jk}R[\gamma]_{iknj}Z^{i}Z^{n}\leq0$ is satisfied leading to
\begin{eqnarray}
Z=0,
\end{eqnarray}
and therefore 
\begin{eqnarray}
\ker(X\mapsto-g^{jk}(\nabla[\gamma]_{j}\nabla[\gamma]_{k}X^{i}+R[\gamma]^{i}_{knj}X^{n}))=\{0\}\nonumber.
\end{eqnarray}
This concludes the proof that $P$ is an isomorphism between $H^{s+1}$ and $H^{s-1}$.\\
Using the previous lemmas, we will prove three additional lemmas which will be crucial towards proving the stability results.\\
\textbf{Lemma 6:} \textit{Let $s>\frac{n}{2}+2$. Let $B_{s,\delta}(\gamma_{0},0)\subset H^{s}\times H^{s-1}$ be a ball of radius $\delta$ centered at $(\gamma_{0},0)$ and $(g,K^{TT})\in B_{s,\delta}(\gamma_{0},0)$. Let $(\tau^{2}-\frac{2n\Lambda}{n-1})>0$, $\partial_{T}=-\frac{\tau^{2}-\frac{2n\Lambda}{n-1}}{\tau}\partial_{\tau}=-\frac{\phi^{2}}{\tau}\partial_{\tau}$, and assume that the CMCSH gauge condition is satisfied. Then there exists a constant $C=C(\delta)>0$ such that the following inequality holds for any $T$ satisfying $-\infty<T_{1}<T<T_{2}<\infty$
\begin{eqnarray}
\label{eq:Nes}
||\frac{N}{n}-1||_{H^{s+1}}&\leq& C||K^{TT}||^{2}_{H^{s-1}}.
\end{eqnarray}  
}
\textbf{Proof}: Let's consider the re-scaled Lapse equation 
\begin{eqnarray}
\Delta_{g}N+(|K^{TT}|^{2}+\frac{1}{n})N&=&1
\end{eqnarray}
and substitute $Q=\frac{N}{n}-1$ i.e., $N=n(1+Q)$. We obtain
\begin{eqnarray}
\label{eq:source}
\Delta_{g}Q+(|K^{TT}|^{2}+\frac{1}{n})Q&=&-|K^{TT}|^{2}.
\end{eqnarray}
Clearly, as $(|K^{TT}|^{2}+\frac{1}{n})>0$, $``\Delta_{g}+(|K^{TT}|^{2}+\frac{1}{n})id "$ is an isomorphism of $H^{s+1}$ onto $H^{s-1}$. Therefore, from the elliptic regularity of $``\Delta_{g}+(|K^{TT}|^{2}+\frac{1}{n})id "$, we may write the following inequality \cite{grigoryan2009heat}
\begin{eqnarray}
||Q||_{H^{s+1}}\leq C||(\Delta_{g}+(|K^{TT}|^{2}+\frac{1}{n})id)Q||_{H^{s-1}},
\end{eqnarray}       
and using equation (\ref{eq:source}), we may immediately write
\begin{eqnarray}
||Q||_{H^{s+1}}\leq C||K^{TT}||^{2}_{H^{s-1}}
\end{eqnarray} 
or 
\begin{eqnarray}
||\frac{N}{n}-1||_{H^{s+1}}\leq C||K^{TT}||^{2}_{H^{s-1}}. 
\end{eqnarray}
We have thus proved the lemma.\\
\textbf{Lemma 7:} \textit{Let $s>\frac{n}{2}+2$. Let $B_{s,\delta}(\gamma_{0},0)\subset H^{s}\times H^{s-1}$ be a ball of radius $\delta$ centered at $(\gamma_{0},0)$ and $(g,K^{TT})\in B_{s,\delta}(\gamma_{0},0)$. Let $(\tau^{2}-\frac{2n\Lambda}{n-1})>0$ and $\partial_{T}=-\frac{\tau^{2}-\frac{2n\Lambda}{n-1}}{\tau}\partial_{\tau}=-\frac{\phi^{2}}{\tau}\partial_{\tau}$, $0<\frac{\phi(\tau)}{\tau}<1$, and assume that the CMCSH gauge condition is satisfied. Then there exists a constant $C=C(\delta)>0$ such that the following inequality holds for any $T$ satisfying $-\infty<T_{1}<T<T_{2}<\infty$
\begin{eqnarray}
\label{eq:Xes}
||X||_{H^{s+1}}&\leq&C(||K^{TT}||_{H^{s-1}}+\frac{\tau}{\phi}||K^{TT}||^{2}_{H^{s-1}}).
\end{eqnarray}  
}
\textbf{Proof:} The elliptic equation (\ref{eq:shift}) for the shift $X$ reads
 \begin{eqnarray}
 \label{eq:shisom}
 \Delta_{g}(\frac{\phi(\tau)}{\tau}X^{i})-R[g]^{i}_{j}(\frac{\phi(\tau)}{\tau}X^{j})+(2\nabla^{j}(\frac{\phi(\tau)}{\tau}X^{k}))(\Gamma[g]^{i}\nonumber_{jk}-\Gamma[\gamma]^{i}_{jk})=\\\nonumber \frac{\phi(\tau)}{\tau}(2NK^{Tjk})(\Gamma[g]^{i}_{jk}-\Gamma[\gamma]^{i}_{jk}) -(2-n)\nabla[g]^{i}(\frac{N}{n}-1)\\\nonumber -\frac{2\phi(\tau)}{\tau}\nabla[g]^{j}NK^{Ti}_{j}+g^{jk}\partial_{T}\Gamma[\gamma]^{i}_{jk},
 \end{eqnarray}
and we have proved in lemma (5) that the operator $P: H^{s+1}\to H^{s-1}$ i.e., 
\begin{eqnarray}
 X^{i}&\mapsto&\Delta_{g}X^{i}-R[g]^{i}_{j}X^{j}+2\nabla^{j}X^{k}(\Gamma[g]^{i}\nonumber_{jk}-\Gamma[\gamma]^{i}_{jk})
\end{eqnarray}
is an isomorphism and thus the following estimate holds
\begin{eqnarray}
\label{eq:estimate}
||X||_{H^{s+1}}&\leq&C||\Delta_{g}X^{i}-R[g]^{i}_{j}X^{j}+2\nabla^{j}X^{k}(\Gamma[g]^{i}\nonumber_{jk}-\Gamma[\gamma]^{i}_{jk})||_{H^{s-1}}.
\end{eqnarray}
Therefore, use of the shift equation yields 
\begin{eqnarray}
\label{eq:estimate2}
||\frac{\phi(\tau)}{\tau}X||_{H^{s+1}}\leq ||g^{mn}\partial_{T}\Gamma[\gamma]^{i}_{mn}||_{H^{s}}.
\end{eqnarray}
Note that every term on the right hand side of the equation (\ref{eq:shisom}) is of second or higher order except the last term $g^{jk}\partial_{T}\Gamma[\gamma]^{i}_{jk}$. Using the estimate (\ref{eq:CRES}), and the re-scaled field equation (\ref{eq:gd1}), and $0<\frac{\phi(\tau)}{\tau}<1$, we obtain 
\begin{eqnarray}
\label{eq:estimatenew}
||g^{mn}\partial_{T}\Gamma[\gamma]^{i}_{mn}||_{H^{s}}\leq C(||g-\gamma||_{H^{s}}+||\gamma-\gamma_{0}||_{H^{s}})(||\frac{\phi(\tau)}{\tau\nonumber(T)}X||_{H^{s+1}}\\\nonumber +||\frac{\phi(\tau)}{\tau}K^{TT}||_{H^{s-1}}+||K^{TT}||^{2}_{H^{s-1}})\\\nonumber
\leq C\frac{\phi(\tau)}{\tau}(||g-\gamma||_{H^{s}}+||\gamma-\gamma_{0}||_{H^{s}})(||X||_{H^{s+1}}+||K^{TT}||_{H^{s-1}}+\frac{\tau}{\phi}||K^{TT}||^{2}_{H^{s-1}}),
\end{eqnarray}
which upon substituting in (\ref{eq:estimate2}) leads to the desired estimate 
\begin{eqnarray}
||X||_{H^{s+1}}&\leq&C(||g-\gamma||_{H^{s}}+||\gamma-\gamma_{0}||_{H^{s}})(||X||_{H^{s+1}}+||K^{TT}||_{H^{s-1}}\\\nonumber &&+\frac{\tau}{\phi}||K^{TT}||^{2}_{H^{s-1}}).
\end{eqnarray}
Obviously, we can find a constant $C=C(\delta)$, such that the following holds for sufficiently small $(||g-\gamma||_{H^{s}}+||\gamma-\gamma_{0}||_{H^{s}})$
\begin{eqnarray}
C(||g-\gamma||_{H^{s}}+||\gamma-\gamma_{0}||_{H^{s}})\leq 1.
\end{eqnarray}
Therefore, we have the following estimate for $X$
 \begin{eqnarray}
||X||_{H^{s+1}}&\leq&C(||K^{TT}||_{H^{s-1}}+\frac{\tau}{\phi}||K^{TT}||^{2}_{H^{s-1}}),
\end{eqnarray}
and thus we have proved the lemma. Notice an important fact that the potentially dangerous term $\frac{\tau}{\phi}$ which satisfies $e^{T}$ growth as $T\to\infty$ appears in the lemma. However, as we shall see, this dangerous factor cancels with its inverse in the energy inequalities which will be derived later.\\
Finally, we obtain an estimate on the term $\frac{\phi(\tau)}{\tau}X+Y^{||}$, which is stated in the next lemma.\\  
\textbf{Lemma 8:} \textit{Let $s>\frac{n}{2}+2$. Let $B_{s,\delta}(\gamma_{0},0)\subset H^{s}\times H^{s-1}$ be a ball of radius $\delta$ centered at $(\gamma_{0},0)$ and $(g,K^{TT})\in B_{s,\delta}(\gamma_{0},0)$. Let $(\tau^{2}-\frac{2n\Lambda}{n-1})>0$ and $\partial_{T}=-\frac{\tau^{2}-\frac{2n\Lambda}{n-1}}{\tau}\partial_{\tau}=-\frac{\phi^{2}}{\tau}\partial_{\tau}$, and assume that the CMCSH gauge condition is satisfied. Then there exists a constant $C=C(\delta)>0$ such that the following inequality holds for any $T$ satisfying $-\infty<T_{1}<T<T_{2}<\infty$
\begin{eqnarray}
\label{eq:Yes}
||\frac{\phi(\tau)}{\tau}X+Y^{||}||_{H^{s+1}}&\leq&C(\frac{\phi(\tau)}{\tau}||g-\gamma||_{H^{s}}||K^{TT}||_{H^{s-1}}+||K^{TT}||^{2}_{H^{s-1}}),
\end{eqnarray}
}
\textbf{Proof:}
Now let's consider $\gamma\in \mathcal{N}$. Thus, $T_{\gamma}\mathcal{N}\ni \partial_{T}\gamma$ may be written as 
\begin{eqnarray}
\partial_{T}\gamma&=&h^{TT}+L_{Y^{||}}\gamma,
\end{eqnarray} 
where $h^{TT}$ is a transverse-traceless tensor and $Y^{||}$ solves the following equation 
\begin{eqnarray}
-[\nabla[\gamma]^{m}\nabla[\gamma]_{m}Y^{i}+R[\gamma]^{i}_{m}Y^{m}]+(h^{TT||}+L_{Y^{||}}\gamma)^{mn}(\Gamma[\gamma]^{k}_{mn}-\Gamma[\gamma_{0}]^{k}_{mn})=0.\nonumber
\end{eqnarray}
We have already shown (\ref{eq:dgamma}) that the following equation holds
\begin{eqnarray}
\gamma^{mn}\partial_{T}\Gamma[\gamma]^{i}_{mn}=\gamma^{mn}D\Gamma[\gamma]^{i}_{mn}\partial_{T}\gamma=\gamma^{mn}D\Gamma[\gamma]^{i}_{mn}(h^{TT}+L_{Y^{||}}\gamma)\\
=\gamma^{mn}D\Gamma[\gamma]^{i}_{mn}L_{Y^{||}}\gamma=(\nabla[\gamma]^{n}\nabla[\gamma]_{n}Y^{||i}+R[\gamma]^{i}_{m}Y^{||m})
\end{eqnarray}

Now adding $(-\nabla[\gamma]^{n}\nabla[\gamma]_{n}((\frac{\phi(\tau)}{\tau}X^{i})-R[\gamma]^{i}_{m}(\frac{\phi(\tau)}{\tau}X^{m}))$ to both sides of equation (\ref{eq:shisom}), we obtain 
\begin{eqnarray}
 \Delta_{g}(\frac{\phi(\tau)}{\tau}X^{i})-R[g]^{i}_{j}(\frac{\phi(\tau)}{\tau}X^{j})+(2\nabla^{j}(\frac{\phi(\tau)}{\tau}X^{k}))(\Gamma[g]^{i}\nonumber_{jk}-\Gamma[\gamma]^{i}_{jk})\\\nonumber+(-\nabla[\gamma]^{n}\nabla[\gamma]_{n}((\frac{\phi(\tau)}{\tau}X^{i})-R[\gamma]^{i}_{m}(\frac{\phi(\tau)}{\tau}X^{m}))=\frac{\phi(\tau)}{\tau}(2NK^{Tjk})(\Gamma[g]^{i}_{jk}-\Gamma[\gamma]^{i}_{jk})\\\nonumber -(2-n)\nabla[g]^{i}(\frac{N}{n}-1)-\frac{2\phi(\tau)}{\tau}\nabla[g]^{j}NK^{Ti}_{j}\\\nonumber
 +g^{jk}\partial_{T}\Gamma[\gamma]^{i}_{jk} +(-\nabla[\gamma]^{n}\nabla[\gamma]_{n}((\frac{\phi(\tau)}{\tau}X^{i})-R[\gamma]^{i}_{m}(\frac{\phi(\tau)}{\tau}X^{m}))
\end{eqnarray}

\begin{eqnarray}
-g^{jk}\nonumber(\nabla[\gamma]_{j}\nabla[\gamma]_{k}(\frac{\phi(\tau)}{\tau}X^{i})+R[\gamma]^{i}_{knj}(\frac{\phi(\tau)}{\tau}X^{n}))+(-\nabla[\gamma]^{n}\nabla[\gamma]_{n}(\frac{\phi(\tau)}{\tau}X^{i})\\\nonumber -R[\gamma]^{i}_{m}(\frac{\phi(\tau)}{\tau}X^{m}))
 =\frac{\phi(\tau)}{\tau}(2NK^{Tjk})(\Gamma[g]^{i}_{jk}-\Gamma[\gamma]^{i}_{jk})-(2-n)\nabla[g]^{i}(\frac{N}{n}-1)\\\nonumber -\frac{2\phi(\tau)}{\tau}\nabla[g]^{j}NK^{Ti}_{j}+g^{jk}\partial_{T}\Gamma[\gamma]^{i}_{jk}
 +(-\nabla[\gamma]^{n}\nabla[\gamma]_{n}((\frac{\phi(\tau)}{\tau}X^{i})-R[\gamma]^{i}_{m}(\frac{\phi(\tau)}{\tau}X^{m}))
\end{eqnarray}

\begin{eqnarray}
(-\nabla[\gamma]^{n}\nabla[\gamma]_{n}(\frac{\phi(\tau)}{\tau}X^{i})-R[\nonumber \gamma]^{i}_{m}(\frac{\phi(\tau)}{\tau}X^{m}))=\frac{\phi(\tau)}{\tau}(2NK^{Tjk})(\Gamma[g]^{i}_{jk}-\Gamma[\gamma]^{i}_{jk})\\\nonumber
 -(2-n)\nabla[g]^{i}(\frac{N}{n}-1)-\frac{2\phi(\tau)}{\tau}\nabla[g]^{j}NK^{Ti}_{j}+g^{jk}\partial_{T}\Gamma[\gamma]^{i}_{jk}\\\nonumber +(\gamma^{mn}-g^{mn})(-\nabla[\gamma]_{m}\nabla[\gamma]_{n}((\frac{\phi(\tau)}{\tau}X^{i})-R[\gamma]^{i}_{mjn}(\frac{\phi(\tau)}{\tau}X^{j})).
\end{eqnarray}
Now adding $-(\nabla[\gamma]^{n}\nabla[\gamma]_{n}Y^{||i}+R[\gamma]^{i}_{m}Y^{||m})=-\gamma^{mn}\partial_{T}\Gamma[\gamma]^{i}_{mn}$ to the both sides of previous equation, we obtain 
\begin{eqnarray}
(-\nabla[\gamma]^{n}\nabla[\gamma]_{n}(\frac{\phi(\tau)}{\tau}X^{i}+Y^{||i})-R[\gamma]^{i}_{m}(\frac{\phi(\tau)}{\tau}X^{m}+Y^{||m}))\\\nonumber =\frac{\phi(\tau)}{\tau}(2NK^{Tjk})(\Gamma[g]^{i}_{jk}-\Gamma[\gamma]^{i}_{jk})
 -(2-n)\nabla[g]^{i}(\frac{N}{n}-1)-\frac{2\phi(\tau)}{\tau}\nabla[g]^{j}NK^{Ti}_{j}\\\nonumber +g^{mn}\partial_{T}\Gamma[\gamma]^{i}_{mn}+\nonumber(\gamma^{mn}-g^{mn})(-\nabla[\gamma]_{m}\nabla[\gamma]_{n}((\frac{\phi(\tau)}{\tau}X^{i})\\\nonumber
 -R[\gamma]^{i}_{mjn}(\frac{\phi(\tau)}{\tau}X^{j}))-\gamma^{mn}\partial_{T}\Gamma[\gamma]^{i}_{mn}
\end{eqnarray}
\begin{eqnarray}
=>(-\nabla[\gamma]^{n}\nabla[\gamma]_{n}(\frac{\phi(\tau)}{\tau}X^{i}+Y^{||i})-R[\gamma]^{i}_{m}(\frac{\phi(\tau)}{\tau}X^{m}+Y^{||\nonumber m}))\\\nonumber 
=\frac{\phi(\tau)}{\tau}(2NK^{Tjk})(\Gamma[g]^{i}_{jk}-\Gamma[\gamma]^{i}_{jk})-(2-n)\nabla[g]^{i}(\frac{N}{n}-1)-\frac{2\phi(\tau)}{\tau}\nabla[g]^{j}NK^{Ti}_{j}\\\nonumber +(\gamma^{mn}-g^{mn})(-\nabla[\gamma]_{m}\nabla[\gamma]_{n}((\frac{\phi(\tau)}{\tau}X^{i})-R[\gamma]^{i}_{mjn}(\frac{\phi(\tau)}{\tau}X^{j}))\\\nonumber 
+(g^{mn}-\gamma^{mn})\partial_{T}\Gamma[\gamma]^{i}_{mn}.
\end{eqnarray}
Now after applying estimate of $\frac{N}{n}-1$, $X$, smoothing operation by shadow gauge $||\frac{\partial \gamma}{\partial T}||_{C^{\infty}}\lesssim ||\frac{\partial g}{\partial T}||_{H^{s-1}}$, the elliptic regularity of the operator $P$, and the algebra property of the space $H^{s}$ for $s>\frac{n}{2}$, we note that every term in the right hand side of the previous equation contributes at least to a second order. Trivial power counting of $\frac{\phi}{\tau}$ yields 
\begin{eqnarray}
||\frac{\phi(\tau)}{\tau}X+Y^{||}||_{H^{s+1}}&\leq&C(\frac{\phi(\tau)}{\tau}||g-\gamma||_{H^{s}}||K^{TT}||_{H^{s-1}}+||K^{TT}||^{2}_{H^{s-1}}),
\end{eqnarray}
for some $C=C(\delta)>0$. This concludes the proof of the lemma. \\
In order to construct a Lyapunov function and to establish its decay property, we need the evolution equations for the small data $(g-\gamma, K^{TT})$. Through the following lemmas and utilizing equations (\ref{eq:gd1})-(\ref{eq:fd1}), we arrive at the final set of evolution equations required to define an energy functional  (Lyapunov function for small data) and obtain its estimate. \\ 
\textbf{Lemma 9:} \textit{Let $(g_{0},K^{TT}_{0},N,X)=(\gamma,0,n,0)$ be a fixed point solution of the re-scaled `Einstein-$\Lambda$' equations, where $R(\gamma)=-\frac{n-1}{n^{2}}\gamma$. Define $u=g-\gamma, v=2nK^{TT}$, and $w=\frac{N}{n}$. The `Einstein-$\Lambda$'  evolution equations are equivalent to the following system
\begin{eqnarray}
\partial_{T}u_{ij}=\frac{\phi}{\tau}wv_{ij}-\frac{\phi}{\tau}X^{m}\nabla[\gamma]_{m}u_{ij}-h^{TT||}_{ij}+2(w-1)(u_{ij}+\gamma_{ij})\\\nonumber -(L_{\frac{\phi}{\tau}X+Y^{||}}\gamma)_{ij}
-\frac{\phi}{\tau}(u_{im}\nabla[\gamma]_{j}X^{m}+u_{mj}\nabla[\gamma]_{i}X^{m}),\\
\partial_{T}v_{ij}=-(n-1)v_{ij}-\frac{\phi}{\tau}n^{2}w\mathcal{L}_{g,\gamma}u_{ij}-\frac{\phi}{\tau}X^{m}\nabla[\gamma]_{m}\nonumber v_{ij}-2\frac{\phi}{\tau}n^{2}w(R[g]_{ij}\\\nonumber
+\frac{n-1}{n^{2}}g_{ij}-\alpha_{ij}) +\frac{2\phi}{\tau}n^{2}\nabla_{i}\nabla_{j}w
+\frac{\phi}{\tau}wv_{im}v^{m}_{j}-2\frac{\phi}{\tau}(w-1)(u_{ij}+\gamma_{ij})\\\nonumber-(n-2)(w-1)v_{ij} -\frac{\phi}{\tau}(v_{im}\nabla[\gamma]_{j}X^{m}
+v_{mj}\nabla[\gamma]_{i}X^{m})+8\frac{\phi(\tau)}{\tau}n^{3}wv_{im}v^{m}_{j},
\end{eqnarray} 
where $\phi^{2}=\tau^{2}-\frac{2n\Lambda}{n-1}>0$, $v^{m}_{j}=g^{ml}v_{lj}$, and $\partial_{T}=-\frac{\phi^{2}}{\tau}\partial_{\tau}.$}\\
\textbf{Proof}: A direct calculation after substituting the transformed variables $u=g-\gamma$, $v=2nK^{TT}$, and $N=nw$ along with the fact that $0\neq\frac{\partial{\gamma}}{\partial{T}}\in T_{\gamma}\mathcal{N}$ and thus $\frac{\partial{\gamma}}{\partial{T}}=h^{TT||}+L_{Y^{||}}\gamma$, we thereby obtain the evolution equation for $u$.\\
Now, we need the following lemma.\\
\textbf{Lemma 10 \cite{andersson2011einstein}:} \textit{The following holds for $R_{ij}[g]$
\begin{eqnarray}
R_{ij}[g]-\alpha_{ij}+\frac{n-1}{n^{2}}g_{ij}&=&\frac{1}{2}\mathcal{L}_{g,\gamma}(g-\gamma)_{ij}+J_{ij},
\end{eqnarray}
where $\alpha_{ij}=\frac{1}{2}(L_{V}g)_{ij}$, $\mathcal{L}_{g,\gamma}h_{ij}=\Delta_{g,\gamma}h_{ij}-2R[\gamma]_{ikjl}h^{kl}$, $\Delta_{g,\gamma}h_{ij}=-\frac{1}{\mu_{g}}\nabla[\gamma]_{m}(g^{mn}\mu_{g}\nabla[\gamma]_{n}h_{ij})$, and $J_{ij}$ satisfies the following estimate
\begin{eqnarray}
||J||_{H^{s-2}}&\leq&C||g-\gamma||^{2}_{H^{s}}.
\end{eqnarray} }
\textbf{Proof:} A direct calculation using the definitions of $\mathcal{L}_{g,\gamma}$ and $\Delta_{g,\gamma}$ yields the result. Note that $\mathcal{L}_{\gamma,\gamma}$ is just $\mathcal{L}$ defined in section (4.1).\\
Using this lemma (10), the evolution equations follow from a direct calculation.\\
\textbf{Lemma 11:} \textit{The evolution equations for $u$ and $v$ are equivalent to the following system
\begin{eqnarray}
\label{eq:uevol}
\partial_{T}u&=&\frac{\phi}{\tau}wv-\frac{\phi}{\tau}X^{m}\nabla[\gamma]_{m}u-h^{TT||}+2(w-1)\gamma+\mathcal{F}_{u},\\
\label{eq:vevol}
\partial_{T}v&=&-(n-1)v-\frac{\phi}{\tau}n^{2}w\mathcal{L}_{g,\gamma}u-\frac{\phi}{\tau}X^{m}\nabla[\gamma]_{m}v+\mathcal{F}_{v},
\end{eqnarray}  
where $(\mathcal{F}_{u})_{ij}=2(w-1)u_{ij}-(L_{\frac{\phi}{\tau}X+Y^{||}}\gamma)_{ij}-\frac{\phi}{\tau}(u_{im}\nabla[\gamma]_{j}X^{m}+u_{mj}\nabla[\gamma]_{i}X^{m})$ and $(\mathcal{F}_{v})_{ij}=-2\frac{\phi}{\tau}n^{2}wJ_{ij}+\frac{2\phi}{\tau}n^{2}\nabla_{i}\nabla_{j}w+\frac{\phi}{\tau}wv_{im}v^{m}_{j}-2\frac{\phi}{\tau}(w-1)(u_{ij}+\gamma_{ij})-(n-2)(w-1)v_{ij}-\frac{\phi}{\tau}(v_{im}\nabla[\gamma]_{j}X^{m}+v_{mj}\nabla[\gamma]_{i})+8\frac{\phi(\tau)}{\tau}n^{3}w v_{im}v^{m}_{j}$, and they satisfy the following estimates 
}\\
\textbf{Proof:} Trivial.
\subsection{\textbf{Linearization}}
Even though we have already established the linearized stability, we may quickly reprove the result using the dynamical equations obtained for the perturbations. Here, we construct an energy functional (Lyapunov function) for the linearized equations, which will motivate the construction of the energy functional for the fully nonlinear stability problem. At this point, we have dynamical equations for perturbations both parallel and perpendicular to $\mathcal{N}$. However, we will see shortly that the parallel component of the perturbation is trivially stable. Once again the fixed points satisfy 
\begin{eqnarray}
w_{0}=\frac{N}{n}=1, X^{i}_{0}=0, u_{0}=\gamma, R(\gamma)=-\frac{n-1}{n^{2}}\gamma, v_{0}=2nK^{TT}_{0}=0.
\end{eqnarray} 
Linearization about these fixed points preserving the gauges and constraints i.e.,  
\begin{eqnarray}
\delta u=u^{TT}, \delta v=v^{TT}, \delta w=0, \delta X=0, \delta (\frac{\phi}{\tau}X+Y^{||})=0, h^{TT||}=\frac{\phi}{\tau}v^{||}
\end{eqnarray}
together with the field equations yield
\begin{eqnarray}
\partial_{T}u^{\perp}&=&\frac{\phi}{\tau}v^{\perp},\\
\partial_{T}v^{||}&=&-(n-1)v^{||},\\
\partial_{T}v^{\perp}&=&-(n-1)v^{\perp}-\frac{\phi(\tau)}{\tau}n^{2}\mathcal{L}_{\gamma,\gamma}u^{\perp},
\end{eqnarray}
where we have used the $L^{2}$ orthogonal decomposition $u^{TT}=u^{TT||}+u^{TT\perp}, v^{TT}=v^{TT||}+v^{TT\perp}$, and $u^{TT||}=0$ (at the linear level, $u$ is $L^{2}$ orthogonal to $\mathcal{N}$). We immediately obtain as $T\to\infty$
\begin{eqnarray}
v^{||}(T)=e^{-(n-1)(T-T_{0})}v^{||}(T_{0})
\end{eqnarray}
The linearized equation for the $L^{2}$ orthogonal component satisfies the following pdes (let's write $u^{\perp}=u$ and $v^{'\perp}=v$ for simplicity)
\begin{eqnarray}
\label{eq:evolutionode1}
\partial_{T}u&=&\frac{\phi}{\tau}v,\\
\label{eq:evolutionode2}
\partial_{T}v&=&-(n-1)v-\frac{\phi(\tau)}{\tau}n^{2}\mathcal{L}_{\gamma,\gamma}u,
\end{eqnarray}
where the operator $\mathcal{L}_{\gamma,\gamma}$ satisfies the eigenvalue equation
\begin{eqnarray}
\mathcal{L}_{\gamma,\gamma}\mathcal{X}&=&\lambda \mathcal{X}
\end{eqnarray}
with 
\begin{eqnarray}
\lambda\geq0.
\end{eqnarray}
Note that the eigentensor corresponding to $\lambda=0$ is tangent to the centre manifold $\mathcal{N}$. Such perturbations are trivially stable as evident from equation (164). Since, on the compact manifold, the spectrum of the second order elliptic operator is essentially discrete, we need to focus on the minimum positive eigenvalue of $\mathcal{L}_{\gamma,\gamma}$. Let the positive minimum of the spectrum of $\mathcal{L}_{\gamma,\gamma}$ be $\lambda_{0}>0$ i.e., $\lambda>\lambda_{0}>0~~\forall~\lambda\in Spec(\mathcal{L})$. Clearly, the coupled pde system can be reduced to the following pair of odes
\begin{eqnarray}
\partial_{T}u&=&\frac{\phi}{\tau}v,\\
\partial_{T}v&=&-(n-1)v-\frac{\phi(\tau)}{\tau}n^{2}\lambda u.
\end{eqnarray}
In the linearized analysis section, we have already constructed a Lyapunov function for this system. The most natural energy (Lyapunov function) may be defined as follows
\begin{eqnarray}
\mathcal{E}&=&\frac{1}{2}v^{2}+\frac{n^{2}\lambda}{2}u^{2}.
\end{eqnarray}
The energy $\mathcal{E}$ is positive semi-definite and vanishes precisely when $(u,v)\equiv 0$, that is, at the fixed points.
The time derivative of the energy reads 
\begin{eqnarray}
d_{T}\mathcal{E}=-(n-1)v^{2}<0
\end{eqnarray}
Therefore, we observe that  the energy $\mathcal{E}$ is monotonically decaying. Utilizing well known theorem of dynamical system, strictly monotonic decay of Lyapunov function implies asymptotic stability. As $T\to\infty$, we have the following decay
\begin{eqnarray}
v(T)\lesssim e^{-T}, \partial_{T}u\lesssim e^{-2T},|u-u^{*}|\lesssim e^{-2T}.
\end{eqnarray}
Here we have utilized the fact that $\frac{\phi(\tau)}{\tau}\sim e^{-T}$ as $T\to\infty$. Using the evolution equation, we observe that the re-scaled metric converges to a limit metric (in the space of metrics with constant scalar curvature $-\frac{n-1}{n}$) as $T\to\infty$.

\subsection{\textbf{Non-linear perturbations}}
From here onward, we will focus on fully non-linear perturbations to the background solutions. Let us fix a background metric $\gamma_{0}\in \mathcal{E}in_{-\frac{n-1}{n^{2}}}$. Let $\mathcal{N}$ be the deformation space with respect to $\gamma_{0}$ and assume $\gamma$ is close to $\gamma_{0}$. There exists a harmonic slice $S_{\gamma}\subset \mathcal{M}$ as the solution of the following equation satisfied by the tension field equation i.e., 
\begin{eqnarray}
V^{i}=g^{jk}\left(\Gamma[g]^{i}_{jk}-\Gamma[\gamma]^{i}_{jk}\right)=0.
\end{eqnarray}
We want to consider $(g\in\mathcal{M}, K^{TT})$ which satisfies the constraint equations (\ref{eq:con1}-\ref{eq:con2}) as well as the harmonicity condition that the idenitity map `$id: (M,g)\to (M,\gamma)$' is harmonic. Let us denote this constraint slice by $\mathcal{S}_{c,\gamma}$ corresponding to $\mathcal{S}_{\gamma}$. Following the analysis of \cite{andersson2004future} (see lemma 2.3), we may represent the constraint slice $\mathcal{S}_{c,\gamma}$ as a graph over its tangent space, that is, we may write $(g,K^{TT})\in\mathcal{S}_{c,\gamma}$ in the following form
\begin{eqnarray}
g-\gamma&=&u^{TT}+z,\\
2nK^{TT}&=&v^{TT}+r,
\end{eqnarray}
where $u^{TT}$ and $v^{TT}$ are transverse-traceless with respect to $\gamma$
with $<z|u^{TT}>_{L^{2}}=0,$ $<v^{TT}|r>_{L^{2}}=0$, and $||z||_{H^{s}}\leq C(||u^{TT}||^{2}_{H^{s}}+||v^{TT}||^{2}_{H^{s-1}})$, and $||r||_{H^{s-1}}\leq C(||u^{TT}||^{2}_{H^{s}}+||v^{TT}||^{2}_{H^{s-1}})$ for $C>0$. From here onward, we will write $u$ and $v$ (resp. $v$) for $u^{TT}+z$ and $v^{TT}+r$, respectively for simplicity.

\section{Constructing the Lyapunov functional: definition of Energy}
The spectrum of the self-adjoint operator $\mathcal{L}_{g,\gamma}$ will play a vital role in the definition of the energy. In general for a closed manifold, the spectrum of $\mathcal{L}_{g,\gamma}$ is non-negative (because, we have assumed that the compact negative Einstein spaces are stable) i.e., $\lambda$ satisfying 
\begin{eqnarray}
\mathcal{L}_{g,\gamma}\mathcal{X}=\lambda \mathcal{X}
\end{eqnarray}
also satisfies 
\begin{eqnarray}
\lambda\geq 0. 
\end{eqnarray}
We will observe later that $\lambda=0$ case is trivially stable provided the smallness condition ($H^{s}\times H^{s-1}$ norm) on the initial data is met. Let us re-write down the fully non-linear evolution equations for $(u,v)\in B_{s,\delta}(0,0)$,
\begin{eqnarray}
\partial_{T}u&=&\frac{\phi}{\tau}wv-\frac{\phi}{\tau}X^{m}\nabla[\gamma]_{m}u-h^{TT||}+2(w-1)\gamma_{ij}+\mathcal{F}_{u},\\
\label{eq:vevol}
\partial_{T}v&=&-(n-1)v-\frac{\phi}{\tau}n^{2}w\mathcal{L}_{g,\gamma}u-\frac{\phi}{\tau}X^{m}\nabla[\gamma]_{m}v+\mathcal{F}_{v},
\end{eqnarray}  
where $(\mathcal{F}_{u})_{ij}=2(w-1)u_{ij}-(L_{\frac{\phi}{\tau}X+Y^{||}}\gamma)_{ij}-\frac{\phi}{\tau}(u_{im}\nabla[\gamma]_{j}X^{m}+u_{mj}\nabla[\gamma]_{i}X^{m})$ and $(\mathcal{F}_{v})_{ij}=-2\frac{\phi}{\tau}n^{2}wJ_{ij}+\frac{2\phi}{\tau}n^{2}\nabla_{i}\nabla_{j}w+\frac{\phi}{\tau}wv_{im}v^{m}_{j}-2\frac{\phi}{\tau}(w-1)(u_{ij}+\gamma_{ij})-(n-2)(w-1)v_{ij}-\frac{\phi}{\tau}(v_{im}\nabla[\gamma]_{j}X^{m}+v_{mj}\nabla[\gamma]_{i})+8\frac{\phi(\tau)}{\tau}n^{3}w v_{im}v^{m}_{j}$, and they roughly satisfy a third order estimates. The exact estimates will be derived later (when necessary).\\
Motivated by the energy associated with the linear stability analysis, we define a natural wave equation type of energy (can be read off from the evolution equations) as follows
\begin{eqnarray}
\mathcal{E}_{i}=\frac{1}{2}<v|\mathcal{L}^{i-1}_{g,\gamma}v>_{L^{2}}+\frac{n^{2}}{2}<u|\mathcal{L}^{i}_{g,\gamma}u>_{L^{2}}\\\nonumber
=\frac{1}{2}\int_{M}(v_{ij}\mathcal{L}^{i-1}_{g\gamma}v_{kl})\gamma^{ik}\gamma^{jl}\mu_{g}+\frac{n^{2}}{2}\int_{M}(u_{ij}\mathcal{L}^{i}_{g\gamma}u_{kl})\gamma^{ik}\gamma^{jl}\mu_{g}.
\end{eqnarray}
The lowest order term $\mathcal{E}_{1}$ may be explicitly calculated as follows 
\begin{eqnarray}
\mathcal{E}_{1}=\frac{1}{2}\int_{M}v_{ij}v_{kl}\gamma^{ik}\gamma^{jl}\mu_{g}+\frac{n^{2}}{2}\nonumber\int_{M}(\nabla[\gamma]_{m}u_{ij}\nabla[\gamma]_{n}u_{kl}g^{mn}\gamma^{ik}\gamma^{jl}\\ -2R[\gamma]_{i}~^{m}~_{j}~^{n}u_{mn}u_{kl}\gamma^{ik}\gamma^{jl})\mu_{g}.
\end{eqnarray}

The total energy may be defined by summing all order energies up to $s$
\begin{eqnarray}
\label{eq:energynonlinear}
\mathcal{E}_{s}=\sum_{i=1}^{s}\mathcal{E}_{i}.
\end{eqnarray}
This energy is positive semi-definite and it vanishes only when $(u,v)\equiv0$, that is, on the centre manifold. We will now check the non-negative definiteness of the hessian of the energy functional, which will be used to obtain several useful estimates. The first variation of the energy with $\delta u=h$, and $\delta v=k$ at $(0,0)$ vanishes 
\begin{eqnarray}
D\mathcal{E}_{s}(h,k)&=&0,
\end{eqnarray} 
i.e., $(u,v)=(0,0)$ is a critical point of $\mathcal{E}_{s}$. The second variation about the critical point yields
\begin{eqnarray}
D^{2}\mathcal{E}_{s}((h,k),(h,k))=\sum_{i=1}^{s}<k|\mathcal{L}_{\gamma,\gamma}^{i-1}k>_{L^{2}}+n^{2}\sum_{i=1}^{s}<h|\nonumber\mathcal{L}^{i}_{\gamma,\gamma}h>_{L^{2}}\nonumber 
\end{eqnarray}
and we immediately obtain the positive semi-definiteness of the hessian of energy using the spectrum of $\mathcal{L}_{\gamma,\gamma}$
\begin{eqnarray}
D^{2}\mathcal{E}_{s}((h,k),(h,k))\geq0
\end{eqnarray}
with equality holding if and only if $h=h^{TT||}$ and $k=0$. Therefore $(0,0)$ is a non-degenerate critical point of $\mathcal{E}_{s}$. Once we have established the positive semi-definiteness of the hessian of the energy functional, we use this property to obtain a control of the desired $H^{s}\times H^{s-1}$ norm of the data $(u,v)$ in terms of the energy. The following two lemmas will in fact provide such control of the desired norm.\\ 
\textbf{Lemma 12:} \textit{ Let ($\gamma, g, K^{TT}$) be such that $(g-\gamma)$ satisfies the shadow gauge and $g-\gamma=u$, $2nK^{TT}=v$. There exists a constant $\delta>0$ sufficiently small, and a constant $C=C(\delta)>0$ such that if $(u,v)\in \mathcal{B}_{\delta}(0,0)\in H^{s}\times H^{s-1}$, the following estimate holds 
\begin{eqnarray}
||u^{||}||_{H^{s}}\leq C \left(||u^{\perp}||^{2}_{H^{s}}+||v||^{2}_{H^{s-1}}\right).
\end{eqnarray} 
}
\textbf{Proof:} Following the shadow gauge ($u$ is $L^{2}$-orthogonal to $\mathcal{N}$), we may write 
\begin{eqnarray}
<(g-\gamma),h^{TT||}+L_{Y^{||}}\gamma>_{L^{2}}&=&0\\\nonumber
<u^{||}+u^{\perp}+z,h^{TT||}+L_{Y^{||}}\gamma>_{L^{2}}&=&0\\\nonumber
<u^{||},h^{TT||}>_{L^{2}}+<z,h^{TT||}+L_{Y^{||}}\gamma>_{L^{2}}&=&0
\end{eqnarray}   
where we have used the facts that $<u^{\perp},h^{TT||}>_{L^{2}}=0$ and $<u^{TT},L_{Y^{||}}\gamma>_{L^{2}}=0$. Using the relation obtained, we may say that $u^{||}$ is a smooth function of $z$ which satisfies 
\begin{eqnarray}
||z||_{H^{s}}\leq C(||u^{TT}||^{2}_{H^{s}}+||v^{TT}||^{2}_{H^{s-1}}),
\end{eqnarray} 
and therefore, $u^{||}$ satisfies the following estimate 
\begin{eqnarray}
||u^{||}||_{H^{s}}\leq C\left(||u^{\perp}||^{2}_{H^{s}}+||v||^{2}_{H^{s-1}}\right).
\end{eqnarray}
Using the above definition of the energy together with the positive definiteness of its hessian at $(0,0)$, we have the following crucial lemma which together with the lemma (12) will yield a control of the $H^{s}\times H^{s-1}$ norm of the data $(u,v)$ in terms of the energy.\\
\textbf{Lemma 13:} \textit{Let $s>\frac{n}{2}+2$, $\gamma\in\mathcal{E}in_{-\frac{n-1}{n^{2}}}$, and $\mathcal{E}_{s}$ be the the total energy defined in (\ref{eq:energynonlinear}). Then $\exists~\delta>0, C=C(\delta)>0$, such that $\forall (u,v)\in \mathcal{B}_{\delta}(0,0)\in H^{s}\times H^{s-1}$, the following estimate holds
\begin{eqnarray}
||u^{\perp}||^{2}_{H^{s}}+||v||^{2}_{H^{s-1}}\leq C\mathcal{E}_{s}.
\end{eqnarray} 
}
for $\frac{\phi}{\tau}\geq\delta$.\\
\textbf{Proof:} We have observed the positive semi-definiteness of the hessian of the energy functional while restricted to the subspace $\mathcal{H}^{s}_{TT}=H^{s}_{TT\perp}\times H^{s-1}_{TT}$ (in local co-ordinates $(u^{\perp},v)$). Thus, $D^{2}\mathcal{E}_{s}: \mathcal{H}^{s}_{TT}\to Image(D^{2}\mathcal{E}_{s})$ is an isomorphism leading to the following inequality 
\begin{eqnarray}
||u^{\perp}||^{2}_{H^{s}}+||v||^{2}_{H^{s-1}}\leq C D^{2}\mathcal{E}_{s}\cdot((h,k),(h,k)),
\end{eqnarray}  
for some finite $C>0$. Now, using a version of the Morse lemma (Hilbert space version) on the non-degenerate critical point $(0,0)$, we obtain that there exists a $\delta>0$ such that for variations lying within $B_{\delta}(0,0)$ and restricted to $\mathcal{H}^{s}_{TT}$, the following holds up to a possibly non-linear diffeomorphism $\mathcal{E}_{s}=\mathcal{E}_{s}(0,0)+D^{2}\mathcal{E}_{s}\cdot ((h,k),(h,k))=D^{2}\mathcal{E}_{s}\cdot ((h,k),(h,k))$ (notice that $\mathcal{E}_{s}(0,0)=0$). Therefore, we prove the lemma
\begin{eqnarray}
||u^{\perp}||^{2}_{H^{s}}+||v||^{2}_{H^{s-1}}\leq C\mathcal{E}_{s}.
\end{eqnarray} 

Using the previous two lemmas (12 and 13), we immediately obtain the following crucial result 
\begin{eqnarray}
\label{eq:gestimate}
||u||^{2}_{H^{s}}+||v||^{2}_{H^{s-1}}\leq C\mathcal{E}_{s},
\end{eqnarray}
 which clearly shows that the energy controls the desired norm of the data $(u,v)$.

\section{Decay of the energy (or the Lyapunov function)}
In this section we study the time evolution of the total energy functional. In order to obtain the decay property of the energy, we state several necessary lemmas. Utilizing these lemmas, we compute the time evolution for the lowest order energy and following analogous calculations, the time evolution of higher order energies are obtained. Since $||A||_{H^{s_{1}}}\lesssim ||A||_{H^{s_{2}}}$ for $s_{1}<s_{2}$, we will bound every Sobolev norm of a tensor or a vector field by its maximum available Sobolev norm. Occasionally we will use the Sobolev embedding $||A||_{L^{\infty}}\lesssim ||A||_{H^{a}}$ for $a>\frac{n}{2}$ and the following product estimates 
\begin{eqnarray}
||AB||_{H^{s}}\lesssim (||A||_{L^{\infty}}||B||_{H^{s}}+||A||_{H^{s}}||B||_{L^{\infty}}),s>0,\\
||AB||_{H^{s}}\lesssim ||A||_{H^{s}}||B||_{H^{s}}, s>\frac{n}{2},\\\nonumber
||[P,A]B||_{H^{a}}\lesssim (||\nabla A||_{L^{\infty}}||B||_{H^{s+a-1}}+||A||_{H^{s+a}}||B||_{L^{\infty}}), P\in \mathcal{OP}^{s}, s>0, a\geq 0,
\end{eqnarray}
where $\mathcal{OP}^{s}$ denotes the pseudo-differential operators with symbol in the Hormander class $S^{s}_{1,0}$ (see \cite{Taylor} for details). The first and second inequalities essentially emphasize the algebra property of $H^{s}\cap L^{\infty}$ for $s>0$ and of $H^{s}$ for $s>\frac{n}{2}$, respectively. In addition, we of course use integration by parts, Holder's and Minkowski inequality whenever necessary.\\ 
\textbf{Lemma 14:} \textit{Let $s>\frac{n}{2}+2$, $\gamma\in\mathcal{E}in_{-\frac{n-1}{n^{2}}}$ be the shadow of $g\in\mathcal{M}$, $g-\gamma=u,~2nK^{TT}=v$, and assume there exists a $\delta>0$ such that $(u,v)\in \mathcal{B}_{\delta}(0,0)\subset H^{s}\times H^{s-1}$, then the following estimates hold  
\begin{eqnarray}
(1) |\int_{M}<\mathcal{L}_{g,\gamma}u,h^{TT||}>\mu_{g}|\leq C\frac{\phi}{\tau}||u||^{2}_{H^{s}}||v||_{H^{s-1}},\\\nonumber
(2)\int_{M}<u,\partial_{T}\mathcal{L}_{g,\gamma}u>\mu_{g}= \int_{M}<\partial_{T}u|\mathcal{L}_{g,\gamma}u>\mu_{g}+\mathcal{R}, |\mathcal{R}|\leq C||u||^{2}_{H^{s}}||v||_{H^{s-1}},\\\nonumber
(3) \int_{M}<v,v>g^{ij}\partial_{T}g_{ij}\mu_{g}\leq C(\frac{\phi}{\tau}||v||^{3}_{H^{s-1}}+||v||^{4}_{H^{s-1}}),\\\nonumber
(4)\int_{M}<u,\mathcal{L}_{g,\gamma}u>g^{ij}\partial_{T}g_{ij}\mu_{g})\leq C(\frac{\phi}{\tau}||u||^{2}_{H^{s}}||v||_{H^{s-1}}+||u||^{2}_{H^{s}}||v||^{2}_{H^{s-1}}),\\\nonumber
(5) |\int_{M}\frac{\phi}{\tau}<v,X^{m}\nabla[\gamma]_{m}v>\mu_{g}|\leq C(\frac{\phi}{\tau}||v||^{3}_{H^{s-1}}+||v||^{4}_{H^{s-1}}),\\\nonumber
(6) |\int_{M}\frac{\phi}{\tau}<\mathcal{L}_{g,\gamma}u,X^{m}\nabla[\gamma]_{m}u>\mu_{g}|\leq C(\frac{\phi}{\tau}||u||^{2}_{H^{s}}||v||_{H^{s-1}}+||u||^{2}_{H^{s}}||v||^{2}_{H^{s-1}}),
\end{eqnarray}
and $C=C(\delta)>0$.}\\
\textbf{Proof:} 
(1)
Using the self-adjoint property of $\mathcal{L}_{g,\gamma}$, we may write 
\begin{eqnarray}
\label{eq:2}
\int_{M}<\mathcal{L}_{g,\gamma}u,h^{TT||}>\mu_{g}&=&\int_{M}<u,\mathcal{L}_{g,\gamma}h^{TT||}>\mu_{g}.
\end{eqnarray}
We have 
\begin{eqnarray}
(\Delta_{g,\gamma}h)_{ij}&=&-\frac{1}{\mu_{g}}\nabla[\gamma]_{m}\left(g^{mn}\mu_{g}(\nabla[\gamma]_{n}h_{ij})\right)
\end{eqnarray}
and the definition of $\mathcal{L}_{g,\gamma}$ 
\begin{eqnarray}
\mathcal{L}_{g,\gamma}h_{ij}&=&\Delta_{g,\gamma}h_{ij}-2R[\gamma]_{ikjl}h^{kl},\\\nonumber
&=&-g^{mn}(\nabla[\gamma]_{m}\nabla[\gamma]_{n}h_{ij})-V^{m}\nabla[\gamma]_{m}h_{ij}-2R[\gamma]_{ikjl}h^{kl},\\\nonumber
&=&-(g^{mn}-\gamma^{mn})(\nabla[\gamma]_{m}\nabla[\gamma]_{n}h_{ij})-\gamma^{mn}\nabla[\gamma]_{m}\nabla[\gamma]_{n}h_{ij}\\\nonumber &&-2R[\gamma]_{ikjl}h^{kl},\\\nonumber
&=&-(g^{mn}-\gamma^{mn})(\nabla[\gamma]_{m}\nabla[\gamma]_{n}h_{ij})+\mathcal{L}_{\gamma,\gamma}h_{ij},
\end{eqnarray}
where we have used the identity $\nabla[\gamma]_{m}(\mu_{g}g^{-1})^{mn}=-V^{n}\mu_{g}$, and set $V^{m}=0$. Replacing $h$ by $h^{TT||}$
\begin{eqnarray}
\label{eq:1}
\mathcal{L}_{g,\gamma}h^{TT||}_{ij}=-(g^{mn}-\gamma^{mn})(\nabla[\gamma]_{m}\nabla[\gamma]_{n}h^{TT||}_{ij})
\end{eqnarray}
as a consequence of $\mathcal{L}_{\gamma,\gamma}h^{TT||}=0$. Now we will exploit the shadow gauge condition to obtain an estimate of $h^{TT||}$. The shadow gauge reads
\begin{eqnarray}
<g-\gamma|\frac{\partial \gamma}{\partial q^{\alpha}}>_{L^{2}}&=&0,\\
<u|h^{TT||\alpha}+L_{Y^{||\alpha}}\gamma>_{L^{2}}&=&0
\end{eqnarray}
which upon time differentiation becomes 
\begin{eqnarray}\nonumber
<\partial_{T}u|h^{TT||\alpha}+L_{Y^{||\alpha}}\gamma>_{L^{2}}+second~order~terms =0,\\\nonumber
<\frac{\phi}{\tau}wv-\frac{\phi(\tau)}{\tau}X^{m}\nabla[\gamma]_{m}u-h^{TT||}+\mathcal{F}_{u}|h^{TT||\alpha}+L_{Y^{||\alpha}}\gamma>_{L^{2}}\\\nonumber +second~order~terms =0.
\end{eqnarray}
Now, using the estimates on $(w-1)$, $X$, and $\mathcal{F}_{u}$, and the identity $<A^{TT}|L_{Z}\gamma>_{L^{2}}=0$ for any transverse-traceless tensor $A^{TT}$ and vector field $Z\in \mathfrak{X}(M)$, we immediately obtain
\begin{eqnarray}
<\frac{\phi}{\tau}v-h^{TT||}|h^{TT||\alpha}>_{L^{2}}+second~order~terms=0
\end{eqnarray}
which leads to
\begin{eqnarray}
h^{TT||}=\frac{\phi}{\tau}v^{||}+ second~order~terms,
\end{eqnarray}
where $v^{||}$ is the projection of $v$ onto the subspace of TT tensors belonging to the kernel of $\mathcal{L}_{g,\gamma}$. Now using the equation (\ref{eq:1}), we observe that every term of $<u,\mathcal{L}_{g,\gamma}h^{TT||}>_{L^{2}}$ is of at least third order and the following claim follows
\begin{eqnarray}
|\int_{M}<\mathcal{L}_{g,\gamma}u,h^{TT||}>_{L^{2}}\mu_{g}|\leq C\frac{\phi}{\tau}||u||^{2}_{H^{s}}||v||_{H^{s-1}}.
\end{eqnarray}

(2) We need the estimate for the term $<u|\partial_{T}\mathcal{L}_{g\gamma}u>_{L^{2}}$. Using the explicit expression for $\mathcal{L}_{g,\gamma}$, we may write 
\begin{eqnarray}
\partial_{T}\mathcal{L}_{g,\gamma}u_{ij}&=&\partial_{T}\left(\Delta_{g,\gamma}u_{ij}-2R[\gamma]_{ikjl}u^{kl}\right),\\\nonumber
&=&\partial_{T}\left(-g^{mn}(\nabla[\gamma]_{m}\nabla[\gamma]_{n}u_{ij})-V^{m}\nabla[\gamma]_{m}u_{ij}-2R[\gamma]_{ikjl}u^{kl}\right)
\end{eqnarray}
and imposing spatial harmonic gauge $V^{i}=0$
\begin{eqnarray}
\partial_{T}\mathcal{L}_{g,\gamma}u_{ij}&=&\partial_{T}\left(-g^{mn}(\nabla[\gamma]_{m}\nabla[\gamma]_{n}u_{ij})-2R[\gamma]_{ikjl}u^{kl}\right),\\\nonumber
&=&\partial_{T}(-g^{mn}(\nabla[\gamma]_{m}\nabla[\gamma]_{n}u_{ij})-2(DR[\gamma]_{ikjl}\cdot \partial_{T}\gamma)u^{kl}\\\nonumber
&&-2R[\gamma]_{ikjl}\partial_{T}u^{kl}.
\end{eqnarray}
Let the operator $g^{mn}\nabla[\gamma]_{m}\nabla[\gamma]_{n}$ be denoted as $\mathcal{D}$, then we write 
\begin{eqnarray}
\partial_{T}\mathcal{D}[g,\gamma]u_{ij}=(\frac{\partial\mathcal{D}[g,\gamma]}{\partial g}\cdot\partial_{T}g+\frac{\partial\mathcal{D}[g,\gamma]}{\partial \gamma}\cdot\partial_{T}\gamma)u_{ij}+\mathcal{D}[g,\gamma]\partial_{T}u_{ij}
\end{eqnarray}
which yields 
\begin{eqnarray}
\partial_{T}\mathcal{L}_{g,\gamma}u_{ij}=\mathcal{L}_{g,\gamma}\partial_{T}u_{ij}-(\frac{\partial\mathcal{D}[g,\gamma]}{\partial g}\cdot \partial_{T}g+\frac{\partial\mathcal{D}[g,\gamma]}{\partial \gamma}\cdot\partial_{T}\gamma)u_{ij}\\\nonumber 
-2(DR[\gamma]_{ikjl}\cdot \partial_{T}\gamma) u^{kl}.
\end{eqnarray}
Since the metric $\gamma$ is now time dependent, we need to control the terms involving $\partial_{T}\gamma$. Fortunately, we do have the shadow gauge at our disposal. Utilizing the smoothing operation via shadow gauge one readily bounds $\frac{\partial\gamma}{\partial T}$ in terms of $\frac{\partial g}{\partial T}$ 
\begin{eqnarray}
||\partial_{T}\gamma||_{H^{s}}\leq C||\partial_{T}g||_{H^{s-1}}.
\end{eqnarray}
On the other hand, we of course have the estimate for $\frac{\partial g}{\partial T}$ from the evolution equation
\begin{eqnarray}
||\frac{\partial g}{\partial T}||_{H^{s-1}}\lesssim (\frac{\phi}{\tau}||K^{TT}||_{H^{s-1}}+\frac{\phi}{\tau}||X||_{H^{s+1}}+||K^{TT}||^{2}_{H^{s-1}})
\end{eqnarray}
which yields through $||X||_{H^{s+1}}\lesssim (||K^{TT}||_{H^{s-1}}+\frac{\tau}{\phi}||K^{TT}||^{2}_{H^{s-1}})$
\begin{eqnarray}
||\partial_{T}\gamma||_{H^{s}}\lesssim ||K^{TT}||_{H^{s-1}}.
\end{eqnarray}
In fact, following the $C^{\infty}$ topology of the deformation space $\mathcal{N}$, one may write 
\begin{eqnarray}
||\partial_{T}\gamma||_{C^{\infty}}\lesssim ||K^{TT}||_{H^{s-1}}.
\end{eqnarray}
Therefore, one is led to the conclusion that the term $-(\frac{\partial\mathcal{D}[g,\gamma]}{\partial g}\cdot\partial_{T}g+\frac{\partial\mathcal{D}[g,\gamma]}{\partial \gamma}\cdot\partial_{T}\gamma)u_{ij}-2(DR[\gamma]_{ikjl}\cdot \partial_{T}\gamma) u^{kl}$ satisfies a second order estimate. This estimates together yields the desired result
\begin{eqnarray}
<u|\partial_{T}\mathcal{L}_{g\gamma}u>_{L^{2}}&=&<\partial_{T}u|\mathcal{L}_{g,\gamma}u>_{L^{2}}+C||u||^{2}_{H^{s}}||v||_{H^{s-1}},
\end{eqnarray}
which concludes the proof of the second part.
(3,4) Let us explicitly compute $g^{ij}\frac{\partial g_{ij}}{\partial T}$ using the evolution equation 
\begin{eqnarray}
g^{ij}\frac{\partial g_{ij}}{\partial T}=-2n(1-\frac{N}{n})-\frac{\phi}{\tau}(\nabla[g]_{i}X^{i}).
\end{eqnarray}
Therefore, we obtain the following expressions 
\begin{eqnarray}
\int_{M}<v,v>g^{ij}\partial_{T}g_{ij}\mu_{g}=-2n\int_{M}<v,v>\left((1-\frac{N}{n})\nonumber+\frac{\phi}{\tau}\nabla[g]_{i}X^{i}\right)\mu_{g},\\\nonumber 
\int_{M}<u,\mathcal{L}_{g,\gamma}u>g^{ij}\partial_{T}g_{ij}\mu_{g}=-2n\int_{M}<u,\mathcal{L}_{g,\gamma}u>\left((1-\frac{N}{n})\nonumber+\frac{\phi}{\tau}\nabla[g]_{i}X^{i}\right)\mu_{g}.
\end{eqnarray}
Now we invoke the elliptic equation for the lapse $N$ 
\begin{eqnarray}
\Delta_{g}N+(|K^{TT}|^{2}+\frac{1}{n})N=1.
\end{eqnarray}
An straightforward maximum principle yields the following estimate for $N$
\begin{eqnarray}
N\leq n
\end{eqnarray}
throughout $M$ yielding $1-\frac{N}{n}\geq 0$ and since, $<v,v>$ and $<u,\mathcal{L}_{g,\gamma}u>$ are non-negative definite, we may immediately write 
\begin{eqnarray}
\int_{M}<v,v>g^{ij}\partial_{T}g_{ij}\mu_{g}\leq -2n\frac{\phi}{\tau}\int_{M}<v,v>\nabla[g]_{i}X^{i}\mu_{g},\\\nonumber 
\int_{M}<u,\mathcal{L}_{g,\gamma}u>g^{ij}\partial_{T}g_{ij}\mu_{g}\leq -2n\frac{\phi}{\tau}\int_{M}<u,\mathcal{L}_{g,\gamma}u>\nabla[g]_{i}X^{i}\mu_{g}
\end{eqnarray}
yielding the desired estimates 
\begin{eqnarray}
\int_{M}<v,v>g^{ij}\partial_{T}g_{ij}\mu_{g}\leq C(\frac{\phi}{\tau}||v||^{3}_{H^{s-1}}+||v||^{4}_{H^{s-1}}),\\\nonumber
\int_{M}<u,\mathcal{L}_{g,\gamma}u>g^{ij}\partial_{T}g_{ij}\mu_{g}\leq C(\frac{\phi}{\tau}||u||^{2}_{H^{s}}||v||_{H^{s-1}}+||u||^{2}_{H^{s}}||v||^{2}_{H^{s-1}})
\end{eqnarray}
upon utilizing the estimate of $X$ from lemma 7.  
(4) and (5) are straightforward to obtain using the estimates of $X,N$ in terms of $v$. Importantly, it is clear that each one satisfies a third order estimate. 
The following lemma characterizes the temporal evolution of the lowest order energy. The higher order energy behaviour may be computed in a similar way.\\
\textbf{Lemma 15:} \textit{Let $s>\frac{n}{2}+2$, $\gamma\in\mathcal{E}in_{-\frac{n-1}{n^{2}}}$ be the shadow of $g\in\mathcal{M}$, and assume there exists a $\delta>0$ such that $(u,v)\in \mathcal{B}_{\delta}(0,0)\subset H^{s}\times H^{s-1}$, , then the following holds 
\begin{eqnarray}
\partial_{T}\mathcal{E}_{1}=-(n-1)<v|v>_{L^{2}}+ \mathcal{A}_{1},
\end{eqnarray}
with $\mathcal{A}_{1}$ satisfying 
\begin{eqnarray}
\mathcal{A}_{1}\leq C(||u||_{H^{s}}||v||^{2}_{H^{s-1}}+||u||^{2}_{H^{s}}||v||^{2}_{H^{s}}+||v||^{4}_{H^{s-1}}+\frac{\phi}{\tau}||u||_{H^{s}}||v||^{2}_{H^{s-1}}\\\nonumber +\frac{\phi}{\tau}||u||^{2}_{H^{s}}||v||_{H^{s-1}}+\frac{\phi}{\tau}||v||^{3}_{H^{s-1}}),\nonumber
\end{eqnarray}
and $C=C(\delta)>0$.} \\
\textbf{Proof:} A direct calculation using equations (\ref{eq:uevol})-(\ref{eq:vevol}) yields 
\begin{eqnarray}
\partial_{T}\mathcal{E}_{1}=<v|\partial_{T}v>_{L^{2}}+\frac{n^{2}}{2}<\partial_{T}u|\mathcal{L}_{g,\gamma}u>_{L^{2}}+\frac{n^{2}}{2}<u|\partial_{T}\mathcal{L}\nonumber _{g,\gamma}u>_{L^{2}}\\\nonumber
+\frac{1}{4}\int_{M}((v_{ij}v_{kl})\gamma^{ik}\gamma^{jl}+n^{2}(u_{ij}\mathcal{L}_{g\gamma}u_{kl})\gamma^{ik}\gamma^{jl})g^{mn}\frac{\partial g_{mn}}{\partial T}\mu_{g}\\\nonumber
-\int_{M}((v_{ij}v_{kl})\mu_{g}+n^{2}(u_{ij}\mathcal{L}_{g\gamma}u_{kl}))\gamma^{im}\gamma^{kn}\gamma^{jl}\frac{\partial \gamma_{mn}}{\partial T}\mu_{g}\\\nonumber 
=<v|-(n-1)v-\frac{\phi}{\tau}n^{2}w\mathcal{L}_{g,\gamma}u-\frac{\phi}{\tau}X^{m}\nabla[\gamma]_{m}v+F_{v}>_{L^{2}}\\\nonumber
+\frac{n^{2}}{2}<\frac{\phi}{\tau}wv-\frac{\phi}{\tau}X^{m}\nabla[\gamma]_{m}u-h^{TT||}+2(w-1)\gamma+\mathcal{F}_{u}|\mathcal{L}_{g,\gamma}u>_{L^{2}}\\\nonumber
+\frac{n^{2}}{2}<u|\partial_{T}\mathcal{L}\nonumber _{g,\gamma}u>_{L^{2}}+\mathcal{B}\\\nonumber
=-(n-1)<v|v>_{L^{2}}-\frac{\phi}{\tau}<v|X^{m}\nabla[\gamma]_{m}v>_{L^{2}}-\frac{n^{2}\phi}{2\tau}<X^{m}\nabla[\gamma]_{m}u|\mathcal{L}_{g,\gamma}u>_{L^{2}}\\\nonumber 
-\frac{n^{2}}{2}<h^{TT||}|\mathcal{L}_{g,\gamma}>_{L^{2}}+n^{2}<(w-1)\gamma|\mathcal{L}_{g,\gamma}u>_{L^{2}}-\frac{n^{2}\phi}{2\tau}<wv|\mathcal{L}_{g,\gamma}u>_{L^{2}}\\\nonumber 
+\frac{n^{2}}{2}<u|\partial_{T}\mathcal{L}\nonumber _{g,\gamma}u>_{L^{2}}+<v|\mathcal{F}_{v}>_{L^{2}}+\frac{n^{2}}{2}<\mathcal{F}_{u}|\mathcal{L}_{g,\gamma}u>_{L^{2}}+\mathcal{B}.
\end{eqnarray}
Now utilizing point (2) of lemma (14), we may write 
\begin{eqnarray}
\int_{M}<u,\partial_{T}\mathcal{L}_{g,\gamma}u>\mu_{g}= \int_{M}<\partial_{T}u|\mathcal{L}_{g,\gamma}u>\mu_{g}+\mathcal{R}, |\mathcal{R}|\leq C||u||^{2}_{H^{s}}||v||_{H^{s-1}},\nonumber
\end{eqnarray}
and therefore the time derivative of first order energy (note that $||\frac{\partial\gamma}{\partial T}||_{H^{s}}\lesssim ||\frac{\partial g}{\partial T}||_{H^{s}}$ from shadow gauge) becomes 
\begin{eqnarray}
\partial_{T}\mathcal{E}_{1}=-(n-1)<v|v>_{L^{2}}-\frac{\phi}{\tau}<v|X^{m}\nabla[\gamma]_{m}v>_{L^{2}}\nonumber-\frac{n^{2}\phi}{\tau}<X^{m}\nabla[\gamma]_{m}u|\mathcal{L}_{g,\gamma}u>_{L^{2}}\\\nonumber 
-n^{2}<h^{TT||}|\mathcal{L}_{g,\gamma}>_{L^{2}}+2n^{2}<(w-1)\gamma|\mathcal{L}_{g,\gamma}u>_{L^{2}}-\frac{n^{2}\phi}{2\tau}<wv|\mathcal{L}_{g,\gamma}u>_{L^{2}}\\\nonumber 
+\frac{n^{2}\phi}{2\tau}<wv|\mathcal{L}_{g,\gamma}u>_{L^{2}}+<v|\mathcal{F}_{v}>_{L^{2}}+n^{2}<\mathcal{F}_{u}|\mathcal{L}_{g,\gamma}u>_{L^{2}}+\mathcal{B}\\\nonumber 
=-(n-1)<v|v>_{L^{2}}-\frac{\phi}{\tau}<v|X^{m}\nabla[\gamma]_{m}v>_{L^{2}}\nonumber-\frac{n^{2}\phi}{\tau}<X^{m}\nabla[\gamma]_{m}u|\mathcal{L}_{g,\gamma}u>_{L^{2}}\\\nonumber 
-n^{2}<h^{TT||}|\mathcal{L}_{g,\gamma}>_{L^{2}}+2n^{2}<(w-1)\gamma|\mathcal{L}_{g,\gamma}u>_{L^{2}}+<v|\mathcal{F}_{v}>_{L^{2}}\\\nonumber +n^{2}<\mathcal{F}_{u}|\mathcal{L}_{g,\gamma}u>_{L^{2}}+\mathcal{B}.
\end{eqnarray}
We note that the potentially problematic term $\frac{n^{2}\phi}{2\tau}<wv|\mathcal{L}_{g,\gamma}u>_{L^{2}}$ gets cancelled in the previous expression. Now we will estimate each term in the energy expression utilizing lemma (14) and the basic inequalities stated at the beginning of this section. Once again, we will bound every term by their maximum available Sobolev norm (since $||A||_{H^{s_{1}}}\lesssim ||A||_{H^{s_{2}}}$ for $s_{1}<s_{2}$). The following holds 
\begin{eqnarray}
|<v|X^{m}\nabla[\gamma]_{m}v>_{L^{2}}|\lesssim ||v||^{3}_{H^{s-1}}+\frac{\tau}{\phi}||v||^{4}_{H^{s-1}},|<X^{m}\nonumber\nabla[\gamma]_{m}u|\mathcal{L}_{g,\gamma}u>_{L^{2}}|\\\nonumber
\lesssim ||u||^{2}_{H^{s}}||v||_{H^{s-1}}+\frac{\tau}{\phi}||u||^{2}_{H^{s}}||v||^{2}_{H^{s-1}},
|<h^{TT||}|\mathcal{L}_{g,\gamma}>_{L^{2}}|\lesssim \frac{\phi}{\tau}||u||^{2}_{H^{s}}||v||_{H^{s}},\\\nonumber 
|<(w-1)\gamma|\mathcal{L}_{g,\gamma}u>_{L^{2}}|\lesssim ||u||_{H^{s}}||v||^{2}_{H^{s-1}}, \mathcal{B}\lesssim \frac{\phi}{\tau}||v||^{3}_{H^{s-1}}+||v||^{4}_{H^{s-1}}\\\nonumber 
+\frac{\phi}{\tau}||u||^{2}_{H^{s}}||v||_{H^{s-1}}+||u||^{2}_{H^{s}}||v||^{2}_{H^{s-1}}.
\end{eqnarray}
Now, let us compute the product $<v|\mathcal{F}_{v}>_{L^{2}}$ and $<\mathcal{F}_{u}|\mathcal{L}_{g,\gamma}u>_{L^{2}}$
\begin{eqnarray}
<v|\mathcal{F}_{v}>_{L^{2}}=-2n^{2}\frac{\phi}{\tau}<v|wJ>_{L^{2}}\nonumber+2n^{2}\frac{\phi}{\tau}<v|\nabla\otimes\nabla w>_{L^{2}}+\frac{\phi}{\tau}<v|wv\cdot v>_{L^{2}}\\\nonumber -\frac{2\phi}{\tau}<v|(w-1)(u+\gamma)>_{L^{2}}-(n-2)<v|(w-1)v>_{L^{2}}+8n^{3}\frac{\phi}{\tau}<v|wv\cdot v>_{L^{2}}\\\nonumber 
-\frac{\phi}{\tau}<v|v\circ \nabla X>_{L^{2}},\\\nonumber 
<\mathcal{F}_{u}|\mathcal{L}_{g,\gamma}u>_{L^{2}}=2n^{2}<(w-1)u|\mathcal{L}_{g,\gamma}u>_{L^{2}}-n^{2}<L_{\frac{\phi}{\tau}X+Y^{||}}\gamma|\mathcal{L}_{g,\gamma}u>_{L^{2}}\\\nonumber 
-n^{2}\frac{\phi}{\tau}<u\circ \nabla X|\mathcal{L}_{g,\gamma}u>_{L^{2}},
\end{eqnarray}
where $(\nabla\otimes \nabla w)_{ij}=\nabla_{i}\nabla_{j}w, (v\cdot v)_{ij}=v_{im}v^{m}_{j}$, and $(v\circ \nabla X)_{ij}=v_{im}\nabla_{j}X^{m}+v_{jm}\nabla_{i}X^{m}$. Once again utilizing lemma (14), basic inequalities, and the elliptic estimates for the lapse function and the shift vector field, each term of the above expressions may be estimated as follows 
\begin{eqnarray}
|<v|wJ>_{L^{2}}|\lesssim ||u||^{2}_{H^{s}}||v||_{H^{s-1}}, |<v|\nabla\otimes\nabla w>_{L^{2}}|\lesssim ||v||^{3}_{H^{s-1}},\\\nonumber 
|<v|wv\cdot v>_{L^{2}}|\lesssim ||v||^{3}_{H^{s-1}},|<v|(w-1)(u+\gamma)>_{L^{2}}|\lesssim ||u||_{H^{s}}||v||^{3}_{H^{s-1}},\\\nonumber 
|<v|(w-1)v>_{L^{2}}|\lesssim ||v||^{4}_{H^{s-1}},|<v|wv\cdot v>_{L^{2}}|\lesssim ||v||^{3}_{H^{s-1}},\\\nonumber 
|<v|v\circ \nabla X>_{L^{2}}|\lesssim ||v||^{3}_{H^{s-1}}+\frac{\tau}{\phi}||v||^{4}_{H^{s-1}},|<(w-1)u|\mathcal{L}_{g,\gamma}u>_{L^{2}}|\\\nonumber\lesssim ||u||^{2}_{H^{s}}||v||^{2}_{H^{s-1}},
|<L_{\frac{\phi}{\tau}X+Y^{||}}\gamma|\mathcal{L}_{g,\gamma}u>_{L^{2}}|\lesssim \frac{\phi}{\tau}||u||^{2}_{H^{s}}||v||_{H^{s-1}}+\\\nonumber ||u||^{2}_{H^{s}}||v||^{2}_{H^{s-1}},
|<u\circ \nabla X|\mathcal{L}_{g,\gamma}u>_{L^{2}}|\lesssim ||u||^{2}_{H^{s}}||v||_{H^{s-1}}+\frac{\tau}{\phi}||u||^{2}_{H^{s}}||v||^{2}_{H^{s-1}}.
\end{eqnarray}
Note that the dangerous factor $\frac{\tau}{\phi}$ cancel out in the time derivative of the energy by its inverse $\frac{\phi}{\tau}$.
Utilizing these estimates and carefully counting powers of $\frac{\phi}{\tau}$, we obtain the desired differential equation for the first order energy $\mathcal{E}_{1}$
\begin{eqnarray}
\partial_{T}\mathcal{E}_{1}=-(n-1)<v|v>_{L^{2}}+ \mathcal{A}_{1},
\end{eqnarray}
with $\mathcal{A}_{1}$ satisfying 
\begin{eqnarray}
|\mathcal{A}_{1}|\leq C(||u||_{H^{s}}||v||^{2}_{H^{s-1}}+||u||^{2}_{H^{s}}||v||^{2}_{H^{s}}+||v||^{4}_{H^{s-1}}\\\nonumber +\frac{\phi}{\tau}||u||_{H^{s}}||v||^{2}_{H^{s-1}}+\frac{\phi}{\tau}||u||^{2}_{H^{s}}||v||_{H^{s-1}}+\frac{\phi}{\tau}||v||^{3}_{H^{s-1}})
\end{eqnarray}
This proves the lemma.\\
Now we need to derive the time derivative of the higher order energies in order to obtain a time evolution of the total energy. In order to do so, we need a few additional estimates. The next lemma provides the required estimates. \\
\textbf{Lemma 16:}\textit{ Let $s>\frac{n}{2}+2$, $\gamma\in\mathcal{E}in_{-\frac{n-1}{n^{2}}}$ be the shadow of $g\in \mathcal{M}$, and assume there exists a $\delta>0$ such that $(u,v)\in \mathcal{B}_{\delta}(0,0)\subset H^{s}\times H^{s-1}$, then the following estimates hold
 \begin{eqnarray}
(1) |\int_{M}<\mathcal{L}^{i}_{g,\gamma}u,h^{TT||}>\mu_{g}|\leq C\frac{\phi}{\tau}||u||^{2}_{H^{s}}||v||_{H^{s-1}},\\
(2)\int_{M}<u,\partial_{T}\mathcal{L}^{i}_{g,\gamma}u>\mu_{g}= \int_{M}<\partial_{T}u|\mathcal{L}^{i}_{g,\gamma}u>\mu_{g}\nonumber+\mathcal{R}^{'}, |\mathcal{R}^{'}|\leq C||u||^{2}_{H^{s}}||v||_{H^{s-1}},\\\nonumber
(3) \int_{M}<v,\mathcal{L}^{i-1}_{g,\gamma}v>g^{kl}\partial_{T}g_{kl}\mu_{g}\leq C(\frac{\phi}{\tau}||v||^{3}_{H^{s-1}}+||v||^{4}_{H^{s-1}}),\\\nonumber
(4) \int_{M}<u,\mathcal{L}^{i}_{g,\gamma}u>g^{kl}\partial_{T}g_{kl}\mu_{g})\leq C(\frac{\phi}{\tau}||u||^{2}_{H^{s}}||v||_{H^{s-1}}+||u||^{2}_{H^{s}}||v||^{2}_{H^{s-1}}),\\\nonumber
(5) |\int_{M}\frac{\phi}{\tau}<\mathcal{L}^{i-1}_{g,\gamma}v,X^{m}\nabla[\gamma]_{m}v>\mu_{g}|\leq C(\frac{\phi}{\tau}||v||^{3}_{H^{s-1}}+||v||^{4}_{H^{s-1}}),\\\nonumber
(6) |\int_{M}\frac{\phi}{\tau}<\mathcal{L}^{i}_{g,\gamma}u,X^{m}\nabla[\gamma]_{m}u>\mu_{g}|\leq C(\frac{\phi}{\tau}||u||^{2}_{H^{s}}||v||_{H^{s-1}}+||u||^{2}_{H^{s}}||v||^{2}_{H^{s-1}}), 
\end{eqnarray}
for $2\leq i\leq s$ and $C=C(\delta)>0$.}\\
\textbf{Proof:} Following a calculation analogous to that of lemma (14) and using the formula for the higher order estimates provided in the beginning of this section (i.e., the product estimates and the commutator estimates) for $s>\frac{n}{2}+2$, each of the claims follows.\\
Now that we have the necessary estimates, we may obtain the time derivative of the higher order energies. The following lemma states the time derivative of the higher order energies.\\
\textbf{Lemma 17:} \textit{Let $s>\frac{n}{2}+2$, $\gamma\in\mathcal{E}in_{-\frac{n-1}{n^{2}}}$ be the shadow of $g\in \mathcal{M}$, and assume there exists a $\delta>0$ such that $(u,v)\in \mathcal{B}_{\delta}(0,0)\subset H^{s}\times H^{s-1}$, then the following holds for $1<i\leq s$ 
\begin{eqnarray}
\partial_{T}\mathcal{E}_{i}=-(n-1)<v|\mathcal{L}^{i-1}_{g,\gamma}v>_{L^{2}}+\mathcal{A}_{i},
\end{eqnarray}
with $\mathcal{A}_{i}$ satisfying 
\begin{eqnarray}
|\mathcal{A}_{i}|\leq C(||u||_{H^{s}}||v||^{2}_{H^{s-1}}+||u||^{2}_{H^{s}}||v||^{2}_{H^{s}}+||v||^{4}_{H^{s-1}}\\\nonumber +\frac{\phi}{\tau}||u||_{H^{s}}||v||^{2}_{H^{s-1}}+\frac{\phi}{\tau}||u||^{2}_{H^{s}}||v||_{H^{s-1}}+\frac{\phi}{\tau}||v||^{3}_{H^{s-1}}),
\end{eqnarray}
for $C=C(\delta)>0$.} \\
\textbf{Proof:} A calculation analogous to that of lemma (15) and the higher order estimates from lemma (16) directly yield the desired result.\\ 
Now that we have concluded with the proofs of the important lemmas, we will study the time evolution of the total energy. The time derivative of the total energy may be written using the lemma (15) and (17) as follows
\begin{eqnarray}
\frac{\partial \mathcal{E}}{\partial T}=-(n-1)\sum_{i=1}^{s}<v|\mathcal{L}^{i-1}_{g,\gamma}v>_{L^{2}}+\sum_{i=1}^{s}\mathcal{A}_{i},
\end{eqnarray}  
where each $\mathcal{A}_{i}$ satisfies higher order estimate
\begin{eqnarray}
\mathcal{A}_{i}\leq C(||u||_{H^{s}}||v||^{2}_{H^{s-1}}+||u||^{2}_{H^{s}}||v||^{2}_{H^{s}}+||v||^{4}_{H^{s-1}}\\\nonumber +\frac{\phi}{\tau}||u||_{H^{s}}||v||^{2}_{H^{s-1}}+\frac{\phi}{\tau}||u||^{2}_{H^{s}}||v||_{H^{s-1}}+\frac{\phi}{\tau}||v||^{3}_{H^{s-1}})
\end{eqnarray}
The above expression may be rewritten as follows 
\begin{eqnarray}
\frac{\partial\mathcal{E}}{\partial T}\leq -\left((n-1)-C(||u||_{H^{s}}+||u||^{2}_{H^{s}}+||v||^{2}_{H^{s}})\right)\sum_{i=1}^{s}<v|\mathcal{L}^{i-1}_{g,\gamma}v>_{L^{2}}\\\nonumber
+C\frac{\phi}{\tau}||v||_{H^{s-1}}\mathcal{E},\nonumber
\end{eqnarray}
using the fact that $||v||^{2}_{H^{s-1}}\lesssim \sum_{i=1}^{s}<v|\mathcal{L}^{i-1}_{g,\gamma}v>_{L^{2}}, ||u||^{2}_{H^{s}}+||v||^{2}_{H^{s-1}}\lesssim \mathcal{E}$. Also note that $||u||_{H^{s}},||v||_{H^{s-1}}\lesssim \mathcal{E}^{1/2}$.
Now notice that zero eigenvalues of $\mathcal{L}_{g,\gamma}$ would correspond to the trivially stable case for small data. If $(u,v)\in \ker(\mathcal{L}_{g,\gamma})$, then 
\begin{eqnarray}
\label{eq:tangent1}
\frac{\partial\mathcal{E}}{\partial T}\leq-2(n-1)\mathcal{E}+C\frac{\phi}{\tau}\mathcal{E}^{3/2}
\end{eqnarray}
and if the initial data is sufficiently small in $H^{s}\times H^{s-1}$, then $\mathcal{E}(T)\lesssim e^{-2(n-1)(T-T_{0})}$ as $T\to\infty$. Therefore, we will focus on perturbations $(u,v)$, satisfying $<v|\mathcal{L}^{i-1}_{g,\gamma}v>_{L^{2}}>0~\forall 1\leq i\leq s$. Note an extremely important fact that the possible feedback term (which might lead to energy growth) $||v||_{H^{s-1}}\mathcal{E}$ is multiplied by a factor of $\frac{\phi}{\tau}$. Noting that $\frac{\phi}{\tau}\sim e^{-T}$ as $T\to\infty$, one immediate guess would be that the energy at least remains bounded if the initial energy is chosen to be sufficiently small. Now the time derivative of the energy may be further reduced to the following form
\begin{eqnarray}
\label{eq:diffinequal}
\frac{\partial \mathcal{E}}{\partial T}\leq -\left((n-1)-C\sqrt{\mathcal{E}}\right)\sum_{i=1}^{s}<v|\mathcal{L}^{i-1}_{g,\gamma}v>_{L^{2}}\\
+C\frac{\phi}{\tau}||v||_{H^{s-1}}\mathcal{E}.
\end{eqnarray}
Now, if the energy is sufficiently small (i.e., $(u,v)\in \mathcal{B}_{\delta}(0,0)\in H^{s}\times H^{s-1}$ such that $\mathcal{E}<(\frac{n-1}{2C})^{2}$ for example), then the first term contributes a negative factor and for such small data limit, the linear terms $||v||_{H^{s-1}}$ are absorbed in the constant and the energy satisfies 
\begin{eqnarray}
\frac{\partial \mathcal{E}}{\partial T}\leq C\frac{\phi}{\tau}\mathcal{E},
\end{eqnarray}
integration of which yields 
\begin{eqnarray}
\mathcal{E}(T)\leq \mathcal{E}(T_{0})e^{C\ln|\frac{e^{-T_{0}}+\sqrt{e^{-2T_{0}}+\frac{2n\Lambda}{n-1}}}{e^{-T}+\sqrt{e^{-2T}+\frac{2n\Lambda}{n-1}}}|}=\mathcal{E}(T_{0})\left(\frac{e^{-T_{0}}+\sqrt{e^{-2T_{0}}+\frac{2n\Lambda}{n-1}}}{e^{-T}+\sqrt{e^{-2T}+\frac{2n\Lambda}{n-1}}}\right)^{C}.
\end{eqnarray} 
Therefore, we have the boundedness of energy at the limit $T\to\infty$ ($\lim_{T\to\infty}(e^{-T}+\sqrt{e^{-2T}+\frac{2n\Lambda}{n-1}})=\sqrt{\frac{2n\Lambda}{n-1}}$)
\begin{eqnarray}
\lim_{T\to\infty}\mathcal{E}(T)\leq C^{'}\mathcal{E}(T_{0}).
\end{eqnarray}
In order to close the argument, the following must be satisfied 
\begin{eqnarray}
\sqrt{\mathcal{E}(T)}<\frac{n-1}{2C}
\end{eqnarray}
to ensure that the first term in the inequality (\ref{eq:diffinequal}) contributes a negative term for all time (which we assumed at the beginning). We enforce the following condition which will impose necessary smallness condition on the initial energy $\mathcal{E}(T_{0})$ 
\begin{eqnarray}
\label{eq:initial}
\lim_{T\to\infty}\mathcal{E}(T)\leq C^{'}\mathcal{E}(T_{0})<\frac{(n-1)^{2}}{4C^{2}}. 
\end{eqnarray}
It yields the following smallness of the initial data 
\begin{eqnarray}
\mathcal{E}(T_{0})<\frac{(n-1)^{2}}{4C^{2}C^{'}}.
\end{eqnarray}
 Therefore, by ensuring that the initial data is small enough such that (\ref{eq:initial}) holds, we ensure that the first term in the differential inequality (\ref{eq:diffinequal}) always contributes a negative term. Therefore, we close the argument and obtain that the suitable norm of the data remains bounded  by the initial data as $T\to\infty$ if the initial data is chosen small enough i.e., 
\begin{eqnarray}
\mathcal{E}(T)\lesssim \mathcal{E}(T_{0}),
\end{eqnarray}
or 
\begin{eqnarray}
||u(T)||^{2}_{H^{s}}+||v(T)||^{2}_{H^{s-1}}\lesssim ||u(T_{0})||^{2}_{H^{s}}+||v(T_{0})||^{2}_{H^{s-1}}. 
\end{eqnarray}
Now, since the time interval of existence depends on the size of $||u||_{H^{s}}$ and $||v||_{H^{s-1}}$ from the local existence theorem, we obtain the global existence.\\
Even though we proved the boundedness of the suitable norm of the dynamical variables yielding global existence, this is not satisfactory. We want to prove the attractor property of the Einstein-$\Lambda$ flow and as such we need to establish decay property of suitable norms. \cite{andersson2011einstein} showed the attractor property of the Einstein flow with $\Lambda=0$. Now, with $\Lambda>0$ included, one would naturally expect that the accelerated expansion should kill away the perturbations and drive the flow towards the centre manifold (extended center manifold to be precise). However, we only seem to obtain a boundedness of the energy without any decay. This is mainly due to the fact that we have underestimated the large damping term $-(n-1)\sum_{i=1}^{s}<v|\mathcal{L}^{i-1}_{g,\gamma}v>_{L^{2}}$. Using an iterated scheme, we will now achieve the sharp decay which does agree with the linear analysis as $T\to\infty$ (as it should).\\    
The following analysis holds in the limit $T\to\infty$. We will simply compute the time derivative of $\mathcal{E}_{v}:=\sum_{i=1}^{s-1}<v|\mathcal{L}^{i-1}_{g,\gamma}v>_{L^{2}}, s>\frac{n}{2}+2$
\begin{eqnarray}
\frac{d\mathcal{E}_{v}}{dT}\leq-(n-1)\sum_{i=1}^{s-1}<v|\mathcal{L}^{i-1}_{g,\gamma}v>_{L^{2}}+C||v||^{4}_{H^{s-2}}\\\nonumber 
+\frac{\phi}{\tau}(||v||^{3}_{H^{s-2}}+||u||_{H^{s-1}}||v||^{3}_{H^{s-2}}+||u||_{H^{s}}||v||_{H^{s-2}}).
\end{eqnarray}
Note an important fact that we loose one degree of regularity because, in the computation of the time derivative of $\mathcal{E}_{v}:=\frac{1}{2}\sum_{i=1}^{s}<v|\mathcal{L}^{i-1}_{g,\gamma}v>_{L^{2}}$ alone, the dangerous term $\sum_{i=1}^{s}<\mathcal{L}_{g,\gamma}u|\mathcal{L}^{i-1}_{g,\gamma}v>_{L^{2}}$ does not get cancelled unlike the case of the time derivative of the total energy $\mathcal{E}$. Therefore, we loose one order of regularity. However, since we have $s>\frac{n}{2}+2$, we will still be able to obtain the desired pointwise decay estimate. The previous inequality may also be expressed as follows 
\begin{eqnarray}
\frac{d\mathcal{E}_{v}}{dT}\leq-((n-1)-C||v||^{2}_{H^{s-2}})\sum_{i=1}^{s-1}<v|\mathcal{L}^{i-1}_{g,\gamma}v>_{L^{2}}\\\nonumber 
+C\frac{\phi}{\tau}(||v||^{3}_{H^{s-2}}+||u||_{H^{s-1}}||v||^{3}_{H^{s-2}}+||u||_{H^{s}}||v||_{H^{s-2}})
\end{eqnarray}
Recalling $||v||^{2}_{H^{s-2}}\lesssim\delta^{2}<\frac{n-1}{2C}$ and $||u||_{H^{s}}\lesssim\delta,$ we obtain 
\begin{eqnarray}
\frac{d\mathcal{E}_{v}}{dT}\leq-\frac{n-1}{2}\sum_{i=1}^{s-1}<v|\mathcal{L}^{i-1}_{g,\gamma}v>_{L^{2}}+C\frac{\phi}{\tau}||v||^{2}_{H^{s-2}}(1+||u||_{H^{s-1}})\\\nonumber+C\frac{\phi}{\tau}||u||_{H^{s}}||v||_{H^{s-2}}\\\nonumber 
=-(n-1)\mathcal{E}_{v}+C\frac{\phi}{\tau}||v||^{2}_{H^{s-2}}(1+||u||_{H^{s-1}})\nonumber+C\frac{\phi}{\tau}||u||_{H^{s}}||v||_{H^{s-2}}
\end{eqnarray}
integration of which yields ($n\geq3$)
\begin{eqnarray}
\mathcal{E}_{v}(T)\lesssim e^{-(n-1)(T-T_{0})}\mathcal{E}(T_{0})+Ce^{-T},
\end{eqnarray}
i.e., 
\begin{eqnarray}
\mathcal{E}_{v}\lesssim e^{-T}~or~||v||_{H^{s-2}}\lesssim e^{-T/2}.
\end{eqnarray}
Using $s>\frac{n}{2}+2,$ we obtain applying Sobolev embedding on compact domain 
\begin{eqnarray}
||v||_{L^{\infty}}\lesssim e^{-T/2}.
\end{eqnarray}
Now, note that this decay is not optimal. An iteration scheme would yield a better decay rate. We go back to the expression of $\frac{d\mathcal{E}_{v}}{dT}$ and observe 
\begin{eqnarray}
\label{eq:improve}
\frac{d\mathcal{E}_{v}}{dT}\leq -(n-1)\sum_{i=1}^{s-1}<v|\mathcal{L}^{i-1}_{g,\gamma}v>_{L^{2}}+C||v||^{4}_{H^{s-2}}\\\nonumber 
+C\frac{\phi}{\tau}(||v||^{3}_{H^{s-2}}+||u||_{H^{s-1}}||v||^{3}_{H^{s-2}}+||u||_{H^{s}}||v||_{H^{s-2}})
\end{eqnarray}
to obtain 
\begin{eqnarray}
\frac{d\mathcal{E}_{v}}{dT}\leq -2(n-1)\mathcal{E}_{v}+Ce^{-(1+1/2)T}+Ce^{-2T}
\end{eqnarray}
since $||v||^{4}_{H^{s-2}}\lesssim e^{-2T}, \frac{\phi}{\tau}(||v||^{3}_{H^{s-2}}+||u||_{H^{s-1}}||v||^{3}_{H^{s-2}}+||u||_{H^{s}}||v||_{H^{s-2}})\lesssim e^{-(1+1/2)T/2}$ using the previous estimate. Therefore, the previous differential inequality yields
\begin{eqnarray}
\mathcal{E}_{v}\lesssim e^{-(1+1/2)T}~or~||v||_{H^{s-2}}\lesssim e^{-\frac{(1+1/2)T}{2}}.
\end{eqnarray}
Notice that we have gained a factor of $1/4$ in the decay estimate. Using this estimate, in the next iteration, we obtain 
\begin{eqnarray}
\frac{d\mathcal{E}_{v}}{dT}\leq -2(n-1)\mathcal{E}_{v}+Ce^{-\{1+1/2(1+1/2)\}T}
\end{eqnarray}
and subsequently 
\begin{eqnarray}
\mathcal{E}_{v}\lesssim e^{-\{1+1/2(1+1/2)\}T}, ~or~
||v||_{H^{s-2}}\lesssim e^{-\frac{1}{2}\{1+1/2(1+1/2)\}T}.
\end{eqnarray}
If we continue to iterate, we obtain the decay rate to be the sum of an infinite series, that is, the final decay rate is computed to be 
\begin{eqnarray}
1+\frac{1}{2}(1+\frac{1}{2}(1+\frac{1}{2}(1+\frac{1}{2}(.......=\sum_{k=0}^{\infty}\frac{1}{2^{k}}=\frac{1}{1-1/2}=2.
\end{eqnarray} 
Therefore, the final estimate reads 
\begin{eqnarray}
\mathcal{E}_{v}\lesssim e^{-2T}~or~ ||v||_{H^{s-2}}\lesssim e^{-T}.
\end{eqnarray}
Note that this is optimal in a sense that substituting this decay back into the equation (\ref{eq:improve}) returns the same decay. 
Now we compute the time derivative of $\mathcal{E}_{s}:=\frac{n^{2}}{2}\sum_{i=1}^{s-1}<u|\mathcal{L}^{i}_{g,\gamma}u>_{L^{2}}+\frac{1}{2}\sum_{i=1}^{s-1}<v|\mathcal{L}^{i-1}_{g,\gamma}v>_{L^{2}}$ which yields 
\begin{eqnarray}
\frac{d\mathcal{E}_{s}}{dT}\leq-(n-1)\sum_{i=1}^{s-1}<v|\mathcal{L}^{i-1}_{g,\gamma}v>_{L^{2}}+C(||u||_{H^{s-1}}+||u||^{2}_{H^{s-1}}\\\nonumber 
+||v||^{2}_{H^{s-2}})||v||^{2}_{H^{s-2}}
+C\frac{\phi}{\tau}||v||_{H^{s-2}}\mathcal{E}_{s}.\nonumber
\end{eqnarray}
Utilizing the estimate $||v||_{^{s-2}}\lesssim e^{-T}$ and $||u||_{H^{s-1}}<\delta$, we observe that every term decays exponentially as $T\to\infty$ i.e., $C(||u||_{H^{s-1}}+||u||^{2}_{H^{s-1}}
+||v||^{2}_{H^{s-2}})||v||^{2}_{H^{s-2}}\lesssim e^{-2T}, \frac{\phi}{\tau}||v||_{H^{s-2}}\mathcal{E}_{s}\lesssim e^{-2T}$ and $\sum_{i=1}^{s-1}<v|\mathcal{L}^{i-1}_{g,\gamma}v>_{L^{2}}\lesssim e^{-2T}$   yielding
\begin{eqnarray}
\frac{d\mathcal{E}_{s}}{dT}\lesssim e^{-2T},
\end{eqnarray}
i.e., $\frac{d\mathcal{E}_{s}}{dT}$ decays as$T\to\infty$. In fact, we may show that $\mathcal{E}_{s}$ actually decays as $T\to\infty$. Using the decay estimate $||v||_{H^{s}}\lesssim e^{-T}$ and noting the order of each term 
\begin{eqnarray}
\frac{d\mathcal{E}_{s}}{dT}\leq -(n-1)\delta^{2}e^{-2T}+C\delta^{3}e^{-2T}+C\delta^{3}e^{-2T}\\\nonumber 
=-\left( (n-1)-2C\delta\right)\delta^{2}e^{-2T}.
\end{eqnarray}
Now of course there exists a $\delta>0$ such that $\delta<\frac{n-1}{2C}$ yielding 
\begin{eqnarray}
\frac{d\mathcal{E}_{s}}{dT}<0
\end{eqnarray}
and $\frac{d\mathcal{E}_{s}}{dT}=0$ if and only if $||v||_{H^{s-2}}\equiv 0$ since every term in the right hand side is multiplied by a factor of $||v||_{H^{s-2}}$. This indicates that the re-scaled metric converges to a limit metric. 
Now, we go back to the evolution equation of the metric since, we have not yet established a sharp decay of the metric. The evolution equation for the metric reads 
\begin{eqnarray}
\partial_{T}g_{ij}=\frac{2\phi(\tau)}{\tau}NK^{TT}_{ij}-2(1-\frac{N}{n})g_{ij}-\frac{\phi(\tau)}{\tau}(L_{X}g)_{ij}.
\end{eqnarray}
Now, we have estimated $||v||_{H^{s-2}}=2n||K^{TT}||_{H^{s-2}}\lesssim e^{-T}$ which yields through the elliptic estimate (lemma 6 and 7)
\begin{eqnarray}
(||\frac{N}{n}-1)||_{H^{s}}\lesssim e^{-2T},||X||_{H^{s}}\lesssim e^{-T}.
\end{eqnarray}
Therefore, utilizing these estimates ($s>\frac{n}{2}+2$) we obtain 
\begin{eqnarray}
||\partial_{T}g_{ij}||_{H^{s-2}}\lesssim e^{-2T}~or~||\partial_{T}g_{ij}||_{L^{\infty}}\lesssim e^{-2T}
\end{eqnarray}
which implies that the re-scaled metric $g$ decays to a limit metric $\gamma$ i.e., utilizing the evolution equation $||g-\gamma^{\dag}||_{H^{s-1}}\lesssim e^{-2T}$ or equivalently $||g-\gamma^{\dag}||_{L^{\infty}}\lesssim e^{-2T}$ from Sobolev embedding on compact domain. The question that remains is what are these limit metrics? Invoking the Hamiltonian constraint    
\begin{eqnarray}
R+\frac{n-1}{n}=|K^{TT}|^{2},
\end{eqnarray}
one obtains
\begin{eqnarray}
\lim_{T\to\infty}(R+\frac{n-1}{n})=\lim_{T\to\infty}|K^{TT}|^{2}=\lim_{T\to\infty}e^{-2T}=0.
\end{eqnarray}
This precisely implies that the re-scaled metric $g_{ij}$ converges to an element of the space $\mathcal{M}^{\epsilon}_{-\frac{n-1}{n}}\cap \mathcal{S}_{\gamma}$ i.e., the extended centre manifold. The decay estimate is as follows
\begin{eqnarray}
||g-\gamma^{\dag}||_{H^{s-1}}\lesssim e^{-2T},
\end{eqnarray}
where $\gamma^{\dag}\in \mathcal{M}^{\epsilon}_{-\frac{n-1}{n}}\cap \mathcal{S}_{\gamma}$.
In summary, the metric $g$ along with $(K^{TT},X,N)$ satisfies 
\begin{eqnarray}
\lim_{T\to\infty}||\frac{\partial g_{ij}}{\partial T}||_{L^{\infty}}=0, \lim_{T\to\infty}||K^{TT}||_{L^{\infty}}=0,\\\nonumber \lim_{T\to\infty}||N||_{L^{\infty}}=n, \lim_{T\to\infty}||X||_{L^{\infty}}=0,
\lim_{T\to\infty}R[g]=-\frac{n-1}{n},
\end{eqnarray}
that is
\begin{eqnarray}
\lim_{T\to\infty}g(T)=\gamma^{\dag}~(in~ H^{s-1}~ topology)
\end{eqnarray}
with $\gamma^{\dag}\in\mathcal{M}^{\epsilon}_{-\frac{n-1}{n}}\cap \mathcal{S}_{\gamma}$. This implies that metrics lying in $\mathcal{M}^{\epsilon}_{-\frac{n-1}{n}}\cap \mathcal{S}_{\gamma}$ simply evolves to another point in $\mathcal{M}^{\epsilon}_{-\frac{n-1}{n}}\cap \mathcal{S}_{\gamma}$ and metrics lying sufficiently close to $\mathcal{N}$ yet not in $\mathcal{M}^{\epsilon}_{-\frac{n-1}{n}}\cap \mathcal{S}_{\gamma}$ exponentially converges to $\mathcal{M}^{\epsilon}_{-\frac{n-1}{n}}\cap \mathcal{S}_{\gamma}$ in infinite time.
This completes the proof of the attractor property of the centre manifold (extended to be precise) under the Einstein flow. Theorem $5$ formally summarizes the result.\\
Once we have obtained the estimates for the perturbations to the primary dynamical variables $(u,v)$, the estimates for the lapse function and the shift vector field follow from lemma (6) and (7), respectively 
\begin{eqnarray}
||\frac{N}{n}-1||_{H^{s}}\lesssim e^{-2T},\\
||X||_{H^{s}}\lesssim e^{-T}.
\end{eqnarray}
and thus 
\begin{eqnarray}
||\frac{N}{n}-1||_{L^{\infty}}\lesssim e^{-2T},\\
||X||_{L^{\infty}}\lesssim e^{-T}
\end{eqnarray}
following Sobolev embedding theorem on compact domain with $s>\frac{n}{2}+2$.
We proved earlier that $h^{TT||}=\frac{\phi}{\tau}v^{||}$ up to a second order correction and $Y^{||}$ is estimated by lemma (8). Therefore, the following holds for $h^{TT||}$ and $Y^{||}$
\begin{eqnarray}
||h^{TT}||_{H^{s-1}}\lesssim e^{-2T},~or~ ||h^{TT}||_{L^{\infty}}\lesssim e^{-2T}\\
||Y^{||}||_{H^{s}}\lesssim e^{-2T}~or~||Y^{||}||_{L^{\infty}}\lesssim e^{-2T}.
\end{eqnarray}
Therefore, we summarize the following decay estimates for the perturbations $L^{2}$ orthogonal to the deformation space $\mathcal{N}$
\begin{eqnarray}
||\frac{\partial g}{\partial T}||_{L^{\infty}}\lesssim e^{-2T}, ||g-\gamma^{\dag}||_{L^{\infty}}\lesssim e^{-2T}, ||K^{TT}||_{L^{\infty}}\lesssim e^{-T},\\\nonumber ||\frac{N}{n}-1||_{L^{\infty}}\lesssim e^{-2T},
||X||_{L^{\infty}}\lesssim e^{-T}, ||h^{TT}||_{L^{\infty}}\lesssim e^{-2T}, ||Y^{||}||_{L^{\infty}}\lesssim e^{-2T}.
\end{eqnarray}
For purely tangential perturbations (tangential to the deformation space $\mathcal{N}$), the following decay estimates hold using (\ref{eq:tangent1}), the elliptic estimates (lemma 6, 7, and 8), and the evolution equations
\begin{eqnarray}
||\frac{\partial g}{\partial T}||_{L^{\infty}}\lesssim e^{-nT}, ||g-\gamma^{\dag}||_{L^{\infty}}\lesssim e^{-nT}, ||K^{TT}||_{L^{\infty}}\lesssim e^{-(n-1)T},\\\nonumber ||\frac{N}{n}-1||_{L^{\infty}}\lesssim e^{-2(n-1)T},
||X||_{L^{\infty}}\lesssim e^{-(n-1)T}, ||h^{TT}||_{L^{\infty}}\lesssim e^{-nT},\\\nonumber
||Y^{||}||_{L^{\infty}}\lesssim e^{-2nT}.
\end{eqnarray}
Note an important fact that the asymptotic decay estimates of $\frac{\partial g}{\partial T}$, $(g-\gamma^{\dag})$, and $K^{TT}$ match with the linear decay estimates as expected. Utilizing these asymptotic decay estimates for the relevant fields, we therefore obtain the following theorem regarding the attractor property of the Einstein-$\Lambda$ flow\\
\textbf{Theorem 5:} \textit{Let $(g_{0},K^{TT}_{0})\in B_{\delta}(\gamma_{0},0)\subset H^{s-1}\times H^{s-2}, s>\frac{n}{2}+2$ with $\gamma_{0}\in\mathcal{N}$ and assume the triple $(\gamma_{0},g_{0},K^{TT}_{0})$ satisfies the shadow gauge condition. The Newtonian like time $-\infty<T<\infty$ is defined as the solution of the equation $\partial_{T}=-\frac{\tau^{2}-\frac{2n\Lambda}{n-1}}{\tau}\partial_{\tau}$, with $\tau\in(-\infty,-\sqrt{\frac{2n\Lambda}{n-1}})$ and $\Lambda>0$ being the mean extrinsic curvature (constant) of the Cauchy hypersurface $M$ and the cosmological constant, respectively. Let $T \mapsto (\gamma(T), g(T), K^{TT}(T))$ be the maximal development of the Cauchy problem for the system (\ref{eq:timedef}, \ref{eq:gd1}, \ref{eq:fd1}, \ref{eq:con1}, \ref{eq:con2}, \ref{eq:lapseE},\ref{eq:shiftE}) with shadow gauge condition imposed and initial data $(\gamma_{0},g_{0},K^{TT}_{0})$. Then there exists a $\gamma^{*}\in \mathcal{N}$ and a $\gamma^{\dag}\in \mathcal{M}^{\epsilon}_{-\frac{n-1}{n}}\cap \mathcal{S}_{\gamma}$ such that the triple  $(\gamma, g, K^{TT}))$ flows toward $(\gamma^{*},\gamma^{\dag},0)$ in the limit of infinite time that is 
\begin{eqnarray}
\lim_{T\to\infty}(\gamma(T),g(T),K^{TT}(T))=(\gamma^{*},\gamma^{\dag},0).
\end{eqnarray}
and moreover either $\gamma^{\dag}=\gamma^{*}$ or  $\mathcal{P}\gamma^{\dag}=\gamma^{*}$.
Here $\mathcal{P}$ is the projection operator defined in (\ref{eq:projection}).
}\\
In the limit of infinite time (infinite expansion of the physical metric), the complete solution satisfies 
\begin{eqnarray}
\lim_{T\to\infty}(\gamma(T), g(T), K^{TT}(T), N(T), X(T))=(\gamma^{*},\gamma^{\dag},0,n,0).
\end{eqnarray}
In order to establish the future completeness of the spacetime, we need to show that the length of a timelike geodesic goes to infinity. In other words, the solution of the geodesic equation must exist for an infinite interval of the affine parameter. Let's designate the timelike geodesic by $\mathcal{C}$. The tangent vector $\alpha=\frac{d\mathcal{C}}{d\lambda}=\alpha^{\mu}\partial_{\mu}$ to $\mathcal{C}$ for the affine parameter $\lambda$ satisfies $\hat{g}(\alpha,\alpha)=-1$, where $\hat{g}$ is the spacetime metric. As $\mathcal{C}$ is causal, we may parametrize it as $(T,\mathcal{C}^{i})$, $i=1,2,3$. We must show that $\lim_{T\to\infty}\lambda(T)=+\infty$, that is, 
\begin{eqnarray}
\lim_{T\to\infty}\int_{T_{0}}^{T}\frac{d\lambda}{dT^{'}}dT^{'}=+\infty.
\end{eqnarray}
Noting that $\alpha^{0}=\frac{dT}{d\lambda}$, we must show 
\begin{eqnarray}
\lim_{T\to\infty}\int_{T_{0}}^{T}\frac{1}{\alpha^{0}}dT^{'}=+\infty.
\end{eqnarray}
We follow the method of \cite{andersson2004future} to achieve this. Showing that $|\bar{N}\alpha^{0}|$ is bounded and therefore $\lim_{T\to\infty}\int_{T_{0}}^{T}\bar{N}dT^{'}=+\infty$ is enough to ensure the geodesic completeness. We first show that $|\bar{N}\alpha^{0}|$ is bounded. Let's consider a co-vector field $Z_{\mu}=\bar{N}\delta^{0}_{\mu}$ in local coordinates using the $n+1$ decomposition, where $\delta^{\nu}_{\mu}$ is the Kronecker delta. This shows that $\alpha$ may be expressed as 
\begin{eqnarray}
\alpha=\bar{N}\alpha^{0} Z+W,
\end{eqnarray}
where $W\in\mathfrak{X}(M)$. Noting $\hat{g}(\alpha,\alpha)=-1$, we obtain 
\begin{eqnarray}
|W|^{2}_{\tilde{g}}=\bar{N}^{2}(\alpha^{0})^{2}-1.
\end{eqnarray}
Here $\tilde{g}$ is the induced Riemannian metric on $M$. Clearly we have 
\begin{eqnarray}
|W|^{2}_{\tilde{g}}<\bar{N}^{2}(\alpha^{0})^{2}.
\end{eqnarray}
Let us compute the entity $\frac{d(\bar{N}^{2}(\alpha^{0})^{2})}{dT}$ as follows 
\begin{eqnarray}
\frac{d(\bar{N}^{2}(\alpha^{0})^{2})}{dT}=\frac{d}{dT}(\hat{g}(\alpha,Z))^{2}=\frac{2}{\alpha^{0}}\hat{g}(\alpha,Z)\hat{g}(\alpha,\nabla[\hat{g}]_{\alpha}Z),
\end{eqnarray}
where, we have used the fact that for $\mathcal{C}$ being a geodesic, $\nabla[\hat{g}]_{\alpha}\alpha=0$ and appealed to the Koszul formula for the derivative. Using $\hat{g}(\alpha,Z)=-\bar{N}\alpha^{0}$ and writing the covariant derivative of the spacetime metric as a direct sum of its projection onto the tangent space of $M$ and the second fundamental form of $M$, we obtain
\begin{eqnarray}
\frac{d(\bar{N}^{2}(\alpha^{0})^{2})}{dT}=-2\bar{N}(\alpha^{0}\nabla_{W}\bar{N}-\tilde{K}_{ij}W^{i}W^{j}).
\end{eqnarray}
Decomposing $\tilde{K}=\tilde{K}^{TT}+\frac{\tau}{n}\tilde{g}$ and noting that $\tau<0$, we obtain
\begin{eqnarray}
|\frac{d}{dT}(\ln(\bar{N}^{2}(\alpha^{0})^{2}))|\leq||\nabla\bar{N}||_{L^{\infty};\tilde{g}}+||\bar{N}\tilde{K}^{TT}||_{L^{\infty};\tilde{g}}.
\end{eqnarray}
Now, in the time coordinate $\frac{d\tau}{dt}=1$ and $\frac{-\phi^{2}}{\tau}\frac{d}{d\tau}=\frac{d}{dT}$, the spacetime metric reads 
\begin{eqnarray}
\hat{g}=-\tilde{N}^{2}dt\otimes dt+\tilde{g}_{ij}(dx^{i}+\tilde{X}^{i}dt)\otimes(dx^{j}+\tilde{X}^{j}dt)\\
=-\frac{\tilde{N}^{2}\phi^{4}}{\tau^{2}}dT\otimes dT+\tilde{g}_{ij}(dx^{i}-\frac{X^{i}\phi^{2}}{\tau}dT)(dx^{j}-\frac{X^{j}\phi^{2}}{\tau}dT),
\end{eqnarray}
and therefore $\bar{N}=\frac{\tilde{N}^{2}\phi^{4}}{\tau^{2}}$. Now, we utilize the estimate obtained for the lapse function and the transverse-traceless second fundamental form. Note that these fields are not dimensionless and therefore we need to multiply them with suitable powers of $\frac{1}{\phi}\sim e^{T}$(\ref{eq:scaling}) to extract the dimensionless part. We obtain, $\bar{N}=\frac{\tilde{N}\phi^{2}}{|\tau|}=\frac{N}{|\tau|}$ and $\tilde{K}^{TT}=\frac{1}{\phi}K^{TT}$, where $N$ and $K^{TT}$ are dimensionless. Utilizing the estimates $||\frac{N}{n}-1||_{H^{s}}\lesssim e^{-2T}$ and $||K^{TT}||_{H^{s-2}}\lesssim e^{-T}$ (for $s>\frac{n}{2}+2$, from Sobolev embedding on a compact domain, bounded $H^{s}$ (resp. $H^{s-2}$) norm of N (resp. $K^{TT}$) implies bounded $L^{\infty}$ norm, $||\nabla \bar{N}||_{L^{\infty};\tilde{g}}:=\sup_{M}\sqrt{\tilde{g}^{ij}\nabla_{i}\bar{N}\nabla_{j}\bar{N}}$, $||K^{TT}||_{L^{\infty};\tilde{g}}:=\sup_{M}\sqrt{\tilde{g}^{ij}\tilde{g}^{kl}K^{TT}_{ik}K^{TT}_{jl}}$), we observe that the following holds 
\begin{eqnarray}
|\frac{d}{dT}(\ln(\bar{N}^{2}(\alpha^{0})^{2}))|\lesssim e^{-2T}
\end{eqnarray}
as $T\to\infty$ and therefore $\bar{N}^{2}(\alpha^{0})^{2}$ is bounded, i.e.,
\begin{eqnarray}
\bar{N}^{2}(\alpha^{0})^{2}(T)\leq C
\end{eqnarray}
for some $C<\infty$. Therefore, we need to show that $\lim_{T\to\infty}\int_{T_{0}}^{T}\bar{N}dT^{'}=+\infty$ in order to finish the proof of timelike geodesic completeness. 
Once again using $\bar{N}=\frac{N}{|\tau|}$, the estimate 
(\ref{eq:lapseestimate}) 
\begin{eqnarray}
0<\frac{1}{\sup(|K^{TT}|^{2})+\frac{1}{n}}\leq N\leq n,
\end{eqnarray}
and $||K^{TT}||_{H^{s-1}}\lesssim e^{-T}$ as $T\to\infty$, we clearly see that $N$ is bounded from below by a strictly positive number and therefore 
\begin{eqnarray}
\lim_{T\to\infty}\int_{T_{0}}^{T}\bar{N}dT^{'}=+\infty.
\end{eqnarray}
Therefore, the solution of the geodesic equation must exist for a  semi-infinite  interval  of  the  affine  parameter. This completes the proof timelike geodesic completeness. The case of null geodesics can be handled exactly the same way.\\ 
This proves the future completeness of this family of spacetimes. Previous analysis together with the attractor property stated in theorem $5$ yields the following global existence theorem.\\
\textbf{Theorem 6:} \textit{Let $\mathcal{N}$ be the integrable deformation space of $\gamma_{0}$. Then $\exists \delta>0$ such that for any $(g(T_{0}),K^{TT}(T_{0}))\in B_{\delta}(\gamma_{0},0)\subset H^{s}\times H^{s-1}$, with the triple $(\gamma_{0},g(T_{0}),K^{TT}(T_{0}))$ satisfying the shadow gauge condition, the Cauchy problem for the re-scaled Einstein-$\Lambda$ system with constant mean extrinsic curvature (CMC) and spatial harmonic (SH) gauge is globally well posed to the future and the space-time is future complete.}\\
One might recall for $\Lambda=0$ case that in order to obtain a sharp decay of the energy, \cite{andersson2011einstein} added a correction term to the ordinary wave equation type energy. One could naturally ask whether introduction of such a correction term is necessary, that is, can one just initially bound the energy and later use the iteration scheme to yield the improved decay. However, there is a major difference between the $\Lambda=0$ and $\Lambda>0$ case. In $\Lambda>0$ case, we have the `good' term $\frac{\phi}{\tau}$ (which decays as $e^{-T}$ as $T\to\infty$) which plays an extremely important role in obtaining the decay estimate. Roughly, one sees in the second iteration from equation (\ref{eq:improve}) that given the boundedness of $||u||_{H^{s-1}}$, and $||v||_{H^{s-2}}\lesssim e^{-T/2}$, the terms multiplied by $\frac{\phi}{\tau}$ adds an extra factor of $1$ in the decay estimate of $||v||^{2}_{H^{s-2}}$. In case of $\Lambda=0$, the absence of $\frac{\phi}{\tau}$ would lead to a circular argument in the first step and prohibit one to obtain a decay estimate. Therefore, introduction of a corrected energy becomes essential. 
Physically, this behaviour is expected since addition of a positive cosmological constant yields an accelerated expansion which is expected to destroy the perturbations. Sufficiently small data and the positivity of the spectrum of the operator $\mathcal{L}_{g,\gamma}$ are sufficient to establish the asymptotic stability. 

\begin{center}
\begin{figure}
\begin{center}
\includegraphics[width=13cm,height=60cm,keepaspectratio,keepaspectratio]{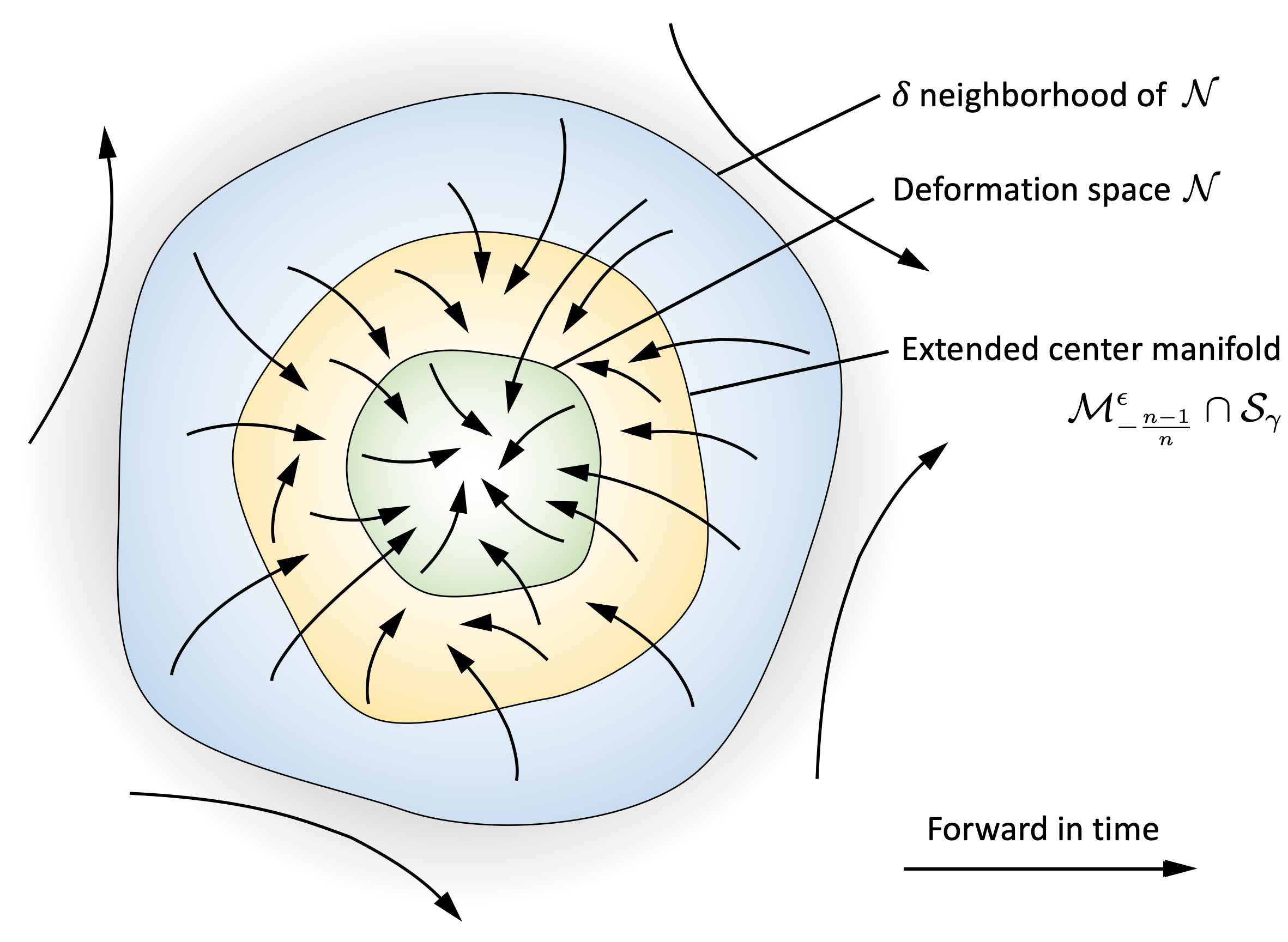}
\end{center}
\begin{center}
\caption{Schematics depicting the dynamics of the Einstein-$\Lambda$ flow in the present context. Metrics lying sufficiently close to the center manifold flow in infinite time towards the extended centre manifold i.e., $||g-\gamma||\lesssim e^{-2T}$ as $t\to\infty$. For perturbations tangent to the centre manifold trivially decays such that the metrics simply flow from one point to another in the centre manifold. Sufficiently large perturbations are not controlled by the Einstein-$\Lambda$ flow in the present context.}
\label{fig:pdf}
\end{center}
\end{figure}
\end{center}
\section{Concluding remarks}
We have proved a global existence theorem for sufficiently small however fully nonlinear perturbations of a family of background solutions (conformal spacetimes)  derived from Einstein's equations in the presence of a positive cosmological constant. However, our current result assumes the initial data to be within a small neighbourhood (in the proper function space setting) of the background solutions, and as such is very far from stating a global result for arbitrarily large data. Nevertheless, several interesting physically relevant features are revealed through the analysis. Firstly, consider the case of $n=3$ i.e., the physical spacetimes. Following Mostow rigidity, the Einstein moduli space consists of a single point that corresponds to the hyperbolic geometry. Therefore, the background spacetime is essentially foliated by compact hyperbolic manifolds which are locally homogeneous and isotropic thereby conforming to the cosmological principle (astronomical observations that motivate the cosmological principle are local). Notice that if we start with an inhomogeneous anisotropic initial spatial metric sufficiently close to the background (in suitable function space settings), this metric does not only remain within a bounded neighbourhood of the background, it actually approaches the space of metrics with constant scalar curvature sufficiently close and containing the Einstein structure. In addition, recent astronomical observations support the claim that the spatial slice of the physical universe is indeed negatively curved (slightly). This notion together with our result opens up the possibility of a rather exotic spatial topology of the universe (hyperbolic 3-manifolds are topologically rich). Of course, our result can only provide an indication of such a claim being true. A complete analysis would entail inclusion of suitable matter sources on the one hand and treating arbitrarily large data perturbations on the other. While \cite{rodnianski2009stability} proved the non-linear stability of the small perturbations to the FLRW background solutions in the presence of irrotational perfect fluid and a positive cosmological constant on $T^{3}\times R$ (and therefore flat spatial topology), such non-linear stability of spacetimes foliated by compact hyperbolic manifolds (which is of physical interest) is still open and the flat model is not likely a viable candidate for the physical universe. A linear stability of spacetimes foliated by compact hyperbolic spatial slices in the presence of a perfect fluid and $\Lambda>0$ (compact variants of the $k=-1$ FLRW model) is under preparation by the current author. \cite{mondal2019asymptotic} has recently studied the asymptotic behavior of a universe filled with matter sources satisfying suitable energy conditions and a positive cosmological constant based on a monotonic decay property of a suitably constructed Lyapunov function. While such a Lyapunov function can treat arbitrarily large data, it can only control the lowest order norm ($H^{1}\times L^{2}$) of the data and therefore, is unable to state a result regarding global existence (at the physically interesting classical solutions we are interested). In a sense, any definite result regarding the evolution of the physical universe requires global existence (or blow up in finite time) for large data perturbations.

This question of global existence is far from obvious and an extremely important (and difficult) open problem in classical general relativity. Global existence is known to be violated for some known examples of spacetime via formation of black holes. These include Schwarzschild, Reissner-Nordstr\"om, Kerr spacetimes, where a true curvature singularity occurs within the event horizon of the black hole. Even in the vacuum case, pure gravity could `blow up' (curvature concentration) i.e., gravitational singularities could prevent global existence or the spacetime could simply lose the global hyperbolicity through formation of Cauchy horizons (Taub-NUT spacetimes for example). Of course, such issues lead to the fundamental question of the \textbf{Cosmic Censorship} conjecture \cite{penrose1999question}, which still remains open. Available results related to the global existence address rather special cases such as spacetimes with non-trivial symmetry groups \cite{choquet2003nonlinear, choquet2001future, chrusciel1990strong} (and which therefore are not generic) or where a certain smallness condition on the initial data is assumed \cite{andersson2011einstein,andersson2004future}. There is however a rather ambitious program under development by Moncrief to control the pointwise ($L^{\infty}$ norm) behaviour of the spacetime curvature through the use of \textit{light cone estimates} \cite{moncrief2005integral}. This method is recently applied to establish the global existence of Yang-Mills and Klein-Gordon fields in curved spacetimes by the current author and Moncrief. Application of this light cone estimates to establish the small data global existence for certain background solutions is currently under investigation by the current author.\\
An interesting question which arises through our result is what role can these non-trivial exotic topologies ($H^{3}/\Gamma,~\Gamma\in SO^{+}(3,1)$ proper, discrete, and torsion free) play in answering the question of global existence or finite time blow up. In a sense, can the topological properties of these interesting manifolds have any control on the fundamental question of large data global existence (or finite time blow up)? Can the interrelation between the dynamics and topology provide crucial information to handle the issue of global existence or finite time blow up, at least in $n=3$ dimensions?
\section{Acknowledgement}
P.M would like to thank Prof. Vincent Moncrief for numerous useful discussions related to this project and for his help improving the manuscript. P.M would also like to thank the anonymous referees for pointing out important points. This work was supported by Yale university.

\author{Puskar Mondal$^{1}$}
\address{$^1$ Department of Planetary Sciences, Yale University}
\ead{puskar.mondal@yale.edu}

\end{document}